\documentclass[a4paper,11pt]{article}
\usepackage{imakeidx}
\makeindex[columns=2, title=Alphabetical Index, intoc]
\usepackage{jheppub}
\usepackage{lineno}
\usepackage{amsmath}
\usepackage{amssymb}
\usepackage{nicefrac}
\usepackage[dvipsnames]{xcolor}
\usepackage{mathtools}
\usepackage{cleveref}
\usepackage{slashed}
\usepackage{multirow}
\usepackage{subcaption}
\usepackage{placeins}
\usepackage{stmaryrd}
\usepackage{capt-of}
\usepackage{tikz-feynman,contour}
\usepackage{tikz}
\usepackage{wasysym}
\usepackage{fontawesome}
\usepackage[export]{adjustbox}
\usepackage{rotating}
\usepackage{comment}
\usepackage{tabularx}
\usepackage{booktabs}
\usepackage{ragged2e}
\usepackage{array}
\usepackage{xparse}

\newcolumntype{Y}{>{\centering\arraybackslash}X}
\newcolumntype{T}{>{\centering\arraybackslash}p{3.4cm}}

\usetikzlibrary{decorations.pathmorphing,arrows.meta,positioning}
\usetikzlibrary{decorations.markings, arrows.meta}
\usetikzlibrary{matrix, positioning, calc, decorations.pathreplacing}

\tikzset{
  edgelabel/.style={
    circle,
    fill=black,
    text=white,
    font=\tiny,
    minimum size=4pt,
    inner sep=0pt
  }
}

\newcommand{\ie}{\textit{i.e.}~}
\newcommand{\eg}{\textit{e.g.}~}
\newcommand{\cfit}{\textit{cf.} }

\newcommand{\as}{\alpha_s}
\newcommand{\asb}{\alpha_{s,b}}
\newcommand{\asnf}{\alpha_{s;n_f}}
\newcommand\MSbar{\overline{\rm MS}}
\newcommand\Sep{{S_\epsilon}}
\newcommand\Ca{C_A}
\newcommand\Cf{C_F}
\newcommand\tr{T_R}
\newcommand\nf{n_{f}}
\newcommand\nh{n_{h}}
\newcommand{\xbj}{x^\mathrm{bj}}
\newcommand{\ep}{\epsilon}
\newcommand{\Dp}{\mathcal{D}}
\newcommand{\DD}{D}
\newcommand{\sd}[1]{\slashed{#1}}
\newcommand{\fvec}{\text{\boldmath$f$}}
\newcommand{\gvec}{\text{\boldmath$g$}}
\newcommand{\hvec}{\text{\boldmath$h$}}
\newcommand{\Avec}{\text{\boldmath$A$}}
\newcommand{\Bvec}{\text{\boldmath$B$}}
\newcommand{\Hvec}{\text{\boldmath$H$}}
\newcommand{\Li}{\mathrm{Li}}
\newcommand{\LqNSs}{\mathcal L^{(s)}_{q,{\rm NS}}}
\newcommand{\LqORqbNSs}{\mathcal L^{(s)}_{q(\bar q),{\rm NS}}}
\newcommand{\LqbORqNSs}{\mathcal L^{(s)}_{\bar q(q),{\rm NS}}}
\newcommand{\LgHQ}{\mathcal L^{(c)}_{g}}
\newcommand{\Op}[2]{\mathrm{#1}\!\left[#2\right]}

\definecolor{darkgreen}{rgb}{0,0.4,0}
\definecolor{darkred}{rgb}{0.8,0,0}

\makeatletter
\newcommand\ps@indexpagestyle{
  \renewcommand\@oddfoot{\hfill-- \thepage\ --\hfill}
  \renewcommand\@oddhead{}
}

\preprint{
\begin{flushright}
OUTP-26-04P
\end{flushright}
}

\allowdisplaybreaks

\makeatother
\indexsetup{firstpagestyle=indexpagestyle}

\def\OX{Rudolf Peierls Centre for Theoretical Physics, University of Oxford, Clarendon Laboratory, Parks
Road, Oxford OX1 3PU}
\def\WAD{Wadham College, University of Oxford, Parks Road, Oxford OX1 3PN, UK}
\def\ETH{Institute for Theoretical Physics, ETH Zurich, 8093 Z\"urich, Switzerland}

\author[a,b]{Fabrizio Caola,}
\author[c]{Giulio Gambuti,}
\author[a,b]{Martin Link}

\emailAdd{fabrizio.caola@physics.ox.ac.uk}
\emailAdd{ggambuti@phys.ethz.ch}
\emailAdd{martin.link@physics.ox.ac.uk}

\affiliation[a]{\OX}
\affiliation[b]{\WAD}
\affiliation[c]{\ETH}

\title{ Analytic results for heavy-quark contributions to
  charged-current DIS at NNLO }

\abstract{ We present analytic results for the
  next-to-next-to-leading-order QCD corrections to heavy-quark
  production in charged-current deep-inelastic scattering, retaining
  the exact dependence on the charm quark mass. We compute the
  complete partonic coefficient functions for the structure functions
  $F_2$, $F_L$, and $F_3$ in the quark and gluon channels, including
  contributions with up to three heavy quarks in the final state.
  Working within the reverse-unitarity framework, we use
  integration-by-parts and canonical differential-equations techniques
  to express all contributions with at most two final-state heavy
  quarks in terms of manifestly real Goncharov polylogarithms which
  allow for a robust and efficient numerical evaluation. The
  three-heavy-quark contribution involves elliptic structures for
  which we give a general representation in terms of Chen iterated
  integrals, as well as expressions in terms of rapidly convergent
  expansions that are valid in the perturbative $Q\gtrsim 5~{\rm
    GeV}$ region and also allow for a flexible and fast numerical
  evaluation. We validate our results against known exact results at
  lower orders, massless NNLO coefficient functions, and existing
  leading-power expansions in the asymptotic limit where the virtuality
  $Q$ is much larger than the charm mass.}

\begin{document}

\maketitle
\raggedbottom
\allowdisplaybreaks



\section{Introduction}
\label{sec:intro}
Deep Inelastic Scattering (DIS) has played a pivotal role in probing
the partonic structure of matter, providing crucial tests of Quantum
Chromodynamics (QCD) and precise determinations of parton distribution
functions (PDFs).  Accurate PDFs are a critical ingredient for
essentially all analyses at the Large Hadron Collider (LHC), where
they often constitute one of the dominant theoretical uncertainties,
see \eg refs~\cite{LH23,Chiefa:2025loi} for a recent discussion.  A
central objective of the global PDFs program in the coming years is to
reduce these uncertainties to the percent level or below, a
requirement that is essential both for high-precision Standard Model
(SM) measurements~\cite{mariarev,Azzi:2019yne,Cepeda:2019klc} and for
maximising the discovery potential of searches for physics beyond the
Standard Model (BSM)~\cite{mandyrev}. To achieve this, a lot of effort
has been devoted to improving current PDFs determinations on several
fronts, including both PDF-fitting methodologies and theoretical
developments, see \eg refs~\cite{NNPDF-rev,CT-rev,MSHT-rev,ABMP-new} for a
recent review.

Although LHC data have become increasingly important in modern PDFs
determinations, DIS measurements continue to play a key role, see \eg
refs~\cite{Forte:2013wc,Gao:2017yyd}.  In particular, charged-current
(CC) DIS data provide clean access to quark flavour separation, and an
important handle on the strange-quark distribution.  The strange PDF
directly impacts the extraction of key SM parameters like the $W$ mass
and the Weinberg angle~\cite{Bagnaschi:2019mzi} and is also relevant
for BSM searches in which precise control of flavour-dependent
backgrounds is required. The strange distribution can be constrained
by neutrino-induced DIS experiments (such as CCFR~\cite{CCFR},
CHORUS~\cite{CHORUS}, NuTeV~\cite{NuTeV:2001dfo,NuTeV},
NOMAD~\cite{NOMAD}, SND@LHC~\cite{SNDLHC:2023pun,Acampora_2024},
SHiP~\cite{Alekhin:2015byh,SHiP:2015vad},
FASER~\cite{FASER:2022hcn,FASER:2023zcr}, FPF~\cite{Feng:2022inv}),
future EIC measurements~\cite{eic_yellow}, as well as by LHC
measurements of \eg associated $W$+charm production
\cite{ATLAS:2014jkm,CMS:2013wql,CMS:2018dxg,ATLAS:2017irc}, whose
complementarity is by now well established, see \eg
refs~\cite{Faura:2020oom,Cruz-Martinez:2023sdv}.

In this context, good theoretical control of charm production in CC
DIS plays a particularly important role.  This process is
intrinsically sensitive to heavy-quark mass effects, which are
especially relevant in the low- and intermediate-$Q^2$ region,
$(Q^2\lesssim 100~{\rm GeV}^2)$. This region is relevant for the
neutrino DIS data mentioned above. In this regime, robust control
over finite-mass effects is highly desirable.
Over the past decade, significant progress has been achieved in the
understanding of heavy-flavour contributions to DIS, particularly in
the asymptotic region $Q^2 \gg m_c^2$, where mass effects factorise
into logarithmic corrections that can be computed perturbatively
\cite{Buza:1996wv,Bierenbaum:2009mv}. In these approximations,
massive corrections are known to $\mathcal O(\as^3)$~\cite{Ablinger:2010ty,Ablinger:2014lka,
Ablinger:2014nga,Ablinger:2014vwa,Behring:2014eya,Behring:2015roa,
Behring:2015zaa,Behring:2016hpa,Blumlein:2016xcy,
Ablinger:2017xml,Ablinger:2018brx,Ablinger:2019etw,
Ablinger:2019gpu,Ablinger:2020snj,Behring:2021asx,
Blumlein:2021xlc,Ablinger:2022wbb,Ablinger:2023ahe,
Ablinger:2024xtt,Ablinger:2025nnq,Ablinger:2025awb,
Ablinger:2025joi}, see ref.~\cite{Ablinger:2024qxg}
for a recent overview and previous results.

In contrast, comparatively less emphasis has been placed on the
phenomenologically important region of intermediate and low $Q^2$,
which is directly relevant for neutrino DIS observables. While the
exact next-to-leading order (NLO) corrections for massive charm
production in CC DIS have been known for a long time
\cite{Gottschalk:1980rv,Gluck:1996ve,Blumlein:2011zu}\footnote{More
recently, NLO results matched with parton showers became available as
well~\cite{Buonocore:2024pdv,Meinzinger:2025pam}.}, the corresponding
next-to-next-to-leading order (NNLO) corrections have so far only been
available in numerical form
\cite{Berger:2016inr,Gao:2017kkx}.\footnote{ For a discussion of the
exact $\mathcal O(\as^2)$ analytic results for the NC case, see
ref.~\cite{Blumlein:2019qze}.}  Although such results are perfectly
adequate for fixed-order phenomenological studies, their numerical
nature makes them not ideal to incorporate into global PDFs fits and
complicates systematic studies of mass effects and power-suppressed
contributions, see \eg ref.~\cite{Risse:2025smp} for a recent
discussion.

In this work, we address this issue and present a fully analytic
computation of the NNLO QCD corrections to heavy-flavour
charged-current DIS with exact dependence on the heavy-quark mass. Our
results provide reasonably compact analytic expressions for all the
relevant coefficient functions, valid over the full kinematic range in
$Q^2$.
The analytic calculation of massive corrections is challenging, as it
involves multi-scale problems with complicated loop integrals and
functions beyond generalised polylogarithms. Recent progress in
computational techniques for multi-loop Feynman integrals has opened
the door to systematically dealing with these issues, and was
fundamental for our calculation.  While in what follows we will limit
ourselves to mentioning the results that we have explicitly used in
our calculation, we point the reader to \eg
ref.~\cite{Travaglini:2022uwo} and references therein for a broader
overview of recent developments.

The remainder of this paper is organised as follows: in
\cref{sec:defs} we fix our notation by briefly describing the
theoretical framework of our calculation. In \cref{sec:bareci} we
discuss the calculation of the bare coefficient functions through 
NNLO. Though the LO and NLO results are well known, we review them in
some detail to make our discussion self-contained and to present some
of the technical challenges appearing in the NNLO calculation in a
simplified context. In \cref{sec:MIs} we describe in detail the
analytic calculation of the master integrals appearing in our
calculation, and in \cref{sec:ren} we discuss the UV and collinear
renormalisation of our result.  In \cref{sec:results} we present
our results for the renormalised coefficient functions and discuss
their main features. Finally, we conclude in \cref{sec:conclusions}.


\section{Process definition and kinematics}
\label{sec:defs}
\begin{figure}[tb]
  \centering
  \begin{tikzpicture}[line width=1pt, >=stealth]

    \draw[
    -,
    postaction={
      decorate,
      decoration={
        markings,
        mark=at position 0.55 with {\arrow[scale=1.2]{>}}
      }
    }
    ]  (-3,2) -- (0,1);
    \node[above] at (-1.5,.8) {$k$};
    \node[left] at (-3,2) {$\mathbf{l}$};
    
    \draw[
      -,
    postaction={
      decorate,
      decoration={
        markings,
        mark=at position 0.55 with {\arrow[scale=1.2]{>}}
      }
    }
    ] (0,1) -- (3,2);
    \node[above] at (1.5,.8) {$k'$};
    \node[right] at (3,2) {$\mathbf{l}'$};

    \draw[decorate, decoration={snake, amplitude=2pt, segment length=6pt}]
      (0,1) -- (0,-0.5);
    \node[left] at (0,0.3) {$q$};

    \filldraw[black] (0,-0.5) circle (0.3);
    
    \draw[
    -,
    postaction={
      decorate,
      decoration={
        markings,
        mark=at position 0.55 with {\arrow[scale=1.2]{>}}
      }
    }
    ]  (-3,-1.5) -- (0,-0.5);
    \node[above] at (-1.5,-1.5) {$P$};
    \node[left] at (-3,-1.5) {$\mathbf{P}$};

    \draw[-] (0,-0.5) -- (3,0.0);
    \draw[-] (0,-0.5) -- (3,-0.5);
    \draw[-] (0,-0.5) -- (3,-1.0);

    \node[right] at (3,-0.5) {$\Biggr\}\mathbf{X}$};
    \node[above] at (1.6,-1.3) {$P_X$};

  \end{tikzpicture}
  \caption{Kinematic definitions for the DIS process}
  \label{fig:disdia}
\end{figure}
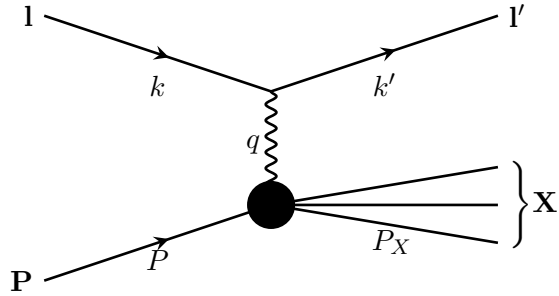

We consider the inclusive DIS process
\begin{equation}
  \mathbf{l}(k) + \mathbf{P}(P) \to \mathbf{l}'(k') + \mathbf{X}(P_X),
  \label{eq:def}
\end{equation}
mediated by a vector boson $\mathbf{V}$. In \cref{eq:def},
$\mathbf{l}$($\mathbf{l}'$) is an incoming (outgoing) lepton,
$\mathbf{P}$ is a proton and $\mathbf{X}$ represents the hadronic
remnants, see fig.~\ref{fig:disdia}.
We define the momentum of the vector boson $q$ as
\begin{equation}
  q = k-k',
\end{equation}
and introduce the standard DIS variables
\begin{equation}
  Q^2 = -q^2,\quad\quad \xbj = \frac{Q^2}{2P\cdot q},\quad\quad
  y = \frac{q\cdot P}{k\cdot P},
  \label{eq:DISvar}
\end{equation}
with $Q^2>0$ and $0<\xbj\le 1$. Neglecting the proton and lepton masses,
the squared centre-of-mass energy ($s$) and the squared invariant mass
of the hadronic final state ($P_X^2$) can be written in terms of these
variables as
\begin{equation}
  s \equiv (P+k)^2 = \frac{Q^2}{\xbj y},
  \quad\quad\quad  P_X^2 = Q^2\frac{(1-\xbj)}{\xbj}.
  \label{eq:s_w}
\end{equation}

As it is standard for DIS, we write the differential cross section $d\sigma$
as the product of a leptonic ($L^{\mu\nu}$) and a hadronic ($W^{\mu\nu}$)
tensor
\begin{equation}
\label{eq:dsigma}
d\sigma = \frac{e^4}{2s} \left[\mathcal{N}_\mathrm{pol} L_{\mu\nu}
  \frac{d^4 k'}{(2\pi)^3}\delta_+(k'^2)
  \right]
  \frac{1}{(Q^2 + m_V^2)^2} d^4q\,\delta^{(4)}(q+k'-k) \times
  \left[ (4\pi) W^{\mu\nu}\right],
\end{equation}
where $m_V$ is mass of the vector boson.
Assuming a generic Feynman rule for vector-boson/fermions interactions
\begin{equation}\label{eq:W_vertex}
  i\Gamma^\mu = ie\gamma^\mu (c_V-c_A\gamma_5),
\end{equation}
with $e$ the electric coupling, the leptonic tensor is defined
as
\begin{equation}
  \label{eq:lepten}
  \begin{split}
  L^{\mu\nu} & = {\rm Tr}[\slashed k' \gamma^{\mu}(c_V-c_A \gamma_5)
    \slashed k \gamma^{\nu}(c_V-c_A\gamma_5)] =
  \\
  &=
  4 (c_V^2+c_A^2)(k^\mu k'^\nu + k^\nu k'^\mu - g^{\mu\nu}k\cdot k')
  - 8 i c_V c_A \ep(k,k',\mu,\nu).
  \end{split}
\end{equation}
For CC DIS, which is the focus of this work, one has
\begin{equation}
  c_V = c_A = \frac{\sqrt{2}}{4\sin\theta_W},
  \label{eq:cvca}
\end{equation}
where $\theta_W$ is the weak mixing angle and where we assumed a diagonal
CKM matrix.\footnote{We will show how to reinstate the full CKM dependence
in \cref{sec:results}.}
In \cref{eq:lepten}, $\ep(\alpha,\beta,\mu,\nu)$ is the Levi-Civita
tensor, which satisfies
\begin{equation}
  {\rm Tr}[\slashed p_1 \slashed p_2 \slashed p_3 \slashed p_4 \gamma_5]=
  4i \ep(p_1,p_2,p_3,p_4).
  \label{eq:g5}
\end{equation}
Note that \cref{eq:lepten} is the leptonic tensor for an incoming
lepton. When considering an incoming anti-lepton (such as $e^+$), we
need to swap $\mu$ and $\nu$, leading to a flipped sign in the
axial-vector component.  The factor $\mathcal{N}_\mathrm{pol}$ in the
first square bracket of \cref{eq:dsigma} is either $1$ or $1/2$,
depending on whether or not we need to average over the initial-state
lepton polarisations. In the case of neutrino-induced DIS
$\mathcal{N}_\mathrm{pol}=1$, in the case of unpolarised incoming
electrons $\mathcal{N}_\mathrm{pol}=1/2$.

The hadronic tensor $W^{\mu\nu}$ is defined as usual as the spin- and
colour-summed/averaged cross section for the hadronic process
$\mathbf{V}(q)+\mathbf{P}(P) \to \mathbf{X}(P_X)$, without any flux
factor and divided by $ 4\pi e^2 $. The $4\pi$ factor is removed from
$W$ for convenience, and accounted for explicitly in
\cref{eq:dsigma}. Since the hadronic tensor is conserved, it can be
expressed in full generality in terms of three form factors
$F_i^{\mathbf{V}}$:
\begin{equation}
  W^{\mu\nu} =
  \big(c_V^2+c_A^2\big)\left[
      T_1^{\mu\nu} F^\mathbf{V}_1 +
      T_2^{\mu\nu} F^\mathbf{V}_2 \right]
  +2 c_V c_A\, T_3^{\mu\nu} F^\mathbf{V}_3.
  \label{eq:wdef}
\end{equation}
The three independent tensor structures $T_i$ are defined as
\begin{equation}
  \begin{split}
    T_1^{\mu\nu} = -g^{\mu\nu} + \frac{q^\mu q^\nu}{q^2},\quad
    T_2^{\mu\nu} = \frac{\overline{P}^\mu \overline{P}^\nu}{P\cdot q},
    \quad T_3^{\mu\nu} = -i \frac{\ep(P,q,\mu,\nu)}{2 P\cdot q},
  \end{split}
  \label{eq:Tdefs}
\end{equation}
where $\overline{P}^\mu = P^\mu - P\cdot q/q^2 \, q^\mu$. In
\cref{eq:wdef}, $F_i^\mathbf{V} \equiv F_i^\mathbf{V}(\xbj,Q^2)$ and
for the remainder of this paper, we will keep the $\xbj$ and $Q^2$
dependence implicit.  The superscript denotes the exchanged vector
boson.  We also find it convenient to work with the longitudinal
structure function $F_L^\mathbf{V}\equiv F_2^\mathbf{V}-2\xbj
F_1^\mathbf{V}$ instead of $F_1^\mathbf{V}$. In terms of these
structure functions, the double-differential DIS cross section for an
incoming lepton (anti-lepton) can be written as
\begin{equation}
  \label{eq:discross}
  \begin{split}
    \frac{d\sigma^{l(\bar{l})}}{d \xbj dQ^2} =& \frac{4 \pi
      \mathcal{N}_\mathrm{pol} \alpha^2}{\xbj (Q^2+m_V^2)^2} \times
    \\ & \left[
      \big(c_V^2+c_A^2\big)^2\left(\big[1+(1-y)^2\big]F_2^\mathbf{V} -
        y^2 F_L^\mathbf{V}\right) \pm 4 c_A^2 c_V^2 \,y(2-y) \xbj
      F_3^\mathbf{V} \right],
  \end{split}
\end{equation}
where $\alpha = {e^2}/{4\pi}$ is the fine-structure constant.
Using the explicit form for the vector and axial couplings \cref{eq:cvca},
we can write the double-differential DIS cross section for
incoming neutrinos/electrons as
\begin{equation}
  \label{eq:discrossccdis}
  \begin{split}
    \frac{d\sigma^{\nu(\bar{\nu})}}{d \xbj dQ^2} =&
    \frac{2\mathcal N_{\rm CC}}{\xbj (Q^2+m_V^2)^2}
    \times\left[\left[1+(1-y)^2\right] F_2^{W^\pm}-y^2  F_L^{W^\pm}
      \pm   y(2-y) \xbj  F_3^{W^\pm}\right],\\
    \frac{d\sigma^{e^-(e^+)}}{d \xbj dQ^2} =&
    \frac{\mathcal N_{\rm CC}}{\xbj (Q^2+m_V^2)^2}
    \times\left[\left[1+(1-y)^2\right] F_2^{W^\mp}-y^2  F_L^{W^\mp}
      \pm   y(2-y) \xbj  F_3^{W^\mp}\right],
  \end{split}
\end{equation}
with $\mathcal N_{\rm CC} =
\pi\alpha^2/(8\sin^4\theta_W)$.\footnote{We note that in the
literature it is often customary to use a different normalisation for
the neutrino-induced DIS structure functions. The relation between the
$F_i^{\nu(\bar\nu)}$ structure functions defined \eg
in~\cite{Ellis:1996mzs} and our $F_i^{W^\pm}$ reads $F_i^{\nu(\bar\nu)} =
2 F_i^{W^\pm}$}

Up to higher-twist corrections, the structure functions $F_i^{W^\pm}$
can be written as a convolution between non-perturbative parton
distribution functions $f_a$ and short-distance coefficient functions
$c_{i;a}$.  To calculate the coefficient functions it is
sufficient to consider the partonic tensor $\hat{W}^{\mu\nu}$, defined
analogously to its hadronic counterpart $W^{\mu\nu}$ as the spin- and
colour-summed/averaged cross section for the partonic process
$\mathbf{V}(q)+\mathbf{f}_a(p) \to \mathbf{X}(p_X)$, divided by $4 \pi
e^2$.
For our calculation, the final state $\mathbf{X}$ contains up to three
massive charm quarks, with mass $m_q$. We find it convenient to parametrise
our kinematics using the following dimensionless quantities
\begin{equation}
  x = \frac{Q^2}{2 p\cdot q}, \quad\quad\quad z \equiv
  \frac{m_q^2}{Q^2}.
\end{equation}
Writing our coefficient functions in terms of these partonic variables
such that $c_{i;a}=c_{i;a}(x,z,Q^2)$, the convolutions between PDFs
and coefficient functions take the explicit form
\begin{equation}
  \label{eq:explconv}
  \left[f_a \otimes c_{i;a}\right](\xbj) =
  \int_{\xbj}^1 \frac{dx}{x} 
  f_a\left(\frac{\xbj}{x},Q^2\right) 
  c_{i;a}(x,z,Q^2)
  .
\end{equation}
The precise relation between structure functions $F_i$ and coefficient
functions $c_i$ reads\footnote{The different treatment of $F_3$ stems
from the extra power of $\xbj$ in front of this structure function in
\cref{eq:discross}.}
\begin{equation}
  F_{2,L} = \xbj\sum_a f_a \otimes  c_{2,L;a} ,
  \quad\quad
  F_{3} =\sum_a f_a \otimes  c_{3;a},
  \label{eq:FfromC}
\end{equation}
see \eg ref.~\cite{Ellis:1996mzs} for details.

When massive quarks are present in the final state, the partonic
invariant mass must be greater than the (multi)-quark threshold,
$(p+q)^2 \ge (n m_q)^2$ where $n$ is the number of final-state massive
quarks. Because of these different thresholds, we find it convenient
to treat contributions with different numbers of massive final-state
quarks separately.
From now on, we denote in square brackets the number of final-state
massive quarks, such that the coefficient functions are written
\begin{equation}
  c_{i;a} = \sum_n c_{i;a}^{[n]}.
\end{equation}
Due to the mass threshold, each of the $c_{i;a}^{[n]}$ vanish unless
\begin{equation}
  x \le \frac{1}{1+n^2 m_q^2/Q^2} = \frac{1}{1+n^2 z}.
  \label{eq:zdef}
\end{equation}
It is then convenient to introduce rescaled variables $x_{r,n}$ and
$\xbj_{r,n}$ 
\begin{equation}
  x^{(\mathrm{bj})}_{r,n} \equiv {x^{(\mathrm{bj})}}\big(1+n^2 z\big),
  \label{eq:xrndef}
\end{equation}
which satisfy
\begin{equation}
  0\le\xbj_{r,n}\le x_{r,n}\le 1.
\end{equation}
From now on, unless specified otherwise we will assume that
$c_{i;a}^{[n]}$ is a function of $z$, $Q^2$, and the \emph{rescaled}
variable $x_{r,n}$, \ie $c^{[n]}_{i;a} = c_{i;a}(x_{r,n},z,Q^2)$. In
terms of these rescaled variables, the relations \cref{eq:FfromC}
read
\begin{equation}
  F_{2,L} = \sum_{n,a} \xbj_{r,n}
  \left[f_a \otimes c_{2,L;a}^{[n]} \right]\left(\xbj_{r,n}\right),
  \quad\quad
  F_{3} = \sum_{n,a}
  \left[f_a \otimes c_{3;a}^{[n]}\right]\left(\xbj_{r,n}\right),
  \label{eq:cdefs}
\end{equation}
with
\begin{equation}
  \label{eq:rescaledconv}
  \left[ f_a \otimes c_{i;a}^{[n]}\right](\xbj) =
  \int_{\xbj_{r,n}}^1 \frac{dx_{r,n}}{x_{r,n}}
  f_a\left(\frac{\xbj_{r,n}}{x_{r,n}},Q^2\right)
  c^{[n]}_{i;a}(x_{r,n},z,Q^2).
\end{equation}
At high-enough $Q^2$, the coefficient functions $c_i$ admit a
perturbative expansion, which we denote in round brackets such that
\begin{equation}
  c_{i;a}^{[n]} = c_{i;a}^{(0),[n]}+
  \left(\frac{\as}{2\pi}\right) c_{i;a}^{(1),[n]}+
  \left(\frac{\as}{2\pi}\right)^2 c_{i;a}^{(2),[n]}+
  \mathcal O(\as^3),
  \label{eq:ci_pert}
\end{equation}
with $\as = \alpha_s(Q^2)$ the strong coupling constant. The main
result of this work is the heavy-flavour contribution to the CC DIS
coefficient functions at NNLO, \ie $c_{i;a}^{ (2),[m]}$, which are
non-vanishing for $m \in \{0,1,2,3\}$.  For our computation, we find
it convenient to also introduce the variable
\begin{equation}
  \label{eq:yrdef}
  y_{r,n} = 1-x_{r,n},
\end{equation}
so that threshold limits correspond to $y_{r,n}\to0$.  Also, since
most of the contributions that we will consider have exactly one
massive quark in the final state, it is convenient to simplify the
notation and define
\begin{equation}
  x_r \equiv x_{r,1} \equiv 1-y_r = {x}(1+z).
  \label{eq:xrdef}
\end{equation}
In the literature, the replacement $x\to x_r$ is referred to as ``slow
rescaling''.  It is often used as a way to approximate mass
effects, as it allows one to retain the heavy-quark kinematic
effects.  In this manuscript we present a calculation which retains
the \emph{exact} $z$ dependence of the CC coefficient functions up to
NNLO. Since we are mostly targeting low-energy neutrino scattering, we
only focus on one type of massive quark in the final state, which we
refer to as ``charm''. We expect that extending our result to the case
of different massive quarks (\eg charm and bottom) would not pose
significant additional challenges, but we postpone such studies to
future work. 

The next sections of this paper are devoted to presenting the details
of our calculation.  We start from a brief discussion of the LO and
NLO results, which provide a simple setting in which we can illustrate
the tools and framework that we used for the NNLO calculation.


\section{Computing the bare coefficient functions}
\label{sec:bareci}
In this section, we illustrate the framework we used to compute the
bare coefficient functions and express them in terms of a minimal set
of Feynman integrals. In \cref{sec:lo,sec:nlo} we discuss in some
detail the simple LO and NLO cases to illustrate the techniques used;
we then describe the full NNLO pipeline in
\cref{sec:nnlo_structure,sec:pipeline}.

\subsection{LO calculation, Larin scheme and projectors}
\label{sec:lo}
At leading order in $\as$, the only diagram contributing to CC DIS is
that of a light quark (anti-quark) scattering off the vector boson to
produce a final-state massive quark (anti-quark). For consistency with
our treatment of the NLO and NNLO cases, we represent the
corresponding contribution to the cross section as a unitarity cut of
a tree-level forward diagram:
\begin{equation*}  
  \begin{tikzpicture}[line width=1pt, >=stealth,baseline=(current bounding box.center)]

    \draw[decorate, decoration={snake, amplitude=2pt, segment length=6pt}]
      (1,0) -- (2,1);

    \draw[decorate, decoration={snake, amplitude=2pt, segment length=6pt}]
      (-1,1) -- (0,0);

    \draw[
    -,double,
    postaction={
      decorate,
      decoration={
        markings,
        mark=at position 0.55 with {\arrow[scale=0.4]{>}}
      }
    }
    ] (0,0) -- (1,0);

    \draw[
    -,
    postaction={
      decorate,
      decoration={
        markings,
        mark=at position 0.55 with {\arrow[scale=0.7]{>}}
      }
    }
    ] (-1,-1) -- (0,0);

    \draw[
    -,
    postaction={
      decorate,
      decoration={
        markings,
        mark=at position 0.55 with {\arrow[scale=0.7]{>}}
      }
    }
    ] (1,0) -- (2,-1);

    \draw[red, thick, dashed] (0.5,0.8) -- (0.5,-0.8);

    \node[left] at (-1,1) {$q$};
    \node[right] at (2,1) {$q$};

    \node[left] at (-1,-1) {$p$};
    \node[right] at (2,-1) {$p$};



\end{tikzpicture}.
\end{equation*}
Here and henceforth we use double lines to represent heavy quarks with
mass $m_q$ and a red dashed line to denote a unitarity cut,
characterised by replacing propagators with delta functions, forcing
the lines to be on-shell:
\begin{align}
  \frac{i}{p^2-m^2} \rightarrow (2 \pi) \delta_+(p^2-m^2),
\end{align}
where we used the standard notation $\delta_+(p^2-m^2) = \theta(p^0)
\delta(p^2-m^2)$.  Evaluating the Feynman diagram above leads to the
following expression for the partonic tensor $\hat W^{\mu\nu}$ (see
the discussion in the previous section):
\begin{equation}
  \begin{split}
    &\hat W^{\mu\nu} =
    \frac{1}{4\pi}\times\frac{1}{2N_c}\\ &
    \quad\times N_c{\rm Tr}
    \left[(\slashed p + \slashed q - m_q)\gamma^\mu (c_V-c_A\gamma_5)\,
      \slashed p\,\gamma^\nu (c_V-c_A\gamma_5)\right]
    (2\pi)\delta_+\left((p+q)^2-m_q^2\right),
  \end{split}
  \label{eq:wpart}
\end{equation}
with $c_{V,A}$ defined in \cref{eq:cvca}. We stress that for the time
being we work with a diagonal CKM matrix, but that we show how to
restore the full CKM dependence in \cref{sec:results}.

Having in mind the NLO and NNLO calculations, we work in $d=4-2\ep$
spacetime dimensions.  As it is well known, doing this is not entirely
straightforward when $\gamma_5$ is involved. Our calculation does not
involve any anomalous diagrams\footnote{Due to the flavour structure
of the charged current coupling, the CC DIS process does not involve
triangle anomalies, whereas neutral-current DIS would.}, so it should
be possible to work with an anticommuting $\gamma_5$. However, in view
of possible future extensions we decided to work in the so-called
Larin scheme. While we refer to ref.~\cite{larin_scheme} for a
detailed explanation of this scheme, here we briefly recap its main
features. In this scheme, one replaces the axial current with
\begin{equation}
  \gamma^\mu\gamma_5 \rightarrow \frac{1}{2}
  \left(\gamma^\mu\gamma_5-\gamma_5\gamma^\mu\right) =
  \frac{i}{6}\ep^{\mu\nu\rho\sigma}\gamma_\nu\gamma_\rho\gamma_\sigma,
\end{equation}
and deals with the product of Levi-Civita tensors using
\begin{equation}
  \ep_{\mu_1\mu_2\mu_3\mu_4}\ep^{\nu_1\nu_2\nu_3\nu_4} =
  -{\rm Det}\left[\delta^{\nu_i}_{\mu_j}\right],
\end{equation}
where $\delta_i^j$ stands for a Kronecker delta. As it is well known,
this procedure leads to a violation of the axial Ward identity, which
must be restored by an additional renormalisation in order to get the
correct final result. We will briefly discuss this in \cref{sec:ren}.

We note that for the parity-even structure functions $F_2$ and $F_L$,
we only need to calculate the parity-even combination of vertex
insertions, \ie vector-vector (VV) and axial-axial (AA).  In fact,
separating the contributions of left- and right-handed quarks, it is
immediate to see that VV and AA have to be equal. Because of this, we
only need to compute the VV contribution for $F_{2,L}$, where
$\gamma_5$ never enters. $F_3$ on the other hand is only sensitive to
the parity-odd vector-axial (VA) cross terms. Hence, in this case we
work in the Larin scheme and perform the finite renormalisation for
the axial current discussed above.  At NLO, we have explicitly checked
that working in the Larin scheme gives the same result as an
anticommuting $\gamma_5$.  At NNLO, we only performed our calculations
in the Larin scheme.  Since the relevant Dirac algebra is relatively
simple, this did not involve any significant overhead. We have
explicitly verified that our result for $F_3$ passes non-trivial
consistency checks, which gives us confidence in the correctness of
our implementation of the Larin scheme.

We now describe how to extract the partonic coefficient functions
$c_i$ from the partonic tensor $\hat W^{\mu\nu}$ 
\cref{eq:wpart}. We introduce projectors $\mathcal P_i$ defined such
that
\begin{equation}
  \mathcal P_{i}^{[n],\mu\nu} \hat{T}_{j,{\mu\nu}} = \kappa_i^{[n]}\delta_{ij},
\end{equation}
with the partonic tensors $\hat{T}_i$ defined as their hadronic
versions in \cref{eq:Tdefs}, but with the proton momentum replaced by
the parton momentum ($P \to p$), and
\begin{equation}
  \kappa^{[n]}_{1,2} = \frac{1}{x_{r,n}(c_V^2+c_A^2)},
  \quad\quad \kappa^{[n]}_3 = \frac{1}{2 c_V c_A}.
\end{equation}
These projectors allow for the direct extraction of the coefficient
functions $c_i$ from the relevant partonic tensor $\hat W$:
\begin{equation}
  c^{[n]}_{i} = \mathcal P_i^{[n],\mu\nu} \hat W^{[n]}_{\mu\nu},
\end{equation}
\cfit \cref{eq:wdef,eq:cdefs}. Here we denote with $\hat W^{[n]}$ the
partonic tensor for a process with exactly $n$ massive final-state
quarks.
The explicit expression for the projectors reads\footnote{Note that in
these definitions the original partonic variable $x$ is used instead
of the rescaled $x_r$ from \cref{eq:xrdef}.}
\begin{equation}
  \label{eq:proj}
  \begin{split}
    &\mathcal P_1^{[n]} = \kappa_1^{[n]}
    \left[\frac{\hat T_1}{d-2} + \frac{2x \hat T_2}{d-2}\right],
    \quad
    \mathcal P_2^{[n]} = \kappa_2^{[n]}
    \left[\frac{2x \hat T_1}{d-2} +
      \frac{4(d-1)x^2 \hat T_2}{d-2}\right],
    \\
    &\mathcal P_L^{[n]} \equiv \mathcal P_2^{[n]} - 2x \mathcal P_1^{[n]} = 
    \kappa_2^{[n]}\left[4 x^2 \hat T_2\right],\quad
    \mathcal P_3^{[n]} = -
    \kappa_3^{[n]}\left[\frac{4 \hat T_3}{(d-2)(d-3)}\right].
  \end{split}
\end{equation}

Applying these projectors to the partonic tensor
\cref{eq:wpart}, we obtain the LO
results for the quark coefficient functions
\begin{equation}
  c_{2;q}^{(0),[1]} = \delta(1-x_r),
  \quad
  c_{L;q}^{(0),[1]} = \frac{z}{1+z}\delta(1-x_r),
  \quad
  c_{3;q}^{(0),[1]} = \delta(1-x_r).
  \label{eq:cq_lo}
\end{equation}
We remind the reader that $x_r = x(1+z)$ and $z=m_q^2/Q^2$.
The coefficient functions for an incoming anti-quark (\ie for $\bar s
\to \bar c$ transitions) can be immediately obtained from these using
\begin{equation}
  \left\{c_{2;\bar q}^{(0),[1]}\,,\,
  c_{L;\bar q}^{(0),[1]}\,,\,
  c_{3;\bar q}^{(0),[1]}\right\}
  =
  \{1,1,-1\}\times\left\{c_{2;q}^{(0),[1]}\,,\,
  c_{L;q}^{(0),[1]}\,,\,
  c_{3;q}^{(0),[1]}\right\}.
  \label{eq:cqb_lo}
\end{equation}
No other channel contributes at LO. We stress that
\cref{eq:cq_lo,eq:cqb_lo} are valid in arbitrary spacetime dimensions,
\ie no $d\to 4$ limit has been taken.

\subsection{NLO calculation and integral notation}
\label{sec:nlo}

\subsubsection*{Quark channel}
At NLO, the quark production channel receives both real and virtual
corrections. For the former, we need to consider the following cut
diagrams:
\begin{equation*}
  \begin{tikzpicture}[line width=1pt, >=stealth,baseline=(current bounding box.center)]

    \draw[decorate, decoration={snake, amplitude=2pt, segment length=6pt}]
      (1,0) -- (2,1);

    \draw[decorate, decoration={snake, amplitude=2pt, segment length=6pt}]
      (-1,1) -- (0,0);

    \draw[
    -,double,
    postaction={
      decorate,
      decoration={
        markings,
        mark=at position 0.55 with {\arrow[scale=0.4]{>}}
      }
    }
    ] (0,0) -- (1,0);

    \draw[
      decorate,
      decoration={coil, aspect=0.8, segment length=4pt, amplitude=1.5pt}
    ] (-0.4,-0.4) .. controls (0.5,-0.4) and (0.5,-0.4) .. (1.4,-0.4);

    \draw[
    -,
    postaction={
      decorate,
      decoration={
        markings,
        mark=at position 0.55 with {\arrow[scale=0.7]{>}}
      }
    }
    ] (-1,-1) -- (0,0);

    \draw[
    -,
    postaction={
      decorate,
      decoration={
        markings,
        mark=at position 0.55 with {\arrow[scale=0.7]{>}}
      }
    }
    ] (1,0) -- (2,-1);

    \draw[red, thick, dashed] (0.5,0.8) -- (0.5,-0.8);



\end{tikzpicture},
  \;\;\;\;
    \begin{tikzpicture}[line width=1pt, >=stealth,baseline=(current bounding box.center)]

    \draw[decorate, decoration={snake, amplitude=2pt, segment length=6pt}]
      (1,0) -- (2,1);

    \draw[decorate, decoration={snake, amplitude=2pt, segment length=6pt}]
      (-1,1) -- (0,0);

    \draw[
    -,double,
    postaction={
      decorate,
      decoration={
        markings,
        mark=at position 0.55 with {\arrow[scale=0.4]{>}}
      }
    }
    ] (0,0) -- (1,0);

    \draw[
      decorate,
      decoration={coil, aspect=0.8, segment length=4pt, amplitude=1.5pt}
    ] (-0.4,-0.4) .. controls (0.5,-0.4) and (0.5,-0.4) .. (0.75,0);

    \draw[
    -,
    postaction={
      decorate,
      decoration={
        markings,
        mark=at position 0.55 with {\arrow[scale=0.7]{>}}
      }
    }
    ] (-1,-1) -- (0,0);

    \draw[
    -,
    postaction={
      decorate,
      decoration={
        markings,
        mark=at position 0.55 with {\arrow[scale=0.7]{>}}
      }
    }
    ] (1,0) -- (2,-1);

    \draw[red, thick, dashed] (0.25,0.8) -- (0.25,-0.8);



  \end{tikzpicture}, 
  \;\;\;\;
    \begin{tikzpicture}[line width=1pt, >=stealth,baseline=(current bounding box.center)]

    \draw[decorate, decoration={snake, amplitude=2pt, segment length=6pt}]
      (1,0) -- (2,1);

    \draw[decorate, decoration={snake, amplitude=2pt, segment length=6pt}]
      (-1,1) -- (0,0);

    \draw[
    -,double,
    postaction={
      decorate,
      decoration={
        markings,
        mark=at position 0.55 with {\arrow[scale=0.4]{>}}
      }
    }
    ] (0,0) -- (1,0);

    \draw[
      decorate,
      decoration={coil, aspect=0.8, segment length=4pt, amplitude=1.5pt}
    ] (0.2,0) .. controls (0.3,-0.3) and (0.7,-0.3) .. (0.8,0);

    \draw[
    -,
    postaction={
      decorate,
      decoration={
        markings,
        mark=at position 0.55 with {\arrow[scale=0.7]{>}}
      }
    }
    ] (-1,-1) -- (0,0);

    \draw[
    -,
    postaction={
      decorate,
      decoration={
        markings,
        mark=at position 0.55 with {\arrow[scale=0.7]{>}}
      }
    }
    ] (1,0) -- (2,-1);

    \draw[red, thick, dashed] (0.5,0.8) -- (0.5,-0.8);



  \end{tikzpicture}, 
  \;\;\;\;
   \begin{tikzpicture}[line width=1pt, >=stealth,baseline=(current bounding box.center)]

    \draw[decorate, decoration={snake, amplitude=2pt, segment length=6pt}]
      (1,0) -- (2,1);

    \draw[decorate, decoration={snake, amplitude=2pt, segment length=6pt}]
      (-1,1) -- (0,0);

    \draw[
    -,double,
    postaction={
      decorate,
      decoration={
        markings,
        mark=at position 0.55 with {\arrow[scale=0.4]{>}}
      }
    }
    ] (0,0) -- (1,0);

    \draw[
      decorate,
      decoration={coil, aspect=0.8, segment length=4pt, amplitude=1.5pt}
    ] (0.25,0) .. controls (0.5,-0.4) and (0.5,-0.4) .. (1.4,-0.4);

    \draw[
    -,
    postaction={
      decorate,
      decoration={
        markings,
        mark=at position 0.55 with {\arrow[scale=0.7]{>}}
      }
    }
    ] (-1,-1) -- (0,0);

    \draw[
    -,
    postaction={
      decorate,
      decoration={
        markings,
        mark=at position 0.55 with {\arrow[scale=0.7]{>}}
      }
    }
    ] (1,0) -- (2,-1);

    \draw[red, thick, dashed] (0.75,0.8) -- (0.75,-0.8);



  \end{tikzpicture}.
\end{equation*}
In principle, it is straightforward to project these diagrams on the
$c_i^{[1]}$ form factors along the lines described in the previous
subsection, and explicitly perform the relevant phase-space integrals
\eg by working in the $q+p$ centre-of-mass frame and parametrising the
kinematics in terms of the angle between the outgoing gluon and
quark. This direct method, however, becomes prohibitively involved at
higher orders. To simplify the calculation, we work within the reverse
unitarity framework: we write real-emission contributions as
(specific) cuts of loop integrals~\cite{Anastasiou:2002yz}. The main
advantage of this method is that one can then use all the tools
developed for multi-loop integrals which are insensitive to whether
the propagator is cut or not.  These include integration-by-parts
identities
(IBPs)~\cite{Tkachov:1981wb,Chetyrkin:1981qh,Laporta:2001dd} to reduce
all the integrals appearing in our calculation to a small set of
master integrals (MIs). This method is standard so we will not discuss
it here. Instead, we briefly discuss its application to introduce the
notation that will be used in \cref{sec:MIs}.

Given a propagator-like structure $i/\Dp$, we denote its corresponding
cut version as
\begin{equation}\label{eq:cutprop_lo}
  \frac{i}{\slashed\Dp}\equiv (2 \pi) \delta_+(\Dp).
\end{equation}
In what follows, we will denote massless propagators with $\Dp_i$,
and add a superscript $m$ for massive ones,
\begin{equation}
  \Dp_i^m \equiv \Dp_i - m_q^2.
\end{equation}
We write our real-emission contribution as a cut forward loop
amplitude (\cfit the diagrams above), and define an ``integral
topology'', \ie a set of linearly independent propagator-like
structures in terms of which one can write all propagators and scalar
products involving the (cut) loop momentum. While in general more than
one topology is necessary, for this simple case the following
``forward box'' configuration is enough:
\begin{equation}
  \mathrm{Top}^\mathrm{fBox} = \{\Dp_1,\Dp_2,\Dp_3\},
\end{equation}
with
\begin{equation}
    \Dp_1 = l^2,\quad\quad\Dp_2 = (p-l)^2,\quad\quad \Dp_3 =
    (p+q-l)^2.
    \label{eq:dpbox}
\end{equation}
As before, $p$ and $q$ are respectively the momenta of the incoming
quark and vector boson, while $l$ and $p+q-l$ are the momenta of the
outgoing gluon and (massive) quark, respectively. Within this
topology, a general (uncut) integral takes the form
\begin{equation}
  {\rm fBox} [i_1,i_2,i_3] = \int \frac{d^d l}{(2\pi)^d}
  \frac{1}{\left[\Dp_1\right]^{i_1}}
  \frac{1}{\left[\Dp_2\right]^{i_2}} \frac{1}{\left[\Dp_3\right]^{i_3}},
  \label{eq:box}
\end{equation}
with $i_j$ either positive or negative.  We add sub- and superscripts
to the respective indices, to denote cuts and masses.  Specifically, we
add a subscript $c$ to denote a cut line, \ie $\slashed{\Dp}_i$, as
well as a superscript $m$ to denote a massive propagator $\Dp_i^m$.
With this notation, all the required phase-space integrals can then be
written as linear combinations of cut integrals
\begin{equation}
  {\rm fBox} [1_c,i_2,1_c^m] = \int \frac{d^d l}{(2\pi)^d}
  \frac{1}{\slashed\Dp_1}
  \frac{1}{\left[\Dp_2\right]^{i_2}}
  \frac{1}{\slashed\Dp_3^m}.
\end{equation}
The key insight of the reverse unitarity method is to use IBP
identities to reduce integrals of this form to a small set of master
integrals. The reduction proceeds as in the non-cut case, only it is
simpler since one can systematically set to zero integrals where
$i_{1,3}< 1$ at any intermediate step, since these vanish in the
on-shell $l^2\to 0$, $(p+q-l)^2\to m_q^2$ limit. For this simple case,
IBP relations allow one to relate all phase-space integrals to the two-body
phase space ${\rm fBox} [1_c,0,1_c^m] $, which can be trivially
computed by direct integration. It reads:
\begin{equation}
  {\rm fBox} [1_c,0,1_c^m]  =
  \frac{Q^{-2\ep}}{8 \pi}
  \frac{(4 \pi)^\ep \Gamma(1-\ep)}{\Gamma(2-2\ep)}
  y_r^{1-2\ep}(1-y_r)^\ep
  (1+z)^{1-2\ep}(y_r+z)^{-1+\ep},
\end{equation}
with $y_r$ defined in \cref{eq:xrdef}.

In the end, we only require the result for the coefficient functions
$c_i$ expanded around $\ep=0$, which makes the calculation of all the
relevant integrals much simpler.  However, for real-emission
contributions some care has to be taken with the $y_r=0$ threshold
region. Indeed, in this region the forward amplitude develops
poles. Schematically,
\begin{equation}
  c_{i;q}^{(1),[1]} \sim \frac{1}{y_r^2} {\rm fBox} [1_c,0,1_c^m] 
  \sim y_r^{-1-2\ep}.
\end{equation} 
This term cannot be trivially expanded in $\ep$, because the $\ep$
dependence in the exponent is necessary to regulate the $y_r = 0$
divergence in the convolution of the partonic coefficient functions
against PDFs. Rather it should be expressed in terms of plus
distributions
\begin{equation}
  y_r^{-1+a\ep} = \frac{1}{a\ep} +
  \sum_{k=0}^{\infty} \frac{(a\ep)^k}{k!}
  \DD_k(y_r),\quad\quad \DD_k(y_r) \equiv
  \left[\frac{\ln^k y_r}{y_r}\right]_+,
  \label{eq:DDdef}
\end{equation}
with the plus prescription defined in general as
\begin{equation}
  \int_0^1 dy_r f(y_r,z) [g(y_r,z)]_+ \equiv \int_0^1 dy_r 
  \left[f(y_r,z)-f(0,z)\right]
  g(y_r,z).
  \label{eq:plus}
\end{equation}
While all of this is of course well known, we highlighted it here to
stress that on top of a standard $\ep$-expansion of the real-emission
master integrals, we also need to extract the $y_r\to 0$ branch cuts,
before expanding in $\ep$. We will review how to do this in full
generality at NNLO in \cref{sec:bcext}.

To complete the calculation of the (bare) coefficient function in the
quark channel, we also need to consider the virtual corrections:
\begin{equation*}  
   \begin{tikzpicture}[line width=1pt, >=stealth,baseline=(current bounding box.center)]

    \draw[decorate, decoration={snake, amplitude=2pt, segment length=6pt}]
      (1,0) -- (2,1);

    \draw[decorate, decoration={snake, amplitude=2pt, segment length=6pt}]
      (-1,1) -- (0,0);

    \draw[
    -,double,
    postaction={
      decorate,
      decoration={
        markings,
        mark=at position 0.55 with {\arrow[scale=0.4]{>}}
      }
    }
    ] (0,0) -- (0.5,0);  

    \draw[
    -,double,
    postaction={
      decorate,
      decoration={
        markings,
        mark=at position 0.55 with {\arrow[scale=0.4]{>}}
      }
    }
    ] (0.5,0) -- (1,0);

    \draw[
      decorate,
      decoration={coil, aspect=0.8, segment length=4pt, amplitude=1.5pt}
    ] (-0.4,-0.4) .. controls (0.25,-0.25) and (0.25,-0.25) .. (0.5,0);

    \draw[
    -,
    postaction={
      decorate,
      decoration={
        markings,
        mark=at position 0.55 with {\arrow[scale=0.7]{>}}
      }
    }
    ] (-1,-1) -- (0,0);

    \draw[
    -,
    postaction={
      decorate,
      decoration={
        markings,
        mark=at position 0.55 with {\arrow[scale=0.7]{>}}
      }
    }
    ] (1,0) -- (2,-1);

    \draw[red, thick, dashed] (0.75,0.8) -- (0.75,-0.8);



  \end{tikzpicture},  
  \;\;\;\;
    \begin{tikzpicture}[line width=1pt, >=stealth,baseline=(current bounding box.center)]

    \draw[decorate, decoration={snake, amplitude=2pt, segment length=6pt}]
      (1,0) -- (2,1);

    \draw[decorate, decoration={snake, amplitude=2pt, segment length=6pt}]
      (-1,1) -- (0,0);

    \draw[
    -,double,
    postaction={
      decorate,
      decoration={
        markings,
        mark=at position 0.55 with {\arrow[scale=0.4]{>}}
      }
    }
    ] (0,0) -- (0.5,0);  

    \draw[
    -,double,
    postaction={
      decorate,
      decoration={
        markings,
        mark=at position 0.55 with {\arrow[scale=0.4]{>}}
      }
    }
    ] (0.5,0) -- (1,0);

    \draw[
      decorate,
      decoration={coil, aspect=0.8, segment length=4pt, amplitude=1.5pt}
    ] (0.5,0) .. controls (0.75,-0.4) and (0.75,-0.4) .. (1.4,-0.4);

    \draw[
    -,
    postaction={
      decorate,
      decoration={
        markings,
        mark=at position 0.55 with {\arrow[scale=0.7]{>}}
      }
    }
    ] (-1,-1) -- (0,0);

    \draw[
    -,
    postaction={
      decorate,
      decoration={
        markings,
        mark=at position 0.55 with {\arrow[scale=0.7]{>}}
      }
    }
    ] (1,0) -- (2,-1);

    \draw[red, thick, dashed] (0.25,0.8) -- (0.25,-0.8);



  \end{tikzpicture}.
\end{equation*}
All the relevant loop integrals can be mapped into the (non-cut) fBox
topology of \cref{eq:box}, and can then be IBP-reduced to a bubble
($\mathrm{fBox}[0,1,1^m]$) and a tadpole ($\mathrm{fBox}[0,0,1^m]$)
integral, which are straightforward to compute. Summing real and
virtual corrections, and performing the required collinear
renormalisation described in \cref{sec:ren} we obtain the NLO quark
coefficient functions $c_{i;q}^{(1),[1]}$. Their explicit forms can be
found in the ancillary files that accompany this work. Results for the
$\bar q$ channel can be obtained with a relation analogous to
\cref{eq:cqb_lo}.

\subsubsection*{Gluon channel}
At NLO, we also need to consider the gluon-initiated channel, given by
diagrams of the form
\begin{equation*}
  \begin{tikzpicture}[line width=1pt, >=stealth,baseline=(current bounding box.center)]

    \draw[decorate, decoration={snake, amplitude=2pt, segment length=6pt}]
      (1,0.4) -- (2,1);

    \draw[decorate, decoration={snake, amplitude=2pt, segment length=6pt}]
      (-1,1) -- (0,0.4);

    \draw[
    -,
    postaction={
      decoration={
        markings,
        mark=at position 0.55 with {\arrow[scale=0.7]{>}}
      }
    }
    ] (1,0.4) -- (0,0.4);

    \draw[
      -, double,
      postaction = {
      decoration={
        markings,
        mark=at position 0.55 with {\arrow[scale=0.4]{>}}
      }
    }
    ] (0,-0.4) -- (1.0,-0.4);

    \draw[
      -, double,
      postaction = {
      decoration={
        markings,
        mark=at position 0.55 with {\arrow[scale=0.4]{>}}
      }
    }
    ] (0,0.4) -- (0,-0.4);

    \draw[
      -, double,
      postaction = {
      decoration={
        markings,
        mark=at position 0.55 with {\arrow[scale=0.4]{>}}
      }
    }
    ] (1,-0.4) -- (1,0.4);

    \draw[
      decorate,
      decoration={coil, aspect=0.8, segment length=4pt, amplitude=1.5pt}
    ] (-1,-1) -- (0,-0.4);
    
    \draw[
      decorate,
      decoration={coil, aspect=0.8, segment length=4pt, amplitude=1.5pt}
    ] (1.0,-0.4) -- (2,-1);

    \draw[red, thick, dashed] (0.5,0.8) -- (0.5,-0.8);



\end{tikzpicture},
  \;\;\;\;
  \begin{tikzpicture}[line width=1pt, >=stealth,baseline=(current bounding box.center)]

    \draw[decorate, decoration={snake, amplitude=2pt, segment length=6pt}]
      (1,0.4) -- (2,1);

    \draw[decorate, decoration={snake, amplitude=2pt, segment length=6pt}]
      (-1,1) -- (0,0.4);

    \draw[
    -,double,
    postaction={
      decoration={
        markings,
        mark=at position 0.55 with {\arrow[scale=0.4]{>}}
      }
    }
    ] (1,0.4) -- (0,0.4);

    \draw[
      -,
      postaction = {
      decoration={
        markings,
        mark=at position 0.55 with {\arrow[scale=0.7]{>}}
      }
    }
    ] (0,-0.4) -- (1.0,-0.4);

    \draw[
      -,
      postaction = {
      decoration={
        markings,
        mark=at position 0.55 with {\arrow[scale=0.7]{>}}
      }
    }
    ] (0,0.4) -- (0,-0.4);

    \draw[
      -,
      postaction = {
      decoration={
        markings,
        mark=at position 0.55 with {\arrow[scale=0.7]{>}}
      }
    }
    ] (1,-0.4) -- (1,0.4);

    \draw[
      decorate,
      decoration={coil, aspect=0.8, segment length=4pt, amplitude=1.5pt}
    ] (-1,-1) -- (0,-0.4);
    
    \draw[
      decorate,
      decoration={coil, aspect=0.8, segment length=4pt, amplitude=1.5pt}
    ] (1.0,-0.4) -- (2,-1);

    \draw[red, thick, dashed] (0.5,0.8) -- (0.5,-0.8);



\end{tikzpicture},
  \;\;\;\;
  \begin{tikzpicture}[line width=1pt, >=stealth,baseline=(current bounding box.center)]

    \draw[decorate, decoration={snake, amplitude=2pt, segment length=6pt}]
      (1,0.4) -- (2,1);

    \draw[decorate, decoration={snake, amplitude=2pt, segment length=6pt}]
      (-1,1) -- (0,0.4);

    \draw[
    -,double,
    postaction={
      decoration={
        markings,
        mark=at position 0.55 with {\arrow[scale=0.4]{>}}
      }
    }
    ] (1,-0.4) -- (0,0.4);

    \draw[
      -,
      postaction = {
      decoration={
        markings,
        mark=at position 0.55 with {\arrow[scale=0.7]{>}}
      }
    }
    ] (0,-0.4) -- (1.0,0.4);

    \draw[
      -,
      postaction = {
      decoration={
        markings,
        mark=at position 0.55 with {\arrow[scale=0.7]{>}}
      }
    }
    ] (0,0.4) -- (0,-0.4);

    \draw[
      -,double,
      postaction = {
      decoration={
        markings,
        mark=at position 0.55 with {\arrow[scale=0.4]{>}}
      }
    }
    ] (1,-0.4) -- (1,0.4);

    \draw[
      decorate,
      decoration={coil, aspect=0.8, segment length=4pt, amplitude=1.5pt}
    ] (-1,-1) -- (0,-0.4);
    
    \draw[
      decorate,
      decoration={coil, aspect=0.8, segment length=4pt, amplitude=1.5pt}
    ] (1.0,-0.4) -- (2,-1);

    \draw[red, thick, dashed] (0.5,0.8) -- (0.5,-0.8);



\end{tikzpicture},
  \;\;\;\;
  \begin{tikzpicture}[line width=1pt, >=stealth,baseline=(current bounding box.center)]

    \draw[decorate, decoration={snake, amplitude=2pt, segment length=6pt}]
      (1,0.4) -- (2,1);

    \draw[decorate, decoration={snake, amplitude=2pt, segment length=6pt}]
      (-1,1) -- (0,0.4);

    \draw[
    -,
    postaction={
      decoration={
        markings,
        mark=at position 0.55 with {\arrow[scale=0.4]{>}}
      }
    }
    ] (1,-0.4) -- (0,0.4);

    \draw[
      -,double,
      postaction = {
      decoration={
        markings,
        mark=at position 0.55 with {\arrow[scale=0.7]{>}}
      }
    }
    ] (0,-0.4) -- (1.0,0.4);

    \draw[
      -,double,
      postaction = {
      decoration={
        markings,
        mark=at position 0.55 with {\arrow[scale=0.7]{>}}
      }
    }
    ] (0,0.4) -- (0,-0.4);

    \draw[
      -,
      postaction = {
      decoration={
        markings,
        mark=at position 0.55 with {\arrow[scale=0.4]{>}}
      }
    }
    ] (1,-0.4) -- (1,0.4);

    \draw[
      decorate,
      decoration={coil, aspect=0.8, segment length=4pt, amplitude=1.5pt}
    ] (-1,-1) -- (0,-0.4);
    
    \draw[
      decorate,
      decoration={coil, aspect=0.8, segment length=4pt, amplitude=1.5pt}
    ] (1.0,-0.4) -- (2,-1);

    \draw[red, thick, dashed] (0.5,0.8) -- (0.5,-0.8);



\end{tikzpicture}.
\end{equation*}
In principle, we can use a pipeline similar to the one for the quark
channel. However, there is a small extra subtlety that we now
discuss. Consider the third diagram above. Its propagator structure
reads
\begin{equation}
  \frac{1}{\slashed\Dp_1}
  \frac{1}{\Dp_{2}}
  \frac{1}{\slashed\Dp^m_3}
  \frac{1}{\Dp^m_{4}},
\end{equation}
where $\Dp_{4}^{m} = (l-q)^2-m_q^2$ and the other propagators defined
as in \cref{eq:dpbox}. Now $l$ denotes the momentum of the light
outgoing quark.  Naively we could try and define a topology with the
four denominators above, as we would do with a normal scattering
amplitude. However, in the forward limit these are not linearly
independent, so this would not be a valid topology. Indeed
\begin{equation}
  \Dp_1-\Dp_2+\Dp^m_3-\Dp^m_{4} = 2p\cdot q.
\end{equation}
However, we can use this equation to partial fraction any
four-denominator structure, \eg
\begin{equation}
  \begin{split}
    \frac{1}{\slashed\Dp_1}
    \frac{1}{\Dp_{2}}
    \frac{1}{\slashed\Dp^m_3}
    \frac{1}{\Dp^m_{4}} =
    &\left(\frac{1}{2p\cdot q}\right)\frac{\Dp_1-\Dp_2+\Dp^m_3-\Dp^m_{4}}
         {\slashed\Dp_1\Dp_{2}\slashed\Dp^m_3\Dp^m_{4}}
         \longrightarrow\\
         &-\frac{1}{2p\cdot q}\left[
           \frac{1}{\slashed\Dp_1}
           \frac{1}{\Dp_{2}}
           \frac{1}{\slashed\Dp^m_3}
           +
           \frac{1}{\slashed\Dp_1}
           \frac{1}{\slashed\Dp^m_3}
           \frac{1}{\Dp^m_{4}}
           \right],
  \end{split}
\end{equation}
where in the last step we omitted terms that vanish on the cut. In
this way, we can express all our results in terms of two independent
topologies $\{\Dp_1,\Dp_2,\Dp_3^m\}$,
$\{\Dp_1,\Dp_3^m,\Dp_{4}^m\}$. The appearance of linearly-dependent
propagators is a well-known feature of forward scattering
amplitudes. Here we illustrated in a nutshell how one deals with them.
In \cref{sec:pipeline} we will describe how we systematised this
approach for the NNLO calculation.

Before concluding this section we note that the gluon channel contains
diagrams with both a massive quark and an antiquark in the final
state. The contributions to $c_{2,L;g}^{(1),[1]}$ coming from quark
and antiquark final states are identical, while the contributions to
$c_{3;g}^{(1),[1]}$ differ by an overall sign. In other words, if we
were to sum over massive quark and antiquark final states, $c_{2,L;g}$
would just be twice the result for the massive quark only, while
$c^{(1),[1]}_{3;g}$ would vanish.

\subsection{Structure of the NNLO result}
\label{sec:nnlo_structure}
At NNLO, we need to consider double-virtual (VV), real-virtual (RV)
and double-real (RR) contributions. As for NLO, we need to consider
both the quark and gluon channel. We start by describing the structure
of the former.

Contrary to the LO and NLO cases, heavy-flavour effects involve
diagrams not only with one final-state massive quark, but also
diagrams with two, three, or zero. Contributions with no massive
quarks in the final state are purely virtual and stem from diagrams of
the form\footnote{We stress that in this section we are considering
bare amplitudes. We will discuss UV and collinear renormalisation in
\cref{sec:ren}.}
\begin{equation*}
   \begin{tikzpicture}[line width=1pt, >=stealth,baseline=(current bounding box.center)]

    \draw[decorate, decoration={snake, amplitude=2pt, segment length=6pt}]
      (1.3,0) -- (2,1);

    \draw[decorate, decoration={snake, amplitude=2pt, segment length=6pt}]
      (-1,1) -- (-0.3,0);

    \draw[
    -,
    postaction={
      decorate,
      decoration={
        markings,
        mark=at position 0.55 with {\arrow[scale=0.7]{>}}
      }
    }
    ] (-0.3,0) -- (0.5,0);  

    \draw[
    -,
    postaction={
      decorate,
      decoration={
        markings,
        mark=at position 0.55 with {\arrow[scale=0.7]{>}}
      }
    }
    ] (0.5,0) -- (1.3,0);

    \draw[
      decorate,
      decoration={coil, aspect=0.8, segment length=4pt, amplitude=1.5pt}
    ] (0.2,-0.45) .. controls (0.3,-0.45) and (0.45,-0.45) .. (0.55,0);

    \draw[
      decorate,
      decoration={coil, aspect=0.8, segment length=4pt, amplitude=1.5pt}
    ] (-0.61,-0.45) -- (-0.31,-0.45);

  \draw[-,
    thick, double,
    bend left=90] (-0.31,-0.45) to (0.2,-0.45);

  \draw[-,
    thick, double,
    bend left=90] (0.2,-0.45) to (-0.31,-0.45);

    \draw[
    -,
    postaction={
      decorate,
      decoration={
        markings,
        mark=at position 0.55 with {\arrow[scale=0.7]{>}}
      }
    }
    ] (-1,-1) -- (-0.3,0);

    \draw[
    -,
    postaction={
      decorate,
      decoration={
        markings,
        mark=at position 0.55 with {\arrow[scale=0.7]{>}}
      }
    }
    ] (1.3,0) -- (2,-1);

    \draw[red, thick, dashed] (0.85,0.8) -- (0.85,-0.8);

  \node at (-0.01,-0.345) {%
  \tikz[baseline=-0.5ex,scale=0.7]
    \draw[-{>}] (0,0) -- +(0.01pt,0);
  };

  \node at (-0.11,-0.635) {%
    \tikz[baseline=-0.5ex,scale=0.7]
      \draw[{<}-] (0,0) -- +(0.01pt,0);
  };



  \end{tikzpicture}.
\end{equation*}
We refer to this class of diagrams as ``VV$_0$''. Contributions with
one massive heavy-flavour in the final state receive RR, RV, and
VV corrections. Some representative diagrams are
\begin{equation*}
  \begin{tikzpicture}[line width=1pt, >=stealth,baseline=(current bounding box.center)]

    \draw[decorate, decoration={snake, amplitude=2pt, segment length=6pt}]
      (1.3,0.4) -- (2,1);

    \draw[decorate, decoration={snake, amplitude=2pt, segment length=6pt}]
      (-1,1) -- (-0.3,0.4);

    \draw[
    -,double,
    postaction={
      decorate,
      decoration={
        markings,
        mark=at position 0.55 with {\arrow[scale=0.4]{>}}
      }
    }
    ] (-0.3,0.4) -- (1.3,0.4);

    \draw[
      decorate,
      decoration={coil, aspect=0.8, segment length=4pt, amplitude=1.5pt}
    ] (-0.3,-0.4) -- (1.3,-0.4);

    \draw[
      decorate,
      decoration={coil, aspect=0.8, segment length=4pt, amplitude=1.5pt}
    ] (-0.3,0) -- (1.3,0);

    \draw[
      -,
      postaction = {
      decorate,
      decoration={
        markings,
        mark=at position 0.55 with {\arrow[scale=0.7]{>}}
      }
    }
    ] (-0.3,-0.4) -- (-0.3,0);

    \draw[
      -,
      postaction = {
      decorate,
      decoration={
        markings,
        mark=at position 0.55 with {\arrow[scale=0.7]{>}}
      }
    }
    ] (-0.3,0) -- (-0.3,0.4);

    \draw[
      -,
      postaction = {
      decorate,
      decoration={
        markings,
        mark=at position 0.55 with {\arrow[scale=0.7]{>}}
      }
    }
    ] (1.3,0.4) -- (1.3,0);

    \draw[
      -,
      postaction = {
      decorate,
      decoration={
        markings,
        mark=at position 0.55 with {\arrow[scale=0.7]{>}}
      }
    }
    ] (1.3,0) -- (1.3,-0.4);

    \draw[
      -, 
      postaction = {
      decorate,
      decoration={
        markings,
        mark=at position 0.55 with {\arrow[scale=0.7]{>}}
      }
    }] (-1,-1) -- (-0.3,-0.4);
    
    \draw[
      -,
      postaction = {
      decorate,
      decoration={
        markings,
        mark=at position 0.55 with {\arrow[scale=0.7]{>}}
      }
    }] (1.3,-0.4) -- (2,-1);

    \draw[red, thick, dashed] (0.5,0.8) -- (0.5,-0.8);



\end{tikzpicture},
  \;\;\;\;
  \begin{tikzpicture}[line width=1pt, >=stealth,baseline=(current bounding box.center)]

    \draw[decorate, decoration={snake, amplitude=2pt, segment length=6pt}]
      (1.3,0.4) -- (2,1);

    \draw[decorate, decoration={snake, amplitude=2pt, segment length=6pt}]
      (-1,1) -- (-0.3,0.4);

    \draw[
    -,double,
    postaction={
      decorate,
      decoration={
        markings,
        mark=at position 0.55 with {\arrow[scale=0.4]{>}}
      }
    }
    ] (-0.3,0.4) -- (1.3,0.4);

    \draw[
      decorate,
      decoration={coil, aspect=0.8, segment length=4pt, amplitude=1.5pt}
    ] (-0.3,-0.4) -- (1.3,-0.4);

    \draw[
      decorate,
      decoration={coil, aspect=0.8, segment length=4pt, amplitude=1.5pt}
    ] (0.3,-0.4) -- (0.3,0.4);

    \draw[
      -,
      postaction = {
      decorate,
      decoration={
        markings,
        mark=at position 0.55 with {\arrow[scale=0.7]{>}}
      }
    }
    ] (-0.3,-0.4) -- (-0.3,0.4);

    \draw[
      -,
      postaction = {
      decorate,
      decoration={
        markings,
        mark=at position 0.55 with {\arrow[scale=0.7]{>}}
      }
    }
    ] (1.3,0.4) -- (1.3,-0.4);

    \draw[
      -, 
      postaction = {
      decorate,
      decoration={
        markings,
        mark=at position 0.55 with {\arrow[scale=0.7]{>}}
      }
    }] (-1,-1) -- (-0.3,-0.4);
    
    \draw[
      -,
      postaction = {
      decorate,
      decoration={
        markings,
        mark=at position 0.55 with {\arrow[scale=0.7]{>}}
      }
    }] (1.3,-0.4) -- (2,-1);

    \draw[red, thick, dashed] (0.8,0.8) -- (0.8,-0.8);



\end{tikzpicture},
  \;\;\;\;
   \begin{tikzpicture}[line width=1pt, >=stealth,baseline=(current bounding box.center)]

    \draw[decorate, decoration={snake, amplitude=2pt, segment length=6pt}]
      (1.3,0) -- (2,1);

    \draw[decorate, decoration={snake, amplitude=2pt, segment length=6pt}]
      (-1,1) -- (-0.3,0);

    \draw[
    -,double,
    postaction={
      decorate,
      decoration={
        markings,
        mark=at position 0.55 with {\arrow[scale=0.4]{>}}
      }
    }
    ] (-0.3,0) -- (0.5,0);  

    \draw[
    -,double,
    postaction={
      decorate,
      decoration={
        markings,
        mark=at position 0.55 with {\arrow[scale=0.4]{>}}
      }
    }
    ] (0.5,0) -- (1.3,0);

    \draw[
      decorate,
      decoration={coil, aspect=0.8, segment length=4pt, amplitude=1.5pt}
    ] (-0.61,-0.45) .. controls (0.1,-0.45) and (0.2,-0.45) .. (0.35,0);

    \draw[
      decorate,
      decoration={coil, aspect=0.8, segment length=4pt, amplitude=1.5pt}
    ] (-0.1,0) .. controls (0,0.4) and (0.4,0.4) .. (0.5,0);

    \draw[
    -,
    postaction={
      decorate,
      decoration={
        markings,
        mark=at position 0.55 with {\arrow[scale=0.7]{>}}
      }
    }
    ] (-1,-1) -- (-0.3,0);

    \draw[
    -,
    postaction={
      decorate,
      decoration={
        markings,
        mark=at position 0.55 with {\arrow[scale=0.7]{>}}
      }
    }
    ] (1.3,0) -- (2,-1);

    \draw[red, thick, dashed] (0.85,0.8) -- (0.85,-0.8);



  \end{tikzpicture}.
\end{equation*}
Analogously to the zero-mass case, we refer to these contributions as
``RR$_1$'', ``RV$_1$'', and ``VV$_1$''. For double-real contributions
we further split our result into a non-singlet (depicted by the
leftmost diagram above) part, where the incoming quark line is
directly connected to the vector boson, and a pure-singlet part, where
it is not. Representative pure-singlet diagrams are
\begin{equation*}
  \begin{tikzpicture}[line width=1pt, >=stealth,baseline=(current bounding box.center)]

    \draw[decorate, decoration={snake, amplitude=2pt, segment length=6pt}]
      (1.3,0.4) -- (2,1);

    \draw[decorate, decoration={snake, amplitude=2pt, segment length=6pt}]
      (-1,1) -- (-0.3,0.4);

    \draw[
    -,double,
    postaction={
      decorate,
      decoration={
        markings,
        mark=at position 0.55 with {\arrow[scale=0.4]{>}}
      }
    }
    ] (-0.3,0.4) -- (1.3,0.4);

    \draw[-,
      postaction = {
      decorate,
      decoration={
        markings,
        mark=at position 0.55 with {\arrow[scale=0.7]{>}}
      }
    }
    ] (-0.3,-0.4) -- (1.3,-0.4);

    \draw[
      -,
      postaction = {
      decorate,
      decoration={
        markings,
        mark=at position 0.55 with {\arrow[scale=0.7]{>}}
      }
    }    
    ] (1.3,0) -- (-0.3,0);

    \draw[
     decorate,
     decoration={coil, aspect=0.8, segment length=4pt, amplitude=1.5pt}
    ] (-0.3,-0.4) -- (-0.3,0.1);

    \draw[
      -,
      postaction = {
      decorate,
      decoration={
        markings,
        mark=at position 0.55 with {\arrow[scale=0.7]{>}}
      }
    }
    ] (-0.3,0) -- (-0.3,0.4);

    \draw[
      -,
      postaction = {
      decorate,
      decoration={
        markings,
        mark=at position 0.55 with {\arrow[scale=0.7]{>}}
      }
    }
    ] (1.3,0.4) -- (1.3,0);

    \draw[
     decorate,
     decoration={coil, aspect=0.8, segment length=4pt, amplitude=1.5pt}
    ] (1.3,-0.4) -- (1.3,0.1);

    \draw[
      -, 
      postaction = {
      decorate,
      decoration={
        markings,
        mark=at position 0.55 with {\arrow[scale=0.7]{>}}
      }
    }] (-1,-1) -- (-0.3,-0.4);
    
    \draw[
      -,
      postaction = {
      decorate,
      decoration={
        markings,
        mark=at position 0.55 with {\arrow[scale=0.7]{>}}
      }
    }] (1.3,-0.4) -- (2,-1);

    \draw[red, thick, dashed] (0.5,0.8) -- (0.5,-0.8);



\end{tikzpicture},
  \;\;\;\;
  \begin{tikzpicture}[line width=1pt, >=stealth,baseline=(current bounding box.center)]

    \draw[decorate, decoration={snake, amplitude=2pt, segment length=6pt}]
      (1.3,0.4) -- (2,1);

    \draw[decorate, decoration={snake, amplitude=2pt, segment length=6pt}]
      (-1,1) -- (-0.3,0.4);

    \draw[
    -,
    postaction={
      decorate,
      decoration={
        markings,
        mark=at position 0.55 with {\arrow[scale=0.7]{>}}
      }
    }
    ] (-0.3,0.4) -- (1.3,0.4);

    \draw[-,
      postaction = {
      decorate,
      decoration={
        markings,
        mark=at position 0.55 with {\arrow[scale=0.7]{>}}
      }
    }
    ] (-0.3,-0.4) -- (1.3,-0.4);

    \draw[
      -, double,
      postaction = {
      decorate,
      decoration={
        markings,
        mark=at position 0.55 with {\arrow[scale=0.4]{>}}
      }
    }    
    ] (1.3,0) -- (-0.3,0);

    \draw[
     decorate,
     decoration={coil, aspect=0.8, segment length=4pt, amplitude=1.5pt}
    ] (-0.3,-0.4) -- (-0.3,0.1);

    \draw[
      -,double,
      postaction = {
      decorate,
      decoration={
        markings,
        mark=at position 0.55 with {\arrow[scale=0.4]{>}}
      }
    }
    ] (-0.3,0) -- (-0.3,0.4);

    \draw[
      -,double,
      postaction = {
      decorate,
      decoration={
        markings,
        mark=at position 0.55 with {\arrow[scale=0.4]{>}}
      }
    }
    ] (1.3,0.4) -- (1.3,0);

    \draw[
     decorate,
     decoration={coil, aspect=0.8, segment length=4pt, amplitude=1.5pt}
    ] (1.3,-0.4) -- (1.3,0.1);

    \draw[
      -, 
      postaction = {
      decorate,
      decoration={
        markings,
        mark=at position 0.55 with {\arrow[scale=0.7]{>}}
      }
    }] (-1,-1) -- (-0.3,-0.4);
    
    \draw[
      -,
      postaction = {
      decorate,
      decoration={
        markings,
        mark=at position 0.55 with {\arrow[scale=0.7]{>}}
      }
    }] (1.3,-0.4) -- (2,-1);

    \draw[red, thick, dashed] (0.5,0.8) -- (0.5,-0.8);



\end{tikzpicture}.
\end{equation*}
We also have non-singlet diagrams with an incoming anti-quark and outgoing
massive quark:
\begin{equation*}
  \begin{tikzpicture}[line width=1pt, >=stealth,baseline=(current bounding box.center)]

    \draw[decorate, decoration={snake, amplitude=2pt, segment length=6pt}]
      (1.3,0.4) -- (2,1);

    \draw[decorate, decoration={snake, amplitude=2pt, segment length=6pt}]
      (-1,1) -- (-0.3,0.4);

    \draw[
    -,double,
    postaction={
      decorate,
      decoration={
        markings,
        mark=at position 0.55 with {\arrow[scale=0.4]{>}}
      }
    }
    ] (-0.3,0.4) -- (1.3,0.4);

    \draw[
      -,
    postaction={
      decorate,
      decoration={
        markings,
        mark=at position 0.55 with {\arrow[scale=0.4]{<}}
      }
    }
    ] (-0.3,-0.4) -- (1.3,0);

    \draw[
       -,
    postaction={
      decorate,
      decoration={
        markings,
        mark=at position 0.55 with {\arrow[scale=0.4]{>}}
      }
    }
    ] (1.3,-0.4) -- (-0.3,0);

    \draw[
     decorate,
      decoration={coil, aspect=0.8, segment length=3pt, amplitude=1.5pt}
    ] (-0.3,-0.4) -- (-0.3,0);

    \draw[
      -,
      postaction = {
      decorate,
      decoration={
        markings,
        mark=at position 0.55 with {\arrow[scale=0.7]{>}}
      }
    }
    ] (-0.3,0) -- (-0.3,0.4);

    \draw[
      -,
      postaction = {
      decorate,
      decoration={
        markings,
        mark=at position 0.55 with {\arrow[scale=0.7]{>}}
      }
    }
    ] (1.3,0.4) -- (1.3,0);

    \draw[
     decorate,
     decoration={coil, aspect=0.8, segment length=3pt, amplitude=1.5pt}
    ] (1.3,0) -- (1.3,-0.4);

    \draw[
      -, 
      postaction = {
      decorate,
      decoration={
        markings,
        mark=at position 0.55 with {\arrow[scale=0.7]{>}}
      }
    }] (-0.3,-0.4) -- (-1,-1);
    
    \draw[
      -,
      postaction = {
      decorate,
      decoration={
        markings,
        mark=at position 0.55 with {\arrow[scale=0.7]{>}}
      }
    }] (2,-1) -- (1.3,-0.4) ;

    \draw[red, thick, dashed] (0.5,0.8) -- (0.5,-0.8);



\end{tikzpicture},
\end{equation*}
which we will refer to as the $\bar q_{{\rm NS}_{\bar q q}}$
contribution. Crucially, this is not the same as the ``standard''
non-singlet channel with $q\to \bar q$.

Contributions with two or three heavy quarks in the final state stem
from diagrams of the form
\begin{equation*}
   \begin{tikzpicture}[line width=1pt, >=stealth,baseline=(current bounding box.center)]

    \draw[decorate, decoration={snake, amplitude=2pt, segment length=6pt}]
      (1.3,0) -- (2,1);

    \draw[decorate, decoration={snake, amplitude=2pt, segment length=6pt}]
      (-1,1) -- (-0.3,0);

    \draw[
    -,
    postaction={
      decorate,
      decoration={
        markings,
        mark=at position 0.55 with {\arrow[scale=0.7]{>}}
      }
    }
    ] (-0.3,0) -- (0.5,0);  

    \draw[
    -,
    postaction={
      decorate,
      decoration={
        markings,
        mark=at position 0.55 with {\arrow[scale=0.7]{>}}
      }
    }
    ] (0.5,0) -- (1.3,0);

    \draw[
      decorate,
      decoration={coil, aspect=0.8, segment length=4pt, amplitude=1.5pt}
    ] (0.2,-0.45) .. controls (0.3,-0.45) and (0.45,-0.45) .. (0.55,0);

    \draw[
      decorate,
      decoration={coil, aspect=0.8, segment length=4pt, amplitude=1.5pt}
    ] (-0.61,-0.45) -- (-0.31,-0.45);

  \draw[-,
    thick, double,
    bend left=90] (-0.31,-0.45) to (0.2,-0.45);

  \draw[-,
    thick, double,
    bend left=90] (0.2,-0.45) to (-0.31,-0.45);

    \draw[
    -,
    postaction={
      decorate,
      decoration={
        markings,
        mark=at position 0.55 with {\arrow[scale=0.7]{>}}
      }
    }
    ] (-1,-1) -- (-0.3,0);

    \draw[
    -,
    postaction={
      decorate,
      decoration={
        markings,
        mark=at position 0.55 with {\arrow[scale=0.7]{>}}
      }
    }
    ] (1.3,0) -- (2,-1);

    \draw[red, thick, dashed] (-0.05,0.8) -- (-0.05,-0.8);

  \node at (-0.01,-0.345) {%
  \tikz[baseline=-0.5ex,scale=0.7]
    \draw[-{>}] (0,0) -- +(0.01pt,0);
  };

  \node at (-0.11,-0.635) {%
    \tikz[baseline=-0.5ex,scale=0.7]
      \draw[{<}-] (0,0) -- +(0.01pt,0);
  };



  \end{tikzpicture},
  \;\;\;\;
   \begin{tikzpicture}[line width=1pt, >=stealth,baseline=(current bounding box.center)]

    \draw[decorate, decoration={snake, amplitude=2pt, segment length=6pt}]
      (1.3,0) -- (2,1);

    \draw[decorate, decoration={snake, amplitude=2pt, segment length=6pt}]
      (-1,1) -- (-0.3,0);

    \draw[
    -,double,
    postaction={
      decorate,
      decoration={
        markings,
        mark=at position 0.55 with {\arrow[scale=0.4]{>}}
      }
    }
    ] (-0.3,0) -- (0.5,0);  

    \draw[
    -,double,
    postaction={
      decorate,
      decoration={
        markings,
        mark=at position 0.55 with {\arrow[scale=0.4]{>}}
      }
    }
    ] (0.5,0) -- (1.3,0);

    \draw[
      decorate,
      decoration={coil, aspect=0.8, segment length=4pt, amplitude=1.5pt}
    ] (0.2,-0.45) .. controls (0.3,-0.45) and (0.45,-0.45) .. (0.55,0);

    \draw[
      decorate,
      decoration={coil, aspect=0.8, segment length=4pt, amplitude=1.5pt}
    ] (-0.61,-0.45) -- (-0.31,-0.45);

  \draw[-,
    thick, double,
    bend left=90] (-0.31,-0.45) to (0.2,-0.45);

  \draw[-,
    thick, double,
    bend left=90] (0.2,-0.45) to (-0.31,-0.45);

    \draw[
    -,
    postaction={
      decorate,
      decoration={
        markings,
        mark=at position 0.55 with {\arrow[scale=0.7]{>}}
      }
    }
    ] (-1,-1) -- (-0.3,0);

    \draw[
    -,
    postaction={
      decorate,
      decoration={
        markings,
        mark=at position 0.55 with {\arrow[scale=0.7]{>}}
      }
    }
    ] (1.3,0) -- (2,-1);

    \draw[red, thick, dashed] (-0.05,0.8) -- (-0.05,-0.8);

  \node at (-0.01,-0.345) {%
  \tikz[baseline=-0.5ex,scale=0.7]
    \draw[-{>}] (0,0) -- +(0.01pt,0);
  };

  \node at (-0.11,-0.635) {%
    \tikz[baseline=-0.5ex,scale=0.7]
      \draw[{<}-] (0,0) -- +(0.01pt,0);
  };



  \end{tikzpicture},
  \;\;\;\;
   \begin{tikzpicture}[line width=1pt, >=stealth,baseline=(current bounding box.center)]

    \draw[decorate, decoration={snake, amplitude=2pt, segment length=6pt}]
      (1.3,0) -- (2,1);

    \draw[decorate, decoration={snake, amplitude=2pt, segment length=6pt}]
      (-1,1) -- (-0.3,0);

    \draw[
    -,double,
    postaction={
      decorate,
      decoration={
        markings,
        mark=at position 0.55 with {\arrow[scale=0.4]{>}}
      }
    }
    ] (-0.3,0) -- (1.10,-0.65);  

    \draw[
    -,double,
    postaction={
      decorate,
      decoration={
        markings,
        mark=at position 0.55 with {\arrow[scale=0.4]{>}}
      }
    }
    ] (-0.10,-0.65) -- (1.3,0);  

    \draw[
    -,double,
    postaction={
      decorate,
      decoration={
        markings,
        mark=at position 0.55 with {\arrow[scale=0.4]{>}}
      }
    }
    ] (1.10,-0.65) -- (-0.10,-0.65);

    \draw[
      decorate,
      decoration={coil, aspect=0.8, segment length=4pt, amplitude=1.5pt}
    ] (1.75,-0.65) -- (1.10,-0.65);;

    \draw[
      decorate,
      decoration={coil, aspect=0.8, segment length=4pt, amplitude=1.5pt}
    ] (-0.75,-0.65) -- (-0.10,-0.65);

    \draw[
    -,
    postaction={
      decorate,
      decoration={
        markings,
        mark=at position 0.55 with {\arrow[scale=0.7]{>}}
      }
    }
    ] (-1,-1) -- (-0.3,0);

    \draw[
    -,
    postaction={
      decorate,
      decoration={
        markings,
        mark=at position 0.55 with {\arrow[scale=0.7]{>}}
      }
    }
    ] (1.3,0) -- (2,-1);

    \draw[red, thick, dashed] (0.5,1.0) -- (0.5,-1.0);



  \end{tikzpicture}.
\end{equation*}
We call them ``RR$_2$'' and ``RR$_3$'', respectively. These
contributions are separately finite in $d=4$. As we will see in the
next section, the result for all but the RR$_3$ contributions can be
expressed in terms of generalised Goncharov
polylogarithms~\cite{goncharov2011multiplepolylogarithmscyclotomymodular}.
RR$_3$, on the other hand, involves elliptic kernels.

The structure of the gluon channel is much simpler. Indeed,
corrections to this channel can be obtained by simply dressing the NLO
diagrams with one additional real or virtual gluon. Doing this yields
NNLO diagrams which, according to the nomenclature above, only fall in
either the RR$_1$ or the RV$_1$ category. Below are representative
diagrams for each:
\begin{equation*}
  \begin{tikzpicture}[line width=1pt, >=stealth,baseline=(current bounding box.center)]

    \draw[decorate, decoration={snake, amplitude=2pt, segment length=6pt}]
      (1,0.4) -- (2,1);

    \draw[decorate, decoration={snake, amplitude=2pt, segment length=6pt}]
      (-1,1) -- (0,0.4);

    \draw[
    -,
    postaction={
      decoration={
        markings,
        mark=at position 0.55 with {\arrow[scale=0.7]{>}}
      }
    }
    ] (1,0.4) -- (0,0.4);

    \draw[
      -, double,
      postaction = {
      decoration={
        markings,
        mark=at position 0.55 with {\arrow[scale=0.4]{>}}
      }
    }
    ] (0,-0.4) -- (1.0,-0.4);

    \draw[
      -, double,
      postaction = {
      decoration={
        markings,
        mark=at position 0.55 with {\arrow[scale=0.4]{>}}
      }
    }
    ] (0,0.4) -- (0,-0.4);

    \draw[
      -, double,
      postaction = {
      decoration={
        markings,
        mark=at position 0.55 with {\arrow[scale=0.4]{>}}
      }
    }
    ] (1,-0.4) -- (1,0.4);

    \draw[
      decorate,
      decoration={coil, aspect=0.8, segment length=4pt, amplitude=1.5pt}
    ] (-1,-1) -- (0,-0.4);
    
    \draw[
      decorate,
      decoration={coil, aspect=0.8, segment length=4pt, amplitude=1.5pt}
    ] (1.0,-0.4) -- (2,-1);

    \draw[
      decorate,
      decoration={coil, aspect=0.8, segment length=4pt, amplitude=1.5pt}
    ] (0,0) -- (1.0,0);

    \draw[red, thick, dashed] (0.5,0.8) -- (0.5,-0.8);



\end{tikzpicture},
  \;\;\;\;
  \begin{tikzpicture}[line width=1pt, >=stealth,baseline=(current bounding box.center)]

    \draw[decorate, decoration={snake, amplitude=2pt, segment length=6pt}]
      (1,0.4) -- (2,1);

    \draw[decorate, decoration={snake, amplitude=2pt, segment length=6pt}]
      (-1,1) -- (0,0.4);

    \draw[
    -,
    postaction={
      decoration={
        markings,
        mark=at position 0.55 with {\arrow[scale=0.7]{>}}
      }
    }
    ] (1,0.4) -- (0,0.4);

    \draw[
      -, double,
      postaction = {
      decoration={
        markings,
        mark=at position 0.55 with {\arrow[scale=0.4]{>}}
      }
    }
    ] (0,-0.4) -- (1.0,-0.4);

    \draw[
      -, double,
      postaction = {
      decoration={
        markings,
        mark=at position 0.55 with {\arrow[scale=0.4]{>}}
      }
    }
    ] (0,0.4) -- (0,-0.4);

    \draw[
      -, double,
      postaction = {
      decoration={
        markings,
        mark=at position 0.55 with {\arrow[scale=0.4]{>}}
      }
    }
    ] (1,-0.4) -- (1,0.4);

    \draw[
      decorate,
      decoration={coil, aspect=0.8, segment length=4pt, amplitude=1.5pt}
    ] (-1,-1) -- (0,-0.4);
    
    \draw[
      decorate,
      decoration={coil, aspect=0.8, segment length=4pt, amplitude=1.5pt}
    ] (0.4,-0.4) -- (0.4,0.4);

    \draw[
      decorate,
      decoration={coil, aspect=0.8, segment length=4pt, amplitude=1.5pt}
    ] (1.0,-0.4) -- (2,-1);

    \draw[red, thick, dashed] (0.7,0.8) -- (0.7,-0.8);



\end{tikzpicture}.
\end{equation*}
The calculation of the bare NNLO coefficient functions follows a
procedure which is analogous -- though much more involved -- to the
one illustrated in the previous section. Due to the large number of
diagrams and contributions, automation is required to some extent. We
describe our NNLO pipeline in the next subsection.

\subsection{Workflow of the NNLO calculation}
\label{sec:pipeline}
The calculation of the NNLO bare coefficient functions proceeds
roughly along the same lines that we described for the LO and NLO
ones. However, due to the inherently large number of terms and
manipulations required, we set up their calculation in an automated
way. Below we briefly report the pipeline that we used.

\begin{itemize}
  \item 
  For a fixed partonic channel, we generate all the two-loop
  $2\rightarrow2$ diagrams of the type seen earlier in the section
  using \texttt{qgraf}~\cite{qgraf}. At this stage, we do not impose
  any forward condition;
  \item 
    we then use an in-house \texttt{Mathematica} script to generate
    a list of all allowed unitarity cuts from the \texttt{qgraf}
    output. At this stage, we remove all ``illegal cuts'',
    corresponding to external-leg corrections;
  \item 
  we define a minimal set of general-kinematics integral families
  $\mathcal{T}$ (which we do not give here as they only appear as an
  intermediate step and were chosen arbitrarily) so that each diagram
  can be mapped to at least one of them.  To identify all diagram
  mappings and loop-momenta shifts, we use \texttt{Reduze
    2}~\cite{Studerus:2009ye,reduze};
  \item 
  once each diagram has been assigned to an integral family, we impose
  forward kinematics, insert the Feynman rules, project onto
  coefficient functions contracting with the appropriate projectors
  \cref{eq:proj} and perform both colour and Dirac algebra using
  in-house \texttt{FORM}~\cite{form} routines. We work in Feynman
  gauge, so we also include the relevant ghost diagrams. As an
  internal self-consistency check, in the gluon channel we perform the
  sum over polarisations of incoming gluons in two different ways:
  $a)$ we use physical polarisation tensors, \ie $\sum_\lambda
  \ep_\lambda(p_i)^\mu\ep_\lambda(p_i)^{*,\nu} = -g^{\mu\nu} +
  (p_i^\mu n^\nu + p_i^\nu n^\mu)/n\cdot p_i$; $b)$ we sum over
  polarisations using $-g^{\mu\nu}$ and remove the contribution of
  longitudinal polarisations by including incoming ghost-like
  contributions. We find perfect agreement between the two
  calculations.  While doing the Dirac algebra, we also tag all
  initial and final states, and keep track of the Dirac and flavour
  structure of the original diagrams to separate singlet and
  non-singlet contributions;

  \item 
  having taken the forward limit in the previous step, many of the
  scalar integrals belonging to the general-kinematics integral
  families $\mathcal{T}$ develop linear relations among their
  propagators.  Indeed, for a general (non-forward) two-loop
  four-point topology one can build 3 scalar products quadratic in the
  loop momenta and 6 scalar products linear in the loop
  momenta. Therefore 9 irreducible scalar products (ISP) are required.
  In forward kinematics however, this number shrinks to 7, meaning
  that any given integral in a 9-ISP topology is mapped in general to
  a linear combination of integrals in various 7-ISP topologies.
  Computing the exact decomposition requires partial fractioning,
  along the lines discussed in the previous subsection. We automate
  this by finding for each integral family $\mathcal{T}$ all possible
  linear relations between its propagators.  We find that all
  relations are of two types: they either can be written in the form
      \begin{align}
        \sum_{i=1}^9 d_i \Dp^{({\mathcal{T}})}_{i} = 1,
        \label{eq:partf}
      \end{align}
  where the coefficients $d_i$ only depend on kinematic invariants, or
  they are trivial identities between two of the ISPs:
  $\Dp^{({\mathcal{T}})}_i = \Dp^{({\mathcal{T}})}_j$.  In the latter
  case we consistently remove one of the two ISPs in favour of the
  other.  This allows us to find all subsets of 7 ISPs which span all
  scalar products appearing in the forward limit of each full
  $2\rightarrow2$ topology $\mathcal{T}$. We call these forward
  topologies $\overline{\mathcal{T}}$ and we give them explicitly in
  \cref{sec:topologies}.  Afterwards, for each integral, we iterate
  over the following process:
  \begin{enumerate}
    \item check if the occurring propagators can already be
      assigned to one of the forward topologies
      $\overline{\mathcal{T}}$. If so, assign the topology and move to the
      next integral;
    \item if not, multiply the integral by
      $1 = \sum_i d_i \Dp^{({\mathcal{T}})}_i$
      (\cfit \cref{eq:partf}) of the relevant
      full topology $\mathcal{T}$, expand the terms in the sum and
      simplify numerators with denominators.  Go back to step 1;
  \end{enumerate}
  
  \item 
  for each forward topology $\overline{\mathcal{T}}$ we express all
  scalar products involving the loop momenta in terms of the ISPs of
  that topology and write the forward amplitude as a linear
  combination of scalar integrals of the form
  $$ 
  {\overline{\mathcal{T}}}[{i_1,\dots,i_7}]
  = 
  \int \frac{d^d l_1}{(2\pi)^d} \frac{d^d l_2}{(2\pi)^d} \frac{1}{
    \left[\Dp^{(\overline{\mathcal{T}})}_{1}\right]^{i_1} 
    \cdots 
    \left[\Dp^{(\overline{\mathcal{T}})}_{7}\right]^{i_7} 
    }, 
  $$
  with coefficients depending on kinematics, $d$ as well as on
  colour and flavour structures;
  \item 
  at this point we perform all allowed unitarity cuts relevant for the
  cross section we are interested in and collect contributions
  according to the total number of final states (RR, RV, VV) and of
  final-state massive quarks (0,1,2,3), as discussed in
  \cref{sec:nnlo_structure}.  Here ``cutting'' a diagram amounts to
  setting to zero all integrals which have no support on the given
  cut, and turning the cut propagators into delta functions, \cfit
  \cref{eq:cutprop_lo};
  \item 
  finally, we collect all needed scalar integrals and IBP reduce them
  to master integrals using \texttt{kira 2}~\cite{kira}.\footnote{The
  full reduction only takes a few hours on a $\mathcal O(50)$~core
  server. Note that version 3 of \texttt{kira} has become available during
  the preparation of this manuscript \cite{kira3}.} 
  After inserting the reduction in our bare cross sections,
  we perform simple algebraic manipulations on the master-integral
  coefficients to simplify the result.
\end{itemize}
We used the same pipeline for the LO, NLO, and NNLO calculations, as
well as for the diagrams involving mass-renormalisation counterterms
(see \cref{sec:ren}).  As a check, we have performed two
fully-independent calculations of all the steps described above, and
found perfect agreement. We have also used the same setup to repeat
the calculation with the heavy-quark mass set to zero as a check, and
found perfect agreement with the NNLO results available in the
literature~\cite{Moch:2008fj,Vermaseren:2005qc}.

\subsection{Integral families}
\label{sec:topologies}
\begin{figure}[tbh]
  \centering
  \begin{subfigure}{0.98\textwidth}
    \centering
    \scalebox{2}{\begin{tikzpicture}[ >=stealth,
  inner sep=0pt, outer sep=0pt,
  shorten <=-0.55pt, shorten >=-0.55pt,
  baseline=(current bounding box.center)]
  \node[inner sep=0pt] (2) at (1.16,1.5) {};
  \node[inner sep=0pt] (e1) at (1.55,1.5) {};
  \node[inner sep=0pt] (5) at (0.7,1.26) {};
  \node[inner sep=0pt] (e2) at (1.55,0.95) {};
  \node[inner sep=0pt] (6) at (1.16,0.95) {};
  \node[inner sep=0pt] (e4u) at (0.38,1.38) {};
  \node[inner sep=0pt] (e4d) at (0.38,1.13) {};
  \draw[-, thick,decorate, decoration={snake, amplitude=.5pt, segment length=3pt}] (2) -- (e1);

  \draw[-, thick] (6) -- (e2);
  \draw[-, thick,decorate, decoration={snake, amplitude=.5pt, segment length=3pt}] (5) -- (e4u);
  \draw[-, thick] (5) -- (e4d);

  \draw[-, thick] (2) to (5);
  \node[edgelabel] at ($ (2)!0.5!(5) $) {$3$};

  \draw[-, thick] (5) -- (6);
  \node[edgelabel] at ($ (5)!0.5!(6) $) {$1$};

  \draw[-, thick] (6) -- (2);
  \node[edgelabel] at ($ (6)!0.5!(2) $) {$2$};

  \end{tikzpicture}}
    \caption{fBox}
  \end{subfigure}\\
  \begin{subfigure}{0.32\textwidth}
    \centering
    \scalebox{2}{\begin{tikzpicture}[>=stealth,
  inner sep=0pt, outer sep=0pt,
  shorten <=-0.55pt, shorten >=-0.55pt,
  baseline=(current bounding box.center)]
    \node[inner sep=0pt] (1) at (1.5,0.54) {};
    \node[inner sep=0pt] (2) at (1.5,0) {};
    \node[inner sep=0pt] (3) at (1,0) {};
    \node[inner sep=0pt] (4) at (0.5,0) {};
    \node[inner sep=0pt] (5) at (0.5,0.54) {};
    \node[inner sep=0pt] (6) at (1,0.54) {};
    \node[inner sep=0pt] (7) at (2,0.54) {};
    \node[inner sep=0pt] (8) at (2,0) {};
    \node[inner sep=0pt] (9) at (0.00,0) {};
    \node[inner sep=0pt] (10) at (0.00,0.54) {};
    \node[inner sep=0pt] (c1) at (0.75,0.74) {};
    \node[inner sep=0pt] (c2) at (1.25,-0.2) {};

    \draw[-, thick] (1) -- (2);
    \node[edgelabel] at ($ (1)!0.5!(2) $) {$2$};

    \draw[-, thick] (2) -- (3);
    \node[edgelabel] at ($ (2)!0.5!(3) $) {$6$};

    \draw[-, thick] (3) -- (4);
    \node[edgelabel] at ($ (3)!0.5!(4) $) {$4$};

    \draw[-, thick] (4) -- (5);
    \node[edgelabel] at ($ (4)!0.5!(5) $) {$1$};

    \draw[-, thick] (5) -- (6);
    \node[edgelabel] at ($ (5)!0.5!(6) $) {$5$};

    \draw[-, thick] (6) -- (1);
    \node[edgelabel] at ($ (6)!0.5!(1) $) {$7$};

    \draw[-, thick] (3) -- (6);
    \node[edgelabel] at ($ (3)!0.5!(6) $) {$3$};

    \draw[-, thick,decorate, decoration={snake, amplitude=.5pt, segment length=3pt}] (1) -- (7);

    \draw[-, thick] (2) -- (8);

    \draw[-, thick] (4) -- (9);

    \draw[-, thick,decorate, decoration={snake, amplitude=.5pt, segment length=3pt}] (5) -- (10);

  \end{tikzpicture}}
    \caption{VB}
  \end{subfigure}\hfill
  \begin{subfigure}{0.32\textwidth}
    \centering
    \raisebox{1cm}{\scalebox{2}{\begin{tikzpicture}[>=stealth,
  inner sep=0pt, outer sep=0pt,
  shorten <=-0.55pt, shorten >=-0.55pt,
  baseline=(current bounding box.center)]
    \node[inner sep=0pt] (1) at (1.5,0.54) {};
    \node[inner sep=0pt] (2) at (1.5,0) {};
    \node[inner sep=0pt] (3) at (1,0) {};
    \node[inner sep=0pt] (4) at (0.5,0) {};
    \node[inner sep=0pt] (5) at (0.5,0.54) {};
    \node[inner sep=0pt] (6) at (1,0.54) {};
    \node[inner sep=0pt] (7) at (2,0.54) {};
    \node[inner sep=0pt] (8) at (2,0) {};
    \node[inner sep=0pt] (9) at (0.00,0) {};
    \node[inner sep=0pt] (10) at (0.00,0.54) {};

    \draw[-, thick] (1) -- (2);
    \node[edgelabel] at ($ (1)!0.5!(2) $) {$7$};

    \draw[-, thick] (2) -- (3);
    \node[edgelabel] at ($ (2)!0.5!(3) $) {$2$};

    \draw[-, thick] (3) -- (4);
    \node[edgelabel] at ($ (3)!0.5!(4) $) {$1$};

    \draw[-, thick] (4) -- (5);
    \node[edgelabel] at ($ (4)!0.5!(5) $) {$4$};

    \draw[-, thick] (5) -- (6);
    \node[edgelabel] at ($ (5)!0.5!(6) $) {$5$};

    \draw[-, thick] (6) -- (1);
    \node[edgelabel] at ($ (6)!0.5!(1) $) {$6$};

    \draw[-, thick] (3) -- (6);
    \node[edgelabel] at ($ (3)!0.5!(6) $) {$3$};

    \draw[-, thick] (1) -- (7);

    \draw[-, thick,decorate, decoration={snake, amplitude=.5pt, segment length=3pt}] (2) -- (8);

    \draw[-, thick] (4) -- (9);

    \draw[-, thick,decorate, decoration={snake, amplitude=.5pt, segment length=3pt}] (5) -- (10);

  \end{tikzpicture}}}
    \caption{VBpq}
  \end{subfigure}\hfill
  \begin{subfigure}{0.32\textwidth}
    \centering
    \raisebox{1cm}{\scalebox{2}{\begin{tikzpicture}[>=stealth,
  inner sep=0pt, outer sep=0pt,
  shorten <=-0.55pt, shorten >=-0.55pt,
  baseline=(current bounding box.center)]
    \node[inner sep=0pt] (1) at (1.5,0.54) {};
    \node[inner sep=0pt] (2) at (1.5,0) {};
    \node[inner sep=0pt] (3) at (1.2,0) {};
    \node[inner sep=0pt] (4) at (0.5,0) {};
    \node[inner sep=0pt] (5) at (0.5,0.54) {};
    \node[inner sep=0pt] (6) at (1.2,.54) {};
    \node[inner sep=0pt] (7) at (2,0.54) {};
    \node[inner sep=0pt] (8) at (2,0) {};
    \node[inner sep=0pt] (9) at (0.00,0) {};
    \node[inner sep=0pt] (10) at (0.00,0.54) {};

    \draw[-, thick] (1) -- (2);
    \node[edgelabel] at ($ (1)!0.5!(2) $) {$2$};

    \draw[-, thick] (2) -- (3);
    \node[edgelabel] at ($ (2)!0.5!(3) $) {$6$};

    \draw[-, thick] (6) -- (1);
    \node[edgelabel] at ($ (6)!0.5!(1) $) {$7$};

    \draw[-, thick] (3) -- (4);
    \node[edgelabel] at ($ (3)!0.5!(4) $) {$4$};

    \draw[-, thick] (4) -- (6);
    \node[edgelabel] at ($ (4)!0.7!(6) $) {$1$};

    \draw[-, thick] (5) -- (6);
    \node[edgelabel] at ($ (5)!0.5!(6) $) {$5$};

    \draw[-, thick] (3) -- (5);
    \node[edgelabel] at ($ (3)!0.3!(5) $) {$3$};

    \draw[-, thick,decorate, decoration={snake, amplitude=.5pt, segment length=3pt}] (1) -- (7);

    \draw[-, thick] (2) -- (8);

    \draw[-, thick] (4) -- (9);

    \draw[-, thick,decorate, decoration={snake, amplitude=.5pt, segment length=3pt}] (5) -- (10);

  \end{tikzpicture}}}
    \caption{VNP}
  \end{subfigure}
  \medskip
  \begin{subfigure}{0.49\textwidth}
    \raggedleft
    \scalebox{2}{\begin{tikzpicture}[>=stealth,
  inner sep=0pt, outer sep=0pt,
  shorten <=-0.55pt, shorten >=-0.55pt,
  baseline=(current bounding box.center)]
    \node[inner sep=0pt] (1) at (1.2,0.8) {};
    \node[inner sep=0pt] (2) at (1.2,0) {};
    \node[inner sep=0pt] (3) at (0.6,0.4) {};
    \node[inner sep=0pt] (4) at (0.6,0) {};
    \node[inner sep=0pt] (5) at (0.6,0.8) {};
    \node[inner sep=0pt] (6) at (1.2,0.4) {};
    \node[inner sep=0pt] (7) at (1.8,0.8) {};
    \node[inner sep=0pt] (8) at (1.8,0) {};
    \node[inner sep=0pt] (9) at (0.00,0) {};
    \node[inner sep=0pt] (10) at (0.00,0.8) {};

    \draw[-, thick] (1) -- (6);
    \node[edgelabel] at ($ (1)!0.5!(6) $) {$1$};

    \draw[-, thick] (6) -- (2);
    \node[edgelabel] at ($ (6)!0.5!(2) $) {$2$};

    \draw[-, thick] (1) -- (5);
    \node[edgelabel] at ($ (1)!0.5!(5) $) {$4$};

    \draw[-, thick] (2) -- (4);
    \node[edgelabel] at ($ (2)!0.5!(4) $) {$7$};

    \draw[-, thick] (3) -- (6);
    \node[edgelabel] at ($ (3)!0.5!(6) $) {$3$};

    \draw[-, thick] (3) -- (4);
    \node[edgelabel] at ($ (3)!0.5!(4) $) {$6$};

    \draw[-, thick] (3) -- (5);
    \node[edgelabel] at ($ (3)!0.5!(5) $) {$5$};

    \draw[-, thick] (1) -- (7);

    \draw[-, thick,decorate, decoration={snake, amplitude=.5pt, segment length=3pt}] (2) -- (8);

    \draw[-, thick] (4) -- (9);

    \draw[-, thick,decorate, decoration={snake, amplitude=.5pt, segment length=3pt}] (5) -- (10);

  \end{tikzpicture}}
    \captionsetup{justification=raggedleft}
    \caption{HB}
  \end{subfigure}\hfill
  \begin{subfigure}{0.49\textwidth}
    \raggedright
    \scalebox{2}{\begin{tikzpicture}[>=stealth,
  inner sep=0pt, outer sep=0pt,
  shorten <=-0.55pt, shorten >=-0.55pt,
  baseline=(current bounding box.center)]
    \node[inner sep=0pt] (1) at (1.2,0.8) {};
    \node[inner sep=0pt] (2) at (1.2,0) {};
    \node[inner sep=0pt] (3) at (0.6,0.25) {};
    \node[inner sep=0pt] (4) at (0.6,0) {};
    \node[inner sep=0pt] (5) at (0.6,0.8) {};
    \node[inner sep=0pt] (6) at (1.2,0.25) {};
    \node[inner sep=0pt] (7) at (1.8,0.8) {};
    \node[inner sep=0pt] (8) at (1.8,0) {};
    \node[inner sep=0pt] (9) at (0.00,0) {};
    \node[inner sep=0pt] (10) at (0.00,0.8) {};

    \draw[-, thick] (1) -- (6);
    \node[edgelabel] at ($ (1)!0.4!(6) $) {$1$};

    \draw[-, thick] (6) -- (2);
    \node[edgelabel] at ($ (6)!0.5!(2) $) {$7$};

    \draw[-, thick] (1) -- (3);
    \node[edgelabel] at ($ (1)!0.7!(3) $) {$4$};

    \draw[-, thick] (2) -- (4);
    \node[edgelabel] at ($ (2)!0.5!(4) $) {$2$};

    \draw[-, thick] (3) -- (5);
    \node[edgelabel] at ($ (3)!0.6!(5) $) {$3$};

    \draw[-, thick] (3) -- (4);
    \node[edgelabel] at ($ (3)!0.5!(4) $) {$6$};

    \draw[-, thick] (6) -- (5);
    \node[edgelabel] at ($ (6)!0.3!(5) $) {$5$};

    \draw[-, thick] (1) -- (7);

    \draw[-, thick,decorate, decoration={snake, amplitude=.5pt, segment length=3pt}] (2) -- (8);

    \draw[-, thick] (4) -- (9);

    \draw[-, thick,decorate, decoration={snake, amplitude=.5pt, segment length=3pt}] (5) -- (10);

  \end{tikzpicture}}
    \captionsetup{justification=raggedright}
    \caption{HNP}
  \end{subfigure}
  \caption{Graphical representation of the forward topologies
    \cref{eq:topodefs} used in our calculation. See text for details.}
  \label{fig:topologies}
\end{figure}
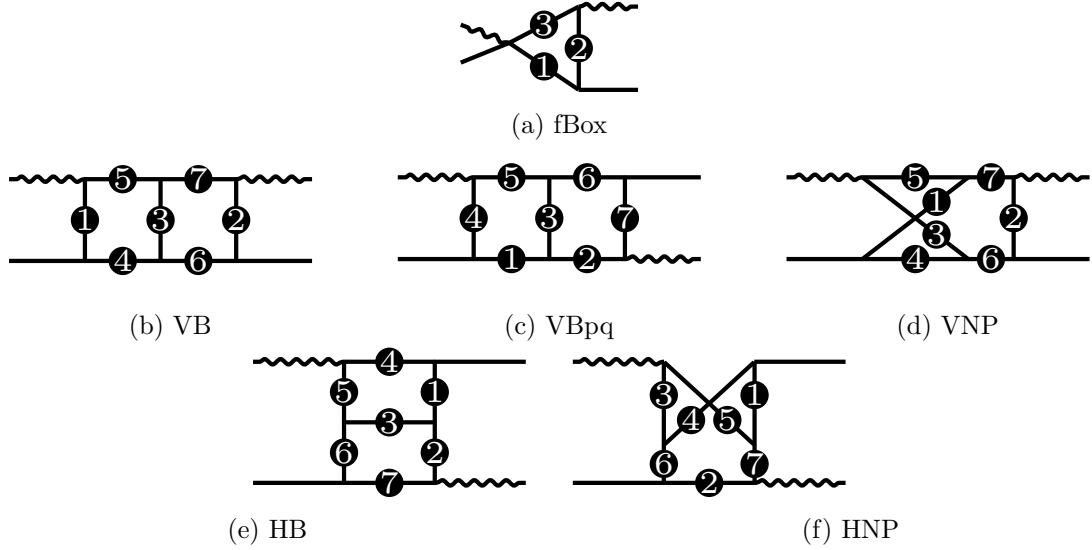

We now discuss the relevant forward topologies
$\overline{\mathcal{T}}$ needed to map all the NNLO diagrams. To do
so, we introduce the following list of propagator-like structures
\begin{equation*}
  \begin{aligned}
    \Dp_1 &= l_1^2 &&
    \Dp_5 = (l_1+q)^2  &&
    \Dp_{9} = (l_2-p)^2 &&
    \Dp_{13} = (l_2+p)^2  \\
    \Dp_2 &= (l_1-p)^2 &&
    \Dp_6 = (l_1+p)^2 &&
    \Dp_{10} = (l_2-p-q)^2 &&
    \Dp_{14} = (l_2+p-q)^2\\
    \Dp_{3} &= (l_1-p-q)^2 &&
    \Dp_7 = (l_1+p-q)^2 &&
    \Dp_{11} = (l_2-q)^2 &&
    \Dp_{15} = (l_1-l_2)^2\\
    \Dp_4 &=  (l_1-q)^2 &&
    \Dp_8 = l_2^2  &&
    \Dp_{12} = (l_2+q)^2 &&
    \Dp_{16} = (l_1-l_2-q)^2
  \end{aligned}
\end{equation*}
in terms of which we can define the following master topologies
spanning all independent scalar products
\begin{equation}
  \label{eq:topodefs}
  \begin{split}
    \mathrm{Top}^\mathrm{fBox} &=
    \{\Dp_1,\Dp_2,\Dp_3\}\\ 
    \mathrm{Top}^\mathrm{VB} &=
    \{\Dp_1,\Dp_8,\Dp_{15},\Dp_2,\Dp_5,\Dp_9,\Dp_{12}\}\\ 
    \mathrm{Top}^\mathrm{HB} &=
    \{\Dp_1,\Dp_8,{\Dp}_{15},{\Dp}_{6},\Dp_{7},\Dp_{14},{\Dp}_{11}\} \\
    \mathrm{Top}^\mathrm{VBpq} &=
    \{{\Dp}_1,\Dp_8,{\Dp}_{15},\Dp_2,\Dp_3,{\Dp}_{10},\Dp_{11}\} \\ 
    \mathrm{Top}^\mathrm{VNP} &=
    \{\Dp_1,\Dp_8,\Dp_{15},\Dp_{2},\Dp_{16},{\Dp}_{9},{\Dp}_{12}\} \\
    \mathrm{Top}^\mathrm{HNP} &=
    \{\Dp_1,{\Dp}_8,\Dp_{15},{\Dp}_{6},{\Dp}_{16},\Dp_{13},\Dp_{12}\}. 
  \end{split}
\end{equation}
These topologies are depicted in fig.~\ref{fig:topologies}.  Each of
them appears with various combinations of cuts and massive
propagators. 

\section{Computing the master integrals}
\label{sec:MIs}
After the procedure described in the previous section, we are left
with 100 different master integrals:
\begin{equation}
\begin{array}{c|cccccccc}
  \text{contribution} & \text{V}_1 & \text{R}_1 &
  \text{VV}_0 & \text{VV}_1 & \text{RV}_1 &
  \text{RR}_1 & \text{RR}_2 & \text{RR}_3 \\
\hline
\text{\# MIs} & 2 & 2 & 4 & 18 & 28 & 30 & 4 & 12
\end{array}\,.
\end{equation}
In this section, we will briefly review how we computed them.

\subsection{Canonical basis for the polylogarithmic and elliptic sectors}
\label{sec:can}
We evaluate all our integrals (both genuine loop integrals and
phase-space integrals) using the method of differential
equations~\cite{Kotikov:1990kg,Remiddi:1997ny,Gehrmann:1999as,Henn:2013pwa}.
We denote the collection of master integrals as $\fvec$, and note that
they depend on the kinematic variables $s_i = \{Q^2,x, z\}$ as well
as the spacetime dimension $d=4-2\ep$.  We use IBP relations to write
derivatives in kinematic scalars $s_i$ as combinations of master
integrals, thus allowing us to find a closed system of differential
equations (DEs) of the form\footnote{From here onward it is
understood that differentials are to be taken only with respect to
kinematic variables, not spacetime dimensions.}
\begin{align}
  d\fvec(s_i) = \sum_i ds_i \cdot \partial_{s_i}\fvec(s_i) =
  d \Avec(s_i,\ep)\cdot\fvec(s_i).
  \label{eq:denoncanon}
\end{align}
In \cref{eq:denoncanon} we suppressed the $\ep$-dependence of the
master integrals (that we treat as parametric) to avoid confusion.
Since $Q^2$ is the only dimensionful quantity, its dependence can be
fully reconstructed from dimensional analysis. We can then rescale all
the integrals to completely remove the $Q^2$ dependence, and are left
with a non-trivial DE in the dimensionless parameters $x$ and $z$.
To deal with it, we first sort our master integrals $\fvec$ in
ascending order of complexity within each occurring contribution
$\mathrm{V}_1$, $\mathrm{R}_1$, $\mathrm{VV}_0$, $\mathrm{VV}_1$,
$\mathrm{RV}_1$, $\mathrm{RR}_1$, $\mathrm{RR}_2$ and $\mathrm{RR}_3$.
This makes the coupling matrix $ d\Avec$ for each of the contributions
lower block-triangular.  Integrals within the same sector (\ie
integrals that share the same propagators) couple to themselves via
the blocks on the diagonal, in addition to being coupled to subsectors
(\ie integrals of a given sector where some propagators are
missing). To solve DEs of the form in \cref{eq:denoncanon} in a series
expansion in $\ep$, it is standard to transform to a suitable basis
via iterated linear transformations
\begin{align}
  \widetilde{\fvec} = \Bvec_r\cdots \Bvec_1 \cdot \fvec
  \equiv \Bvec \cdot \fvec,
\end{align}
such that the differential equations are brought in an
$\ep$-factorised form
\begin{align}
  \label{eq:epfac}
  d \widetilde{\fvec} = \ep \,d\widetilde{ \Avec}\cdot\widetilde{\fvec},
\end{align}
with $\ep \, d\widetilde{ \Avec} = (\Bvec \cdot d\Avec +
d\Bvec)\cdot\Bvec^{-1}$.  If $ d\widetilde{\Avec}$ is also written in
terms of differentials of logarithms (or higher genus generalisations
thereof) we call the basis $\widetilde{\fvec}$
canonical~\cite{Henn:2013pwa}. In this case, the Laurent expansion of
the solution around $\ep=0$ can be immediately written in terms of
Chen iterated integrals~\cite{Chen:1977oja}, see
\cref{sec:chen_and_boundaries}.

To find a canonical basis, we first focused on the homogeneous part of
the differential equation (\ie on the terms on the block diagonal).
Starting with a set of master integrals $\gvec$ within a sector such
that their differential equation can be written as
\begin{align}
  d\gvec = d\Hvec \cdot \gvec + \hvec,
\end{align}
where $d\Hvec$ and $\hvec$ are the homogeneous and inhomogeneous parts
of the DE respectively, we can project on the homogeneous part by
taking the \textit{maximal cut} of the MIs $\gvec\mapsto
\mathrm{Cut}[\gvec]$, \ie by replacing all propagators $\Dp_i$ by
$\sd{\Dp}_i$.  Taking the maximal cut does not change the homogeneous
coupling $d\Hvec$ since cutting commutes with IBP relations, but it
does remove the coupling $\hvec$ to subsectors, since integrals in
$\hvec$ have by construction fewer propagators, so they vanish on the
maximal cut. Schematically, we can write
\begin{align}
  d\mathrm{Cut}[\gvec] = d\Hvec \cdot \mathrm{Cut}[\gvec].
\end{align}
To find good candidates for a canonical basis, we studied the leading
singularity (LS)~\cite{Cachazo:2008vp}\footnote{Since we are working
on the maximal cut, with a slight abuse of notation in this section we
will use the concepts of maximal cut and leading singularity
interchangeably.} of our integrals on the maximal cut, for integer
dimensions, which we computed either by directly solving the maximal
cut equation in $d=4$ (or $d=2$) or by iteratively taking residues in
Baikov
representation~\cite{Baikov:1996iu,Baikov:1996rk,Frellesvig:2017aai,Frellesvig:2024ymq}.

For almost all integrals, the LS turned out to be an algebraic
function. In these cases, dividing by the LS provides a good candidate
for a canonical integral, since the latter -- at least in the
polylogarithmic case -- has unit LS. In practice, some integrals had a
very complex leading singularity. To deal with this, we first explored
different ways of modifying the integral, (\eg by squaring some of the
propagators, which leads to better-behaved integrals in the UV, or by
working in $d=2$ and then using dimensional shift
relations~\cite{Tarasov:1996br,Lee:2009dh} to go back to $d=4$) to
simplify the LS analysis.
Following these ideas, we were able to find a canonical basis for the
homogeneous differential equation of all integrals, apart from one
sector in RR$_3$. After this step was done, it was straightforward to
arrange the coupling to subsectors to obtain a fully-canonical system.

In the case of RR$_3$, the lowest subsector contains the three-mass
sunrise graph
\begin{equation*}
  \scalebox{1.5}{
      \begin{tikzpicture}[
    >=stealth,
  inner sep=0pt, outer sep=0pt,
  shorten <=-0.55pt, shorten >=-0.55pt,
  baseline=(current bounding box.center)]
  \node[inner sep=0pt] (2) at (0.44,0.1) {};
  \node[inner sep=0pt] (5) at (1.44,0.1) {};
  \node[inner sep=0pt] (e2) at (1.84,0.1) {};
  \node[inner sep=0pt] (e4) at (0.00,0.1) {};
  \node[inner sep=0pt] (e2u) at (1.84,0.2) {};
  \node[inner sep=0pt] (e4u) at (0.00,0.2) {};
  \node[inner sep=0pt] (e2d) at (1.84,0.0) {};
  \node[inner sep=0pt] (e4d) at (0.00,0.0) {};

  \draw[-, thick] (e2u) to (5);
  \draw[-, thick] (e4u) to (2);
  \draw[-, thick,decorate, decoration={snake, amplitude=.5pt, segment length=3pt}] (e2d) to (5);
  \draw[-, thick,decorate, decoration={snake, amplitude=.5pt, segment length=3pt}] (e4d) to (2);
  \draw[-, thick,double] (2) to (5);
  \draw[-, thick, double, bend left=90] (5) to (2);
  \draw[-, thick, double, bend left=90] (2) to (5);
  \draw[red, thick, dashed, dash pattern=on 2pt off 1pt]
    let \p1 = (current bounding box.north), \p2 = (current bounding box.south)
    in (0.94,\y2) -- (0.94,\y1);
  \end{tikzpicture}
  },
\end{equation*}
which is famously of elliptic type and cannot be put into
polylogarithmic form. Interestingly, this is the only sector that
appeared in our calculation which is elliptic on its maximal cut.
Fortunately, this sector and its two master integrals are extremely
well-studied.  For our calculation, we closely followed
ref.~\cite{Duhr:2025lbz}. For completeness, we will report here the
main points that are relevant for our calculation.
To highlight the key steps that we performed to obtain a canonical
basis in the elliptic sector, we first discuss a simple,
(non-elliptic) sunrise diagram with one mass. We then discuss the
generalisation to the elliptic case.

\paragraph{Example: one-mass cut sunrise}
We consider the coupled system of the one-mass cut sunrise
\begin{equation*}
  \scalebox{1.5}{
    \begin{tikzpicture}[
    >=stealth,
  inner sep=0pt, outer sep=0pt,
  shorten <=-0.55pt, shorten >=-0.55pt,
  baseline=(current bounding box.center)]
  \node[inner sep=0pt] (2) at (0.44,0.1) {};
  \node[inner sep=0pt] (5) at (1.44,0.1) {};
  \node[inner sep=0pt] (e2) at (1.84,0.1) {};
  \node[inner sep=0pt] (e4) at (0.00,0.1) {};
  \node[inner sep=0pt] (e2u) at (1.84,0.2) {};
  \node[inner sep=0pt] (e4u) at (0.00,0.2) {};
  \node[inner sep=0pt] (e2d) at (1.84,0.0) {};
  \node[inner sep=0pt] (e4d) at (0.00,0.0) {};

  \draw[-, thick] (e2u) to (5);
  \draw[-, thick] (e4u) to (2);
  \draw[-, thick,decorate, decoration={snake, amplitude=.5pt, segment length=3pt}] (e2d) to (5);
  \draw[-, thick,decorate, decoration={snake, amplitude=.5pt, segment length=3pt}] (e4d) to (2);
  \draw[-, thick,double] (2) to (5);
  \draw[-, thick, bend left=90] (5) to (2);
  \draw[-, thick, bend left=90] (2) to (5);
  \draw[red, thick, dashed, dash pattern=on 2pt off 1pt]
    let \p1 = (current bounding box.north), \p2 = (current bounding box.south)
    in (0.94,\y2) -- (0.94,\y1);
  \end{tikzpicture}
  },
\end{equation*}
close to $d = 2$. We multiply integrals in this sector by powers of
$(p+q)^2$ to obtain dimensionless integrals that only depend on the
variable
\begin{equation}\label{eq:wdef2}
  w \equiv \frac{m_q^2}{(p+q)^2} = \frac{x z}{1-x}.
\end{equation}
On the maximal cut, the differential equation for this sector contains
two coupled master integrals $\fvec = (f_1,f_2)$. We choose $f_1$ to
be the phase-space integral rescaled by the appropriate power of
$(p+q)^2$ to make it dimensionless, and $f_2$ to be the derivative of
$f_1$ with respect to $w$, $f_2 = \partial_w f_1$. With this
definition, the differential equation takes the simple form
\begin{align}
  \label{eq:1msun}
  \partial_w\fvec = \left(\Avec^{(w)}+\ep\Bvec^{(w)}(\ep)\right)\cdot\fvec,
\end{align}
with
\begin{equation}
  \Avec^{(w)} = \begin{pmatrix}
    0 & 1 \\
    \alpha(w) & \beta(w)
  \end{pmatrix},
  \quad\quad\quad
  \Bvec^{(w)}(\ep) = \begin{pmatrix}
    0 & 0 \\
    \frac{5+6\ep}{1-w}+\frac{5+6\ep}{w} &
    \frac{4}{1-w}-\frac{1}{w}
  \end{pmatrix},
\end{equation}
and
\begin{equation}
  \alpha(w) = \frac{1}{w(1-w)},\quad
  \beta(w) = \frac{1-3w}{w(w-1)}.
\end{equation}
Neglecting for the time being $\mathcal O(\ep)$ terms, \cref{eq:1msun}
can be recast as a second-order differential equation for $f_2$:
\begin{equation}
  f_2''(w) = \alpha(w) f_2(w) + \beta(w) f_2'(w).
  \label{eq:f2eq}
\end{equation}
This equation admits two independent solutions, $u_0$ and
$u_1$. Indeed, the general solution of \cref{eq:f2eq} reads
\begin{equation}
f_2 = c_1\frac{1}{1-w} + c_2\frac{\ln w}{1-w}.
\end{equation}
We choose $u_0$ to be the solution with $c_1=1$, $c_2=0$, and $u_1$ to
be the one with $c_1=0$, $c_2=1$. Note that with this choice $u_0$ is
purely rational, while $u_1$ diverges like a log at the regular
singular point $w=0$, $u_1 = \ln(w)u_0$. Following
refs~\cite{Broedel:2018qkq,Duhr:2025lbz}, we now study the Wronskian
associated to~\cref{eq:f2eq}
\begin{align}
    \mathbf{W}(w) = \begin{pmatrix}
        u_0(w) & u_1(w) \\
        u_0'(w) & u_1'(w)
        \end{pmatrix},
\end{align}
which satisfies the following differential equation
\begin{equation}
    \partial_w \mathbf{W} (\mathbf{W})^{-1} =  \Avec^{(w)},
\end{equation}
and whose determinant
\begin{align}
  D(w) \equiv \det(\mathbf{W}) = D(w_0) \exp{\left(\int_{w_0}^w
    \beta(s) ds\right)}=\frac{1}{w(1-w)^2}
\end{align}
is algebraic. According to refs~\cite{Broedel:2018qkq,Duhr:2025lbz},
we now split the Wronskian into semi-simple and unipotent
parts\footnote{To make this decomposition unique, we ask the unipotent
part to be upper triangular. See appendix B in
ref.~\cite{Broedel:2019kmn} for a general algorithm to perform this
separation.}, $\mathbf{W} = \mathbf{W}^{\mathrm{ss}}\cdot
\mathbf{W}^{\mathrm{u}}$, which in our case read
\begin{equation}
    \mathbf{W}^{\mathrm{ss}} = \begin{pmatrix}
        u_0 & 0 \\
        u_0' & \frac{\det(\mathbf{W})}{u_0}
    \end{pmatrix},\quad\quad
        \mathbf{W}^{\mathrm{u}} = \begin{pmatrix}
        1 & \frac{u_1}{u_0} \\
        0 & 1
        \end{pmatrix}.
\end{equation}
We note that the semi-simple part does not contain $u_1$, and as a
consequence is purely rational, while the unipotent part is already in
$d\log$ form, $d(u_1/u_0) = d\log(w)$. To obtain a canonical basis we
then rotate away the rational terms in the semi-simple part, \ie we
define a new basis $\tilde \fvec \equiv (\mathbf{W^{ss}})^{-1} \fvec$,
which obeys the differential equation
\begin{equation}
  \partial_w\tilde\fvec =
  \left[\partial_w\mathbf{W^u}\cdot (\mathbf{W^u})^{-1}+\mathcal
    O(\ep)\right]\cdot\tilde\fvec =
  \left[\begin{pmatrix}
      0 & \frac{1}{w} \\
      0 & 0
    \end{pmatrix} + \mathcal O(\ep)\right]\cdot\tilde\fvec.
\end{equation}
This is not yet canonical, first because the $d\log$ contribution is
not multiplied by $\ep$, and second because of the extra $\mathcal
O(\ep)$ terms. To solve the first issue, we simply define a rescaled
solution $\tilde \gvec \equiv {\rm diag}(\ep,1)
\cdot(\mathbf{W^{ss}})^{-1} \cdot\fvec$. Reinstating the $\mathcal
O(\ep)$ terms coming from $\Bvec^{(w)}$ in \cref{eq:1msun}, $\tilde
\gvec$ obeys the full differential equation
\begin{equation}
  \partial_w\tilde\gvec =
  \left[\ep\begin{pmatrix}
        0 & \frac{1}{w} \\
        \frac{6}{1-w} & \frac{4}{1-w}-\frac{1}{w}
    \end{pmatrix}
    +
    \begin{pmatrix}
        0 & 0 \\
        \frac{4}{(1-w)^2} & 0
    \end{pmatrix}
    \right]\cdot\tilde\gvec.
\end{equation}
This new equation is canonical apart from the second matrix in the
bracket. Crucially, this matrix is nilpotent, so it can be removed
with a simple redefinition of the coupling between the two integrals,
without touching the normalisation of the integrals (\ie the terms on
the diagonals). Using an ansatz of the form
\begin{equation}
  \gvec\equiv
  \begin{pmatrix}
        1 & 0 \\
        k(w) & 1
  \end{pmatrix}
  \cdot{\rm Diag}(\ep,1)\cdot(\mathbf{W^{ss}})^{-1}\cdot \fvec,
\end{equation}
we see that $\gvec$ obeys a canonical differential equation if
$k$ is such that
\begin{equation}
  k'(w) + \frac{4}{(1-w)^2}=0\quad \longrightarrow\quad
  k(w) = -\frac{4}{(1-w)} + c,
\end{equation}
for arbitrary $c$. 

While the construction above is overly complicated for finding a
canonical basis in this very simple case, this method can be directly
generalised to the elliptic case, as we now discuss. Once again, we
closely follow ref.~\cite{Duhr:2025lbz}.

\paragraph{Elliptic generalisation of the previous example} We now consider
the equal-mass sunrise diagram
\begin{equation*}
  \scalebox{1.5}{
      \begin{tikzpicture}[
    >=stealth,
  inner sep=0pt, outer sep=0pt,
  shorten <=-0.55pt, shorten >=-0.55pt,
  baseline=(current bounding box.center)]
  \node[inner sep=0pt] (2) at (0.44,0.1) {};
  \node[inner sep=0pt] (5) at (1.44,0.1) {};
  \node[inner sep=0pt] (e2) at (1.84,0.1) {};
  \node[inner sep=0pt] (e4) at (0.00,0.1) {};
  \node[inner sep=0pt] (e2u) at (1.84,0.2) {};
  \node[inner sep=0pt] (e4u) at (0.00,0.2) {};
  \node[inner sep=0pt] (e2d) at (1.84,0.0) {};
  \node[inner sep=0pt] (e4d) at (0.00,0.0) {};

  \draw[-, thick] (e2u) to (5);
  \draw[-, thick] (e4u) to (2);
  \draw[-, thick,decorate, decoration={snake, amplitude=.5pt, segment length=3pt}] (e2d) to (5);
  \draw[-, thick,decorate, decoration={snake, amplitude=.5pt, segment length=3pt}] (e4d) to (2);
  \draw[-, thick,double] (2) to (5);
  \draw[-, thick, double, bend left=90] (5) to (2);
  \draw[-, thick, double, bend left=90] (2) to (5);
  \draw[red, thick, dashed, dash pattern=on 2pt off 1pt]
    let \p1 = (current bounding box.north), \p2 = (current bounding box.south)
    in (0.94,\y2) -- (0.94,\y1);
  \end{tikzpicture}
    }.
\end{equation*}
As in the previous case there are two independent integrals on the
maximal cut, $\fvec = (f_1,f_2)$.  We choose $f_1$ as the
dimensionless\footnote{Again here we rescale by the appropriate power
of $(p+q)^2$.}  phase-space integral and $f_2$ as its derivative
with respect to $w$, $f_2 \equiv \partial_w f_1$. The approach to
finding a canonical basis is entirely analogous to what we described
above in the polylogarithmic case, only now the entries of the ($d=2$)
Wronskian, which we call
\begin{equation}
    \mathbf{W} = \begin{pmatrix}
        w_0 & w_1 \\
        w_0' & w_1'
\end{pmatrix},
\end{equation}
satisfy a more complicated differential equation
\begin{align}
  \label{eq:w0eq}
  w_i''(w) = \left(\frac{3}{4}\frac{1}{1-w}+\frac{3}{w}
  +\frac{81}{4}\frac{1}{1-9w}\right) w_i (w)
  + \left(\frac{1}{1-w}-\frac{1}{w}+\frac{9}{1-9 w}\right) w_i'(w).
\end{align}
Contrary to~\cref{eq:f2eq}, this equation does not admit a purely
algebraic solution. A possible way forward is to work
locally~\cite{Broedel:2018qkq,Duhr:2025lbz}, close to a so-called
maximal unipotent monodromy (MUM) point. These are regular singular
points where the solutions of~\cref{eq:w0eq} diverge like a logarithm.
Specifically, close to a MUM point (which we now take to be $w=0$ for
definiteness), one can write $w_0$ as a Taylor series, and $w_1 =
\ln(w) w_0 + \sum_{i=1}^{\infty} d_i w^i$. Working close to the MUM
point allows one to naturally generalise the polylogarithmic
construction described above, the only difference being that now
$w_1/w_0$ is a logarithmic form only close to the MUM point. To find a
canonical basis we can proceed as above, but this time keep $w_0$
implicit, and perform the semi-simple rotation in terms of $w_0$ and
$w'_0$. When we have to evaluate the resulting function, we work close
to a MUM point and Taylor expand the solution for $w_0$. We note that
in our case, all the regular singular points of~\cref{eq:w0eq} are MUM
points: $w=0,1/9,1,\infty$.

\begin{figure}
  \centering
  \includegraphics[width=10cm]{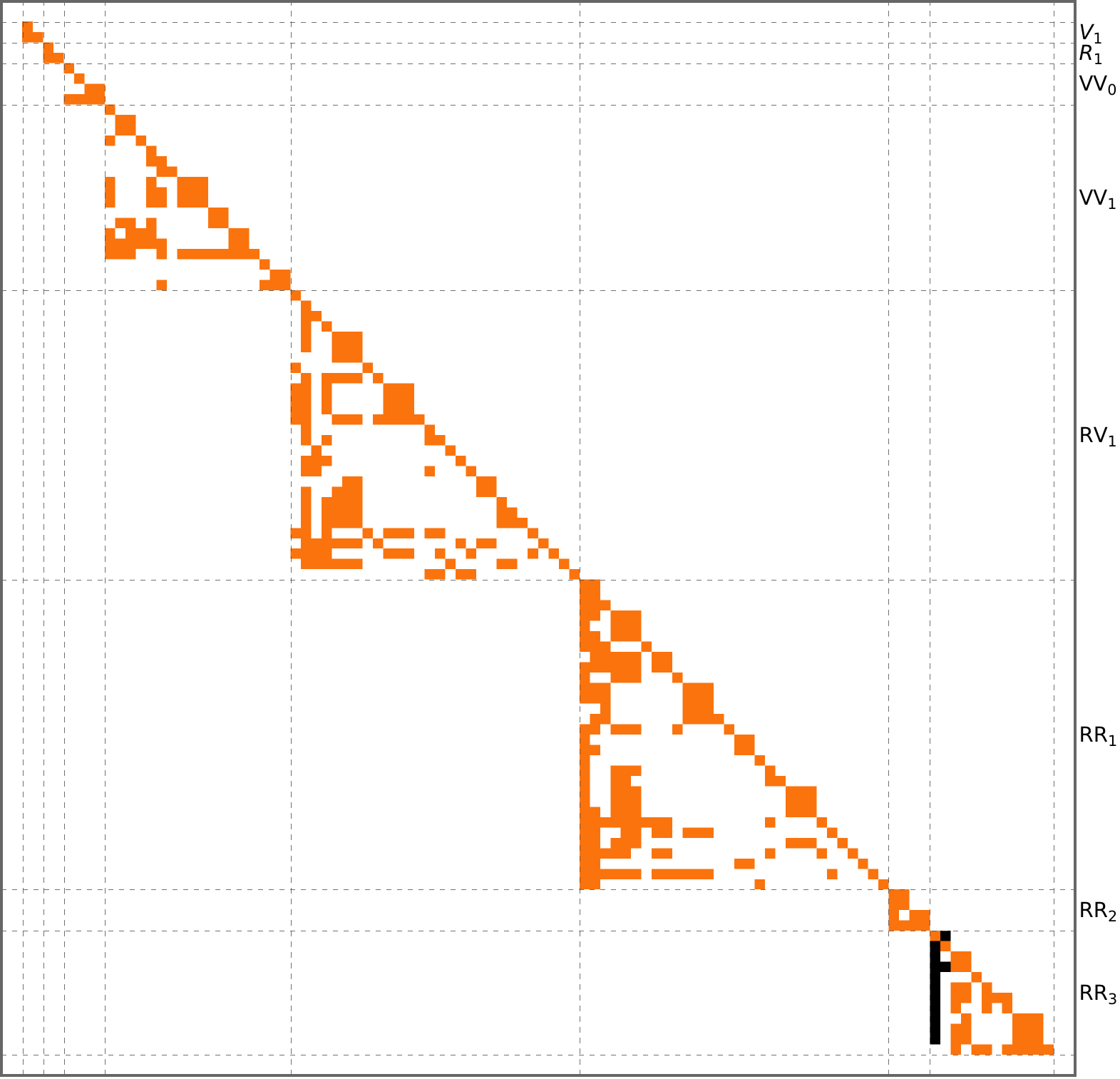}
  \caption{Schematic depiction of the coupling matrix for all NNLO
    canonical master integrals. Elliptic integration kernels are
    denoted in black, while $d\log$ kernels are denoted in orange.}
  \label{fig:desschem}
\end{figure}

Before concluding this section, we note that there is an extra
subtlety that we had to tackle when finding a canonical basis for the
elliptic sector.  Indeed, while the method highlighted above allows
one to find a canonical basis for the homogeneous $2\times2$ sunrise
system, we have to keep in mind that it also couples to all higher
sectors within $\mathrm{RR}_3$.  Most of these couplings could be made
$\ep$ factorised without introducing any other functions.  However, to
fix the three-mass sunrise coupling to the following ice-cone sector
\begin{equation*}
  \scalebox{1.5}{
  \begin{tikzpicture}[
    >=stealth,
  inner sep=0pt, outer sep=0pt,
  shorten <=-0.55pt, shorten >=-0.55pt,
  baseline=(current bounding box.center)]
  \node[inner sep=0pt] (2) at (1.66,1.71) {};
  \node[inner sep=0pt] (e1) at (1.9,2) {};
  \node[inner sep=0pt] (5) at (0.78,1.26) {};
  \node[inner sep=0pt] (e2) at (0.3,1.26) {};
  \node[inner sep=0pt] (e2u) at (0.3,1.37) {};
  \node[inner sep=0pt] (e2d) at (0.3,1.15) {};
  \node[inner sep=0pt] (6) at (1.65,0.80) {};
  \node[inner sep=0pt] (e4) at (1.9,.5) {};
  \draw[-, thick] (2) -- (e1);
  \node[font=\scriptsize] at ($ (2)!0.5!(e1) $) {};

  \draw[-, thick] (5) -- (e2u);
  \node[font=\scriptsize] at ($ (5)!0.5!(e2) $) {};

  \draw[-, thick, decorate, decoration={snake, amplitude=.5pt, segment length=3pt}] (5) -- (e2d);
  \node[font=\scriptsize] at ($ (5)!0.5!(e2) $) {};

  \draw[-, thick,decorate, decoration={snake, amplitude=.5pt, segment length=3pt}] (6) -- (e4);
  \node[font=\scriptsize] at ($ (6)!0.5!(e4) $) {};

  \draw[-, thick, double, bend right=30] (2) to (5);
  \node[font=\scriptsize] at ($ (2)!0.5!(5) $) {};

  \draw[-, thick, double ] (5) -- (6);
  \node[font=\scriptsize] at ($ (5)!0.5!(6) $) {};

  \draw[-, thick] (6) -- (2);
  \node[font=\scriptsize] at ($ (6)!0.5!(2) $) {};

  \draw[-, thick, double, bend left=30] (2) to (5);
  \node[font=\scriptsize] at ($ (2)!0.5!(5) $) {};

  \draw[red, thick, dashed, dash pattern=on 2pt off 1pt]
    let \p1 = (current bounding box.north), \p2 = (current bounding box.south)
    in (1.3,\y2) -- (1.3,\y1);

  \end{tikzpicture}
  },
\end{equation*}
we needed to introduce a second implicit function $g^\mathrm{sc}$,
which depends on $w_0$. It is defined through its differential
\begin{align}
  \label{eq:fdef}
  \begin{split}
    dg^{\mathrm{sc}}(w,z) &\equiv
    \frac{1}{6}(w+z)^{-2}\left(\frac{(z+1)
      \left(-3 w z+w+z^2+z\right)}{w+z}\right)^{-3/2} \\
    \times&\Bigg\{
    dw \bigg[{z (z+1) w_0(w)
        \left(w^2 (9-27 z)+3 w (z (3 z+8)-1)-z (7 z+5)\right)} \\
      &+{(w-1) (9 w-1) z (z+1) \left(-3 w
        z+w+z^2+z\right) w_0'(w)}\bigg] \\
    &+dz\bigg[{(1-w) w (9 w-1) (z+1) \left(-3 w z+w+z^2+z\right) w_0'(w)} \\
      &+  {(1-w) (9 w-1) z w_0(w) \left(-3 w z-w+z^2+z\right)}\bigg]
    \Bigg\}.
  \end{split}
\end{align}
Since the homogeneous differential equation of the three-mass sunrise
is the only one that needs this special treatment, and all the other
homogeneous differential equations for the remaining $\mathrm{RR}_3$
integrals can be put in $d\log$ form,\footnote{We thank Christoph Nega
for helping to confirm that one of the couplings was indeed a
$d\log$.}  the $w_0$ and $g^\mathrm{sc}$ implicit functions are enough
to obtain a canonical basis. Fig.~\ref{fig:desschem} shows a schematic
representation of the coupling matrix of all 100 master integrals
in a canonical basis.
For completeness, we report all our MIs in the original (non-canonical)
basis in \cref{app:appmi}.

\subsection{Solution in terms of Chen iterated integrals}
\label{sec:chen_and_boundaries}
In the previous section we obtained differential equations for all
necessary master integrals in canonical form:
\begin{align} \label{eq:de}
  d\fvec = \ep\, d \Avec \cdot \fvec,
  \quad\quad\quad
  d\Avec = \sum_{i=1} \Avec_i \, d W_{i} ,
\end{align}
with $\Avec_i$ constant matrices and $dW_{i}$ either $d\log$ or
elliptic 1-forms. In the case of $d\log$ forms, we use the notation
\begin{equation}
  dW_i = d\ln(\alpha_i),
\end{equation}
refer to $\alpha_i$ as ``letters'', and to the set of all the letters
as the alphabet of our problem.
In our case, the procedure highlighted in the previous subsection
leads to a large number of kernels, many of which are redundant.  To
reduce to a minimal set, we first find the set of linear relations
among the entries of $d \Avec$ by solving for $c_{ij} \in \mathbb{Q}$
the system of two equations
\begin{equation} 
  \begin{cases}
    \sum_{i,j} c_{ij}
    \left(\frac{\partial \Avec}{\partial x}\right)_{ij} = 0 ,\\
    \sum_{i,j} c_{ij}
    \left(\frac{\partial \Avec}{\partial z}\right)_{ij} = 0 .
  \end{cases}
\end{equation}
This step can easily be performed numerically using the
\texttt{Mathematica} implementation of the PSLQ algorithm
(\texttt{FindIntegerNullVector}) \cite{Ferguson1998APT}.  Once all
relations among the entries are known, they can be solved via Gaussian
elimination, which yields a set of linearly independent integration
kernels $d W_{i}$. We find the total number in our case to be
$N_W=49$\footnote{These also include the $W_i$ needed for the purely
massless calculation that we have performed as a check.}  and we
choose our basis by minimising the complexity of the analytic
expressions of its elements when written in the form $d W_{i} =
W_i^{(x)}\, dx + W_i^{(z)}\, dz$.  In particular, 22 kernels are
$d\log$ of rational functions, 18 are $d\log$ of square roots, and 9
involve elliptic structures. We report the explicit expression for all
the relevant $W_i$ in the ancillary files that accompany this
manuscript.

Eq.~\eqref{eq:de} can be immediately solved in terms of Chen iterated
integrals~\cite{Chen:1977oja}, defined recursively as
\begin{equation}
  I^{(\gamma)}_{1,2,...,n}\equiv \int_\gamma dW_1 I^{(\gamma)}_{2,...,n},
  \label{eq:chen}
\end{equation}
where $\gamma$ is a path connecting a base point $X_0$ (our boundary
condition) to a generic point $X$, and with the recursion starting from
$I\equiv 1$. In what follows, we will omit the reference to $\gamma$
for brevity. In terms of iterated integrals, the solution of
\cref{eq:de} reads
\begin{equation}\label{eq:chen_solution}
  \begin{aligned}
    \fvec(X) = 
    \Bigg[& \mathbb{I} +  \ep \sum_{i_1}  \Avec_{i_1}\, I_{i_1} 
      + \ep^2 \sum_{i_1,i_2}  \Avec_{i_1} \cdot  \Avec_{i_2}\,
      I_{i_1,i_2}
      + \ep^3 \sum_{i_1,i_2,i_3}
      \Avec_{i_1} \cdot  \Avec_{i_2}  \cdot  \Avec_{i_3}\,  
      I_{{i_1},{i_2},{i_3}}\\
      &+\ep^4 \sum_{i_1,i_2,i_3,i_4}
      \Avec_{i_1} \cdot \Avec_{i_2} \cdot \Avec_{i_3} \cdot \Avec_{i_4}\,  
      I_{{i_1},{i_2},{i_3},i_{4}}
      +\mathcal{O}(\ep^5) \Bigg] 
    \times \fvec(X_0)  ,\\
  \end{aligned}    
\end{equation}
where $X=\{y_{r,n},z\}$\footnote{We remind the reader that we use a
different ``natural'' variable $y_{r,n}$ for different contributions,
see \cref{sec:defs}.}, $X_0=\{y^*_{r,n},z^*\}$ is our boundary
point and the boundary value can be expanded as
\begin{equation} \label{eq:chen_boundaries}
  \fvec(X_0) =  \fvec^{(0)}_0 +
  \ep \fvec^{(1)}_0 + \ep^2 \fvec^{(2)}_0 +
  \ep^3 \fvec^{(3)}_0  + \ep^4\fvec^{(4)}_0 + \mathcal{O}(\ep^5) .
\end{equation}
Choosing a value of $X_0$ which corresponds to a MUM point of the
elliptic sector guarantees that the $\fvec^{(k)}_0$ are vectors of
constants of transcendental weight $k$, see \cref{sec:can}.
We note that although each iterated integral depends on the path
$\gamma$ that connects $X_0$ to $X$, the integrability of the
differential equation implies that the full solution does not. In
other words, we can choose any preferred path, as long as we do it
consistently for all iterated integrals.
We also note that if the kernels $W_i$ are linearly independent, then
so are all iterated integrals of fixed transcendental weight.

We note that all the square roots appearing in the $d\log$ part of our
alphabet are rationalisable. Hence, the full solution in
the polylogarithmic sector can be straightforwardly written in terms
of Goncharov polylogarithms (GPLs), which can be efficiently evaluated
numerically~\cite{Vollinga:2004sn}. We illustrate how to do so in
\cref{sec:gpls}. Before discussing this though, we elaborate on how
we fixed the boundary conditions for our solution.

\subsection{Threshold expansion and boundary constants}
\label{sec:thresh_and_boundary}
In order to fix the boundary constants $\fvec^{(k)}_0$ defined in
\cref{eq:chen_boundaries}, we select the base point
\begin{equation}
  \{y^*_{r,n},z^*\} \equiv \{\delta^2,\delta\},\quad\quad\delta\ll 1,
  \label{eq:X0}
\end{equation}
\ie we consider the (massless) threshold limit. This is a regular
singular point for the differential equation, so imposing the right
physical branch cut structure of the result constrains some of the
boundary constants (see \cref{sec:bcext} for a brief discussion of the
branch cuts of our result).  However, we found that this alone was not
enough to immediately fix all the boundary conditions. We then
approached this problem in a more systematic way: we generated a set
of high-precision numerical values for our integrals close to the
threshold with \texttt{AMFlow}~\cite{amflow}, and matched them against a
high-order Frobenius expansion of our solution. This allowed us to fix
all the boundary constants in a systematic and effective way.  We
describe this in more detail below.

Instead of expanding our iterated integrals near threshold, we found
it more convenient to solve the differential equation directly in the
threshold limit.  Near $y_{r,n} = 0$ we can expand the $y_{r,n}$
differential equation in series
\begin{equation} \label{eq:y_n_expanded_eq}
  \partial_{y_{r,n}} \fvec(y_{r,n},z)= \ep
  \left( \partial_{y_{r,n}} \Avec \right) \cdot \fvec(y_{r,n},z) =
  \ep \left(\frac{{\bf X}_{-1}}{y_{r,n}} +
      {\bf X}_0 + {\bf X}_{1}\, y_{r,n} + \dots \right)
      \cdot \fvec(y_{r,n},z),
\end{equation}
with ${\bf X}_i = {\bf X}_i(z)$. 
The absence of higher degree poles is guaranteed for the
polylogarithmic kernels by the fact that they are $d\log$ forms, while
for the elliptic ones by the fact that $y_{r,3} = 0$ is a MUM point
for the relevant sector.  Note that the $\partial W_i / \partial
y_{r,n}$ entering the elliptic sector, which are needed to compute the
relevant matrices ${\bf X}_{i}$ above, depend on the implicit elliptic
functions $w_0$ and $g^\mathrm{sc}$ defined by \cref{eq:w0eq,eq:fdef}.
Therefore, in order to obtain an expansion of the $y_{r,3}$
differential equation, we need to first obtain series expansions for
these auxiliary functions. By definition, both functions can be
expanded in regular Taylor series near the MUM point $y_{r,3} =
0$. However, since $w_0$ is only a function of $w$, see
\cref{eq:wdef2}, we found it convenient to first compute the series
expansion of both $w_0$ and $g^\mathrm{sc}$ in powers of $(w-1/9)$
(corresponding to $y_{r,3}=0$) and then change variables to $y_{r,3}$
and $z$.  In practice we do this by writing the ansatz
\begin{equation}\label{eq:w_f_expansions_w}
  \begin{aligned}
    w_0(w) = \sum_{m=1} c^{(w_0)}_m \, (w-1/9)^m ,
    \quad\quad g^\mathrm{sc}(w,z) = \sum_{m=0}
    c^{(g^\mathrm{sc})}_m(z) \, (w-1/9)^m,
  \end{aligned}
\end{equation} 
whose coefficients can be easily fixed by plugging these expressions
in \cref{eq:w0eq,eq:fdef}, solving the resulting linear constraints
among the $c^{(w_0)}_m$ and $c^{(g^\mathrm{sc})}_m(z)$, and solving
the remaining differential equation in $z$ for the linearly
independent $c^{(g^\mathrm{sc})}_m(z)$.\footnote{Since $g^\mathrm{sc}$ is only
defined through the differential equation \cref{eq:fdef}, we are free
to choose its boundary constant. We entirely fix our function by
imposing the initial condition $c^{(g^\mathrm{sc})}_m(0) = 0$.}
Finally, writing $w$ in terms of $y_{r,3}$ and $z$, we can further
expand \cref{eq:w_f_expansions_w} to obtain Taylor expansions in
$y_{r,3}$ with $z$-dependent coefficients.
With this we can proceed to computing the $y_{r,n}$ expansion of all
canonical master integrals. The general solution to
\cref{eq:y_n_expanded_eq} is a Frobenius expansion of the form
\begin{equation}\label{eq:frobenius_1}
  \fvec(y_{r,n},z) = \sum_{k,m=0}^\infty
  \sum_{l \leq k} \ep^k \, y_{r,n}^m \, \ln^l(y_{r,n}) \,
      {\bf g}_{k,m,l} (z),
\end{equation}
where purely virtual integrals have contributions only to the $m=l=0$
terms.  Consistency with \cref{eq:y_n_expanded_eq} provides linear
relations which allow us to express all the coefficient functions
${\bf g}_{k,m,l} (z)$ in terms of a subset of the ($y_{r,n}$)
leading-power ones: ${\bf g}_{k,0,l} (z)$.\footnote{We note that this
implies that when the boundary constants are fixed at leading-power,
as a by-product of \cref{eq:frobenius_1} we can immediately
obtain a Frobenius expansion to in principle arbitrary orders without
any further input. We have used this feature to compute Frobenius
expansions up to $\mathcal O(y_{r,n}^{20})$, which takes minutes on a
laptop. This provided us with a fast and efficient representation for
all our integrals near the threshold.}

To determine these, we plug the leading power term ($m=0$) of
\cref{eq:frobenius_1} into the $z$ differential equation expanded at
leading power in $y_{r,n}$:
\begin{equation}
  \partial_z \fvec(y_{r,n},z)= \ep \, \partial_z \Avec \cdot
  \fvec(y_{r,n},z) = \ep \, \left({\bf Y}_0(z)
  +\mathcal O(y_{r,n})\right)\cdot \fvec(y_{r,n},z) ,
\end{equation}
which provides us with differential equations for the leading-power
coefficient functions:
\begin{equation}\label{eq:g_z_equation}
  \partial_z {\bf g}_{k,0,l} (z) = {\bf Y}_0(z) \cdot
          {\bf g}_{k-1,0,l} (z), \quad\quad k \in \{1,2,3,\dots\}.
\end{equation}
We found that for all but the elliptic sector RR$_3$, the solution to
\cref{eq:g_z_equation} can be written in terms of very simple GPLs:
\begin{equation}\label{eq:z_fn_space}
  \begin{aligned}
    \text{R}_1,\text{V}_1,\text{RR}_1,\text{RV}_1:&
    \quad\; G(\vec{\ell},z), \quad &&\ell_i \in \{-1,0\}, \\
    \text{VV}_1:&
    \quad\; G(\vec{\ell},z), \quad &&\ell_i \in \{-1,-1/2,0,1\}, \\
    \text{VV}_0:&
    \quad\; G(\vec{\ell},\sqrt{1-4z}), \quad &&\ell_i \in \{-1,0,1\}, \\
    \text{RR}_2:&
    \quad\; G(\vec{\ell},1/\sqrt{1+4z}), \quad &&\ell_i \in \{-1,0,1\}, \\
  \end{aligned}
\end{equation}
and unspecified boundary constants.  We used
\texttt{PolyLogTools}~\cite{Duhr:2019tlz} to manipulate the resulting
GPLs. This, combined with the deep Frobenius expansion for $y_{r,n}$,
provides us with a good approximation of the integrals near the
threshold region, up to the yet unfixed boundary constants.

The solution for $\gvec_{k,0,l}$ in the elliptic sector (RR$_3$) is
much more cumbersome. In this case we proceeded with a Frobenius
expansion in $z$. Following the same lines described above for the
$y_{r,n}$ expansion, we obtain a solution of the form
\begin{equation}
  {\bf g}_{k,0,l} (z) = \sum_{m'=0} \sum_{l'\leq k-l} z^{m'}
  \ln^{l'}(z) \, {\bf h}_{k,l,l',m'},
\end{equation}
where now the ${\bf h}_{k,l,l',m'}$ are simple constants. As for the
$y_{r,n}$ expansion, plugging the ansatz above into
\cref{eq:g_z_equation} yields a system of linear relations among the
${\bf h}_{k,l,l',m'}$, which allows us to express all of them in terms
of the leading-power ones: ${\bf h}_{k,l,l',0}$.

At this point we need to fix the boundary constants of the ${\bf
  g}_{k,0,l} (z)$ solutions, (for instance the ${\bf h}_{k,l,l',0}$
for the elliptic sector).  We do this by using either
\texttt{PolyLogTools} for the non-elliptic integrals or our deep
Frobenius expansions for the elliptic ones to evaluate all master
integrals near the $\{y_{r,n},z\}\to\{0,0\}$ singular point, \eg
$y_{r,n} = 1/100$, $z=1/100$\footnote{Note that our Frobenius
expansion allows us to move away from the strict $y^*_{r,n}\ll z^*$
point.}, and compare them with a numerical evaluation of the integrals
obtained via \texttt{AMFlow} to fix all the constants to numerical
values.\footnote{ We found that AMFlow could compute numerical points
for the elliptic-sector integrals only with the option
\texttt{"AMFMode"} set to \texttt{"Propagator"}.  }
To convert these high-precision numerical results to analytic
transcendental constants we used the PSLQ algorithm in conjunction
with an ansatz containing all expected constants for each
transcendental weight.  In our case we found the following constants
were sufficient for the non-elliptic integrals:
\begin{equation}\label{eq:all_constants}
  \begin{aligned}
    \text{weight 1:} &\quad \pi, \, \ln(2), \\
    \text{weight 2:} &\quad \pi^2, \, \ln^2 (2), \\
    \text{weight 3:} &\quad \pi^3,\, \pi^2 \ln (2),\,
    \ln^3 (2),\, \zeta_3, \\
    \text{weight 4:} &\quad  \pi^4,\, \pi^2 \ln^2 (2),\,
    \ln^4 (2),\, \Li_4(1/2),\, 
    \pi \ \zeta_3,\, \ln(2) \ \zeta_3 .
  \end{aligned}
\end{equation}
For the elliptic sector we limited ourselves to reconstructing only
the constants appearing in the cross section, which we found to be a
single constant of weight 1 proportional to $\sqrt{3} \pi$.  We were
able to verify the accuracy of our analytic reconstructions to a
minimum of 20 digits by comparing to a different point close to
threshold, \eg $y_{r,n} = 1/20$, $z=1/20$.

The outcomes of the procedure highlighted above are a deep Frobenius
expansion around the relevant thresholds for the different sectors,
and a set of boundary values for the canonical integrals $\fvec(X_0)$,
cf.~\cref{eq:chen_boundaries}. Although the Frobenius expansion could
have been skipped entirely for integrals which admit a simple analytic
representation, the method described here is fully general, and it
allowed us to fix all boundary constants in a straightforward
automated fashion.

\subsection{Extraction of the threshold behaviour}
\label{sec:bcext}
As we have illustrated in \cref{sec:nlo}, the coefficient functions
are in general distributions near the $y_{r,n}\to 0$ threshold. At the
level of accuracy relevant to our work, this is only present in the
contributions involving only one heavy quark in the final state, \ie
R$_1$, RV$_1$, RR$_1$. In these cases, one has to extract the relevant
branch-cut structure \emph{before} expanding around $\ep\to 0$. The
canonical differential equation makes this extraction straightforward.

In our calculation, one small extra subtlety appears: individual
contributions have spurious $\sim y_r^{-2+a\ep}$ double poles, which
cancel when everything is combined in the final
result.\footnote{Specifically, there is a cancellation between the RR
and RV contributions and the related one-loop mass counterterms,
described in \cref{sec:ren}.}  To deal with this problem, we have
followed two different approaches. First, we have used the
differential equation to extract the branch-cut structure at
next-to-leading power (NLP) in $y_r$ for the relevant integrals.
Second, we have combined the spurious terms, and written them in terms
of a unique integral basis.\footnote{This required embedding the
product of one-loop integrals and renormalisation constants into
two-loop topologies.} After this was done, the spurious terms dropped
out from the beginning, and a leading power analysis was
sufficient. In the end, we compared the two approaches and found
agreement.  In the remainder of this section, we briefly sketch how we
obtained the required NLP expansion relevant for the first approach.

To do this, we note that near the $y_r\to 0$ branch cut, the solution
to our differential equation for the integrals in the sectors $s \in
\{\text{R}_1, \text{RR}_1,\text{RV}_1\}$ can be expanded up to NLP in
$y_{r}$ as
\begin{equation}\label{eq:branch_cuts_ansatz}
  \fvec^{(s)}(y_{r},z; \ep) \approx  T(\fvec)^{(s)}(y_{r},z; \ep) \equiv
  \sum_{k \in Br(s)} y_{r}^{k\ep} \left( 
  \fvec^{(s,0)}_k(z; \ep) + y_{r}  \, \fvec^{(s,1)}_k(z; \ep) 
  \right).
\end{equation}
To determine the set of integers $Br(s)$, we perform a
method-of-regions analysis~\cite{Beneke:1997zp} of the master
integrals in the threshold limit and read off the coefficient of $\ep$
in the exponent of $y_r$ for each contributing region. Explicitly we
find:
\begin{equation}
  Br(\text{R}_1 ) = \{-2\}, \quad
  Br(\text{RV}_1) = \{-2,-3,-4\}, \quad
  Br(\text{RR}_1) = \{-4\}.
\end{equation}
We then plug the ansatz~\cref{eq:branch_cuts_ansatz} in the
corresponding NLP differential equation
\begin{equation} \label{eq:y_{r}_NLP}
   \partial_{y_{r}} T(\fvec)^{(s)}(y_{r},z)  - \ep \left(\frac{{\bf X}^{(s)}_{-1}(z)}{y_{r}} + {\bf X}_0 (z)  \right) \cdot T(\fvec)^{(s)}(y_{r},z) = 0 ,
\end{equation}
and truncate the result at order $y_{r}^0$, \textit{without} expanding
in $\ep$.  The resulting equation reads
\begin{equation}
  \begin{aligned}  
    \sum_{k \in Br(s)} y_{r}^{k\ep} \Bigg[
      \frac{1}{y_{r}} &\left( 
      k \, \ep\, \fvec^{(s,0)}_k - \ep\,  {\bf X}^{(s)}_{-1}(z)
      \fvec^{(s,0)}_k 
      \right) \\
      &+ \left( 
      (1+k \ep) \, \fvec^{(s,1)}_k-
      \ep\, {\bf X}^{(s)}_{-1}(z) \, \fvec^{(s,1)}_k
      -\ep\, {\bf X}^{(s)}_{0}(z) \, \fvec^{(s,0)}_k            
      \right)
      \Bigg] = 0 ,
  \end{aligned}
\end{equation}
which can be solved power by power in $y_{r}$, yielding the relations
\begin{equation}
  \begin{aligned}
    &k \, \fvec^{(s,0)}_k - \,  {\bf X}^{(s)}_{-1}(z)
    \fvec^{(s,0)}_k  = 0, \\
    &(1+k \ep) \, \fvec^{(s,1)}_k-
    \ep\, {\bf X}^{(s)}_{-1}(z) \, \fvec^{(s,1)}_k
    -\ep\, {\bf X}^{(s)}_{0}(z) \, \fvec^{(s,0)}_k  = 0 ,
  \end{aligned}
\end{equation}
for all values of $k \in Br(s)$. These are linear relations which can
be solved retaining exact dependence in $\ep$ and $z$ and which once
again allow us to express all NLP coefficient functions
$\fvec^{(s,1)}_k(z; \ep)$ in terms of a subset of the leading power
ones, $\fvec^{(s,0)}_k(z; \ep)$.  In order to determine the remaining
leading-power coefficient functions we can further expand
\cref{eq:branch_cuts_ansatz} in $\ep$ and compare with the threshold
expansion obtained in \cref{sec:chen_and_boundaries}. The
$\fvec^{(s,0)}_k(z; \ep)$ can then be solved in terms of GPLs of the
form given in \cref{eq:z_fn_space} and the constants of
\cref{eq:all_constants}. We find that at least up to NLP, only the
$Br(\text{RV}_1)=\{-4,-2\}$ contribute. We also find that not all of
the NLP branch cut structures are already present at leading power. We
note that these results may help inform investigations of the NLP
structure of the threshold expansion in the presence of massive
particles.

With our $\ep$-exact threshold approximations $T(\fvec)^{(s)}$ we can
now rewrite the solutions for each integral which appears in the
coefficient function accompanied by a $y_r$ pole in the following way:
\begin{equation}
    \fvec^{(s)} = T(\fvec)^{(s)} + R(\fvec)^{(s)},
\end{equation}
with 
\begin{equation}
        R(\fvec)^{(s)} = \underbrace{\fvec^{(s)} - T(\fvec)^{(s)}}_{\text{expanded in } \ep} .
\end{equation}
The $T(\fvec)$ term can be used to extract the branch cut structure
and reconstruct the plus distributions along the lines of
\cref{sec:nlo}.  The remainder $R(\fvec)$ vanishes fast enough in
$y_r\to 0$, so it regulates the $y_r$ pole(s) from the coefficient
function and can be straightforwardly expanded in $\ep$.

We reiterate that following the second approach described at the
beginning of this section, the NLP expansion is not necessary and
therefore in that case the procedure above can be truncated to LP.

\subsection{Numerical evaluation: the polylogarithmic case}
\label{sec:gpls}
As we have already mentioned, all integrals in the polylogarithmic
sector (i.e. all contributions apart from RR$_3$) can be expressed in
terms of GPLs. However, this is not immediate due to the presence of
several square roots.  In this section, we briefly illustrate the kind
of functions that enter our result, and describe how we dealt with the
square-root cases.  For completeness, we also give a simple example of
the relation between polylogarithmic iterated integrals and GPLs. We
structure our discussion according to the various contributions that
we need to consider. 

\paragraph{VV$_0$} In this sector, we need to expand 4 master integrals,
up to weight 3. The relevant differential equation can be written in
terms of three logarithmic forms $dW_i$, with
\begin{equation}
  \{W_2,W_8,W_9\} = \left\{\ln(z),
  \ln\left( \frac{1-\sqrt{1-4z}}{1+\sqrt{1-4z}}\right),
  \ln \left(\frac{(1-4z)^2}{z}\right)\right\}.
\end{equation}
Although these letters contain a square root, it is immediate to
rationalise it using the standard transformation
\begin{equation}
  z = \frac{1-t^2}{4},\quad\quad\quad t = \sqrt{1-4z}.
\end{equation}
The resulting alphabet in $t$ only contains the letters
$\{t,t-1,t+1\}$, so it is straightforward to obtain a result in terms
of Harmonic Polylogarithms~\cite{Remiddi:1999ew} or even classical
polylogarithms, evaluated in $t=\sqrt{1-4z}$. These are by far the
simplest functions appearing in our problem.

\paragraph{VV$_1$, RV$_1$, RR$_1$} This case is more involved.
We need to consider 76 master integrals, 34 of which are multiplied by
$1/y_r$ poles in the bare coefficient functions, thus requiring the
procedure described in \cref{sec:bcext}. Fortunately, after inserting
our results in terms of iterated integrals in the coefficient
functions, we find that most of the contributions are very
simple. Indeed, at weight 4 only $d\log$ forms with letters
$\{z,1+z,1-z\}$ appear. Up to weight 3, most of the integrals are
over kernels $W_i = \ln \alpha_i$, with letters
\begin{equation}
  \begin{split}
    & \{\alpha_1,\alpha_2,\alpha_4,\alpha_5,\alpha_6,
    \alpha_7,\alpha_{10},\alpha_{11},\alpha_{12},\alpha_{13}\} =
    \\
    &\quad\left\{
    \frac{1-y_r}{1+z},\,
    z,\,1+z,\,\frac{1+z}{y_r+z},\,\frac{1}{1-z},\,\frac{z}{1+2z},\,
    \frac{1+y_r z}{1+z},\,y_r,\,2-y_r,\,\frac{(1-y_r)z}{1+(2-y_r)z}
    \right\}.
  \end{split}
  \label{eq:somew}
\end{equation}
When this is the case, it is straightforward to re-express the
iterated integrals in terms of GPLs, starting from the
definition~\cref{eq:chen}. In doing so, we have to remember that
although we can choose any path connecting our boundary $\{y_r^*,z^*\}
= \{\delta^2,\delta\}$ to a generic kinematic point to compute our
master integrals, individual iterated integrals are path-dependent. In
other words, we have to use the same path for all iterated integrals
(or at least for integrable combinations thereof), see the discussion
in \cref{sec:chen_and_boundaries}.  In our case, we found it
convenient to choose the following path, that we denote $\gamma$: we
first integrate along the line $\{\delta^2,\delta\}\to
\{\delta^2,z\}$, and then along the line $\{\delta^2,z\}\to
\{y_r,z\}$.  As an example of this procedure, consider the integral
$I_{{12},4,5}$. First, we expand the $d\log$s in the $y_r$ and $z$
differentials, \eg
\begin{equation}
  dW_5 = d\ln\frac{1+z}{y_r+z} = -\frac{dy_r}{y_r+z}+
  \left(\frac{1}{1+z}-\frac{1}{y_r+z}\right)dz,
\end{equation}
see \cref{eq:somew}, and distribute the integration over the different
terms in the kernels:
\begin{equation}
  \begin{aligned}
    &I_{{12},4,5} = \int_{\gamma} d W_{12}
    \int_{\gamma}d W_{4} \int_{\gamma}d W_{5} = 
    \sum_{a,b,c \in \{y_r,z\} }
    \int_{\gamma} \partial_{a} W_{12} \, da
    \int_{\gamma} \partial_{b} W_{4}  \, db
    \int_{\gamma} \partial_{c} W_{5}  \, dc .
  \end{aligned}
\end{equation}
Then, since along the first part of the path $y_r$ is constant, 
we eliminate all terms where a $dy_r$ differential is on the right of a 
$dz$ differential so that for our choice of path
\begin{equation}
  \begin{aligned}
    I_{{12},4,5} &= 
    \int_{\gamma} \partial_{y_r} W_{12}  dy_r
    \int_{\gamma} \partial_{y_r} W_{4}  dy_r
    \int_{\gamma} \partial_{y_r} W_{5} dy_r +
    \int_{\gamma} \partial_{y_r} W_{12}  dy_r
    \int_{\gamma} \partial_{y_r} W_{4}  dy_r
    \int_{\gamma} \partial_z W_{5} dz \\
    & + 
    \int_{\gamma} \partial_{y_r} W_{12} dy_r
    \int_{\gamma} \partial_z W_{4}  dz
    \int_{\gamma} \partial_z W_{5} dz +
    \int_{\gamma} \partial_z W_{12} dz
    \int_{\gamma} \partial_z W_{4}  dz
    \int_{\gamma} \partial_z W_{5} dz.
  \end{aligned}
\end{equation}
Plugging in the explicit kernels and setting $y_r \to \delta^2 \to 0$
in the $dz$ terms\footnote{This limit is legitimate, since in our base
point $y_r\ll z$.} we obtain
\begin{equation}
  \begin{split}
    I_{{12},4,5} &= \int_{\delta^2}^{y_r}\frac{dy}{y-2}\int_\delta^z
    \frac{dz'}{1+z'} \int_\delta^{z'} dz''
    \left(\frac{1}{1+z''}-\frac{1}{z''}\right) = \\ &=
    G(2;y_r)\big[G(-1;z)G(0;\delta) + G(-1,-1;z)-G(-1,0;z)\big],
  \end{split}
\end{equation}
where $G$ are the GPLs, defined recursively as
\begin{equation}
  G\left(a,\vec b;t\right) = \int_0^t \frac{dy}{y-a} G\left(\vec b;y\right),
  \quad\quad G(\,\underbrace{0,...,0}_n;t) = \frac{1}{n!}\ln^n(t).
\end{equation}
We stress the importance of choosing a consistent path among different
iterated integrals by noting that if we had chosen a path where we
first integrate in $y_r$, and then in $z$, $I_{{12},4,5}$ would have
been zero. Along these lines, one can immediately express all the
iterated integrals that do not involve square roots in terms of GPLs.

The situation is more complex for integrals involving a square root.
Square-root letters only appear in the finite part of the coefficient
functions, both for the quark (RR-only) and the gluon (RR and RV)
channels. The part of the result that involves these letters can be
expressed in terms of 5 weight-2 and weight-3 functions,
\begin{equation}
  \left\{T^{(2)}_{q,1},T^{(3)}_{q,2},
  T_{g,1}^{(3)},T_{g,2_{\rm RR}}^{(3)},T_{g,2_{\rm RV}}^{(3)}
  \right\}.
  \label{eq:sqrtT-def}
\end{equation}
Above we have collected in $T_q$ the combination of master integrals
which appear in a non-planar contribution to the quark coefficient
function, and in $T_{g,i}$ those relevant for the gluon channel, which
we have further split according to the different square roots that
they depend on. Their explicit expression in terms of iterated
integrals is given in ancillary files. Here we show how to deal with
the square roots and write the functions in \cref{eq:sqrtT-def} in
terms of GPLs.

We start from $T_q$. We first proceed along the same lines of the GPLs
case illustrated above to express all iterated integrals in the $T_q$
combination that do not involve a square root in terms of GPLs,
choosing the same path $\gamma$ as before. We then focus on the
remaining iterated integrals. Along $\gamma$, it turns out that all
entries that contain $dz$ vanish, so we only have to consider
contributions from the $dy_r$ differential. Furthermore, the rightmost
entry of any leftover iterated integral is always simple: it is either
$dy_r/(1-y_r)$ or $dy_r/(y_r+z)$. Note that neither is singular for
$y_r\to 0$. The other entries contain the square root
\begin{equation}
  r_q = \sqrt{\frac{1+(3-2y_r)^2 z}{1+z}},
\end{equation}
which we rationalise via the transformation
\begin{equation}
  \begin{split}
    & y_r = \frac{t_q^2-2t_q(1+z)-8z(1+z)}{t_q^2-4z(1+z)},
    \\
    &
    t_q = \frac{1+z+
      \sqrt{1+10z-12y_r z+4 y_r^2 z+9z^2-12 y_r z^2+4 y_r^2 z^2}}{1-y_r}.
    \label{eq:tqdef}
  \end{split}
\end{equation}
This transformation makes the ``simple'' letters quadratic in $t_q$:
\begin{equation}
  \frac{dy_r}{1-y_r} = \left(\frac{2t_q}{t_q^2-4z-4z^2}-\frac{1}{t_q+2z}
  \right)dt_q,
\end{equation}
and similarly (but with a different quadratic polynomial) for
$dy_r/(y_r+z)$. This is however not an issue, since at this stage $z$
is just a parameter. We can then simply write
\begin{equation}
  t_q^2-4z-4z^2 =
  \left(t_q-2\sqrt{z(1+z)}\right)\left(t_q+2\sqrt{z(1+z)}\right)
  \equiv (t_q-r_1^+)(t_q-r_1^-),
\end{equation}
and treat $r_1^\pm$ like any other entry of a GPL, \eg
\begin{equation}
  \int_0^{t_q}\frac{dt}{t-r_1^-} = G(r_1^-;t_q).
\end{equation}
Along these lines, one can immediately write a solution in terms of
GPLs.  There is only one last small subtlety. Consider the term $a
\equiv t_q-2\sqrt{z(1+z)}$. Using~\cref{eq:tqdef}, it is simple to see
that $a>0$ in the physical region, so that the integral
\begin{equation}
  \int_{t_0}^{t_q}\frac{dt}{t-2\sqrt{z(1+z)}} =
  \ln\frac{t_q-2\sqrt{z(1+z)}}{t_0-2\sqrt{z(1+z)}},
\end{equation}
with $t_0 = t_q|_{y_r=0}$ never requires an analytic
continuation. However, when integrating in terms of GPLs, the same
expression reads
\begin{equation}
  G(2\sqrt{z(1+z)};t_q)-    G(2\sqrt{z(1+z)};t_0),
\end{equation}
with $t_{0,q}>2\sqrt{z(1+z)}$. Both GPLs here develop an imaginary
part, that cancels between the two. To avoid this, we further change
variable from $t_q$ to
\begin{equation}
  \delta t_q \equiv t_q - t_0 > 0. 
\end{equation}
In terms of $\delta t_q$, the solution only contains GPLs of the form
$ G(\vec b;\delta t_q)$ with all entries of $\vec b$ real and
negative.  These GPLs are manifestly real, and straightforward to
evaluate.

We now move to $T_{g,1}$. The discussion proceeds along lines very
similar to $T_q$. Also in this case all the $dz$ differentials drop if
we choose the path $\gamma$. The $dy_r$ differentials involve the root
\begin{equation}
  r_{g,1}= \sqrt{\frac{4-4y_r+y_r^2(1+z)}{1+z}},
\end{equation}
which is rationalised via
\begin{equation}
  \begin{split}
    & y_r = \frac{1-t_{g,1}^2+z}{1+t_{g,1}+z+t_{g,1}z},
    \\
    & t_{g,1} = \frac{\sqrt{(1+z)(4-4y_r+y_r^2+y_r^2 z)}-y_r(1+z)}{2} > 0.
  \end{split}
\end{equation}
Also in this case, this rationalisation makes the ``simple'' letters
quadratic in $t_{g,1}$, whose factorisation introduces square roots of
polynomials in $z$. Compared to the $T_q$ case, the resulting letters
here are simpler. Indeed, they are of the form $t_{g,1} - a_i$, with
$a_i$ either negative or greater than $t_{g,1}$. This automatically
leads to manifestly real GPLs. There is, however, an additional subtlety
compared to the $T_q$ case. Indeed, here the differential
$dy_{r}/y_r$, which is singular at the boundary, appears. At fixed
value of $z$, this gets remapped to
\begin{equation}
  \frac{dy_r}{y_r} = \left(-\frac{1}{t_{g,1}} + \frac{1}{t_{g,1}-\sqrt{1+z}}
  +\frac{1}{t_{g,1}+\sqrt{1+z}}\right)dt_{g,1}.
\end{equation}
At the boundary, $t_{g,1}\to \sqrt{1+z}$ so in principle we could
obtain logarithmically divergent GPLs in our final result. However, in
this case this does not happen and we can safely take the limit
$t_{g,1}\to \sqrt{1+z}$ when evaluating the boundary terms. Before
moving on to the last case, we note that here, contrary to $T_q$, the
square root only appears in the leftmost letter of the iterated
integral.  As a consequence, two out of three integrations can be
carried out in terms of polylogarithms. This leads to a very compact
one-fold integral representation for $T_{g,1}^{(3)}$:
\begin{equation}
  \begin{split}
    T_{g,1}^{(3)} = \int_0^{y_r} \Bigg\{
    &
    6(6-y)\bigg[\Li_2\left(\frac{y+z}{1+z}\right)-
      \Li_2\left(\frac{z}{1+z}\right)
      +\ln\left(\frac{1-y}{1+z}\right)\ln\left(\frac{y+z}{z}\right)\bigg]
    \\
    &+24y\big[\Li_2(y-1)+\ln(1-y)\ln(2-y)\big]
    -12(y-2)\Li_2\left(-\frac{y}{z}\right)
    \\
    &-15 y\ln^2(1-y) + 2\pi^2 y\Bigg\}
    \frac{dy}{3(1-y)\sqrt{\frac{4-4y+y^2(1+z)}{1+z}}},
  \end{split}
  \label{eq:t1onefold}
\end{equation}
where all the functions are manifestly real for physical kinematics.
We found this representation useful for fast numerical evaluations.

Finally, we move to $T_{g,2}$. This is similar to the other cases,
with some minor differences. First, now the $dz$ differentials do not
decouple. However, they only involve trivial integration kernels so
the $dz$ integration does not pose any challenge. The $dy_r$ integrals
involve the root
\begin{equation}
  \sqrt{\frac{4z - 4y_r z + y_r^2(1+z)}{1+z}},
\end{equation}
which we rationalise using
\begin{equation}
  \begin{split}
    & y_r = \frac{z(1+z)-t_{g,2}^2}{(1+z)(t_{g,2}+z)},
    \\
    & t_{g,2} = \frac{\sqrt{(1+z)(y_r^2+4z-4y_r z + y_r^2 z)}-y_r(1+z)}{2}>0.
  \end{split}
  \label{eq:t2def}
\end{equation}
This transformation makes the simple letters quadratic, but as in the
$T_{g,1}$ case this does not pose any issue because after the
factorisation of the quadratic polynomials one is left with letters of
the form $t_{g,2}-a_i$ with either $a_i < 0$ or $a_i > t_{g,2}$. Also
in this case the $dy_r/y_r$ differential appears, leading to terms of
the form
\begin{equation}
  \frac{dt_{g,2}}{t_{g,2}-\sqrt{z(1+z)}},
\end{equation}
which are singular in the $t_{g,2}\to \sqrt{z(1+z)}$ (\ie $y_r\to 0$)
limit.  Contrary to the $T_{g,1}$ case however, these terms do not drop
from the final result so a proper regulation is necessary. GPLs of the
form $G(a,...;t)$ are singular in the $t\to a$ limit. To extract the
(logarithmic) singularity, one can use the shuffle algebra of the GPLs
(see \eg~\cite{Vollinga:2004sn}). For example, consider $G(a,b;t)$,
with $b\ne a$ (otherwise the result is trivial), in the limit $t\to
a$. We can write it as
\begin{equation}
  G(a,b;t) = -G(b,a;t) + G(b;t) G(a;t),
\end{equation}
with $G(b,a;t)$ and $G(b;t)$ regular in the $t\to a$ limit. Using
relations like this, we can extract all the singularities in the
$t_{g,2}\to \sqrt{z(1+z)}$ limit and write them in terms of
\begin{equation}
  \lim_{t_{g,2}\to t_0}G(\sqrt{z(1+z)};t_{g,2}) =
  \lim_{t_{g,2}\to t_0}
  \ln\left(1-\frac{t_{g,2}}{\sqrt{z(1+z)}}\right),
  \label{eq:divlog}
\end{equation}
where $t_0$ is the boundary point. To proceed, we plug the
$y_r=\delta^2$ boundary value in~\cref{eq:t2def}, to obtain
\begin{equation}
  t_0 \equiv t_{g,2}|_{y_r=\delta^2} = \sqrt{z(1+z)}
  -\frac{1+z+\sqrt{z(1+z)}}{2}\delta^2 + \mathcal O(\delta^3).
\end{equation}
The divergence in~\cref{eq:divlog} then is regulated as
\begin{equation}
  \lim_{t_{g,2}\to t_0}
  \ln\left(1-\frac{t_{g,2}}{\sqrt{z(1+z)}}\right) =
  \ln\delta^2 + \ln\left(\frac{1+z+\sqrt{z(1+z)}}{2\sqrt{z(1+z)}}\right).
\end{equation}
The $\ln\delta^2$ terms cancel against similar terms in the boundary
vectors~\cref{eq:chen_boundaries} to give a final result which is
independent of the chosen boundary point. We conclude the discussion
of $T_{g,2}$ by mentioning that also in this case it is possible to write
the result in terms of a simple one-fold integral. Its expression is,
however, more complicated than the one for $T_{g,1}$~\cref{eq:t1onefold},
so we will not report it here. 
  
\paragraph {RR$_2$} In this case, we only need to compute three
combinations of master integrals, whose expression in terms of
iterated integrals reads
\begin{equation}
  \left\{
  8 I_{W_{19}},2I_{W_1,W_{19}}+2I_{W_5,W_{19}}
  +2 I_{W_{14},W_{29}},-2 I_{W_{29}},
  \right\},
\end{equation}
with\footnote{Note the apparent difference w.r.t.~\cref{eq:somew}.
This is because we first define unambiguously our letters in terms of
the standard Bjorken variable $x$~\cref{eq:DISvar}, to avoid
duplications across different contributions. In a second stage, we
change variable to $y_{r,n}$. So for example
$\exp(W_1)=x=(1-y_{r,1})/(1+z)=(1-y_{r,2})/(1+4z)$.}
\begin{equation}
  \begin{gathered}
    W_1 = \ln\left(\frac{1-y_{r,2}}{1+4z}\right),
    \quad
    W_5 = \ln\left(\frac{1+4z}{y_{r,2}+4z}\right)
    \\
    W_{14} = \ln\left(
    \frac{\sqrt{1+4z}-\sqrt{1+4y_{r,2}z}}{\sqrt{1+4z}+\sqrt{1+4y_{r,2}z}}
    \right),\,
    W_{19} = \ln\left( \frac{\sqrt{y_{r,2}+4z}-\sqrt{y_{r,2}+4
        y_{r,2}z}}{\sqrt{y_{r,2}+4z}+\sqrt{y_{r,2}+4 y_{r,2}z}}
    \right),\\
    W_{29} = \ln\left(
    \frac{\sqrt{1+4 y_{r,2}z}\sqrt{y_{r,2}+4z}-\sqrt{y_{r,2}}(1+4z)}
         {\sqrt{1+4 y_{r,2}z}\sqrt{y_{r,2}+4z}+\sqrt{y_{r,2}}(1+4z)}
         \right).
  \end{gathered}
\end{equation}
Note that all the arguments of the logarithms are real in the physical
region.  Note also that $W_{19,29}$ vanish at the boundary point,
which implies that the weight-1 functions can be integrated by simply
replacing $I_{W_{19,29}}\to W_{19,29}$. The other integral, despite
being only weight two, is complicated due to the many square roots
involved.  We rationalise them with the transformation
\begin{equation}
  \{y_{r,2},z\} =
  \left\{
  \frac{(u-v)^2}{(1+u v)^2},
  \frac{u v}{(1-u^2)(1-v^2)}
  \right\}.
\end{equation}
In terms of the new $\{u,v\}$ variables, the kernels read
\begin{equation}
  \begin{split}
    & \{W_1,W_5,W_{14},W_{19},W_{29}\} =
    \\
    & \left\{
    \ln\left(
    \frac{(1-u^2)(1-v^2)}{(1+uv)^2}
    \right),
    \ln\left(\frac{(1+uv)^2}{(u+v)^2}\right),
    \ln(uv),\ln\left(\frac{u}{v}\right),
    \ln\left(\frac{v(1-u^2)}{u(1-v^2)}\right)
    \right\},
  \end{split}
  \label{eq:lettuv}
\end{equation}
which makes writing the result in terms of GPLs straightforward.
However, the inverse relation between $\{u,v\}$ and $\{y_{r,2},z\}$ is
non-trivial. We then decided to write our solution as a linear
combination of $\Li_2(a_i(u,v))$ and $\ln(b_i(u,v))$ where the $a_i$,
$b_i$ arguments are functions of $u$ and $v$ which have a
simple-enough representation in terms of $y_{r,2}$ and $z$. To do so,
we follow the standard procedure~\cite{Duhr:2011zq}. We start from an
ansatz that only contains weight-2 \emph{functions}, \ie $\Li_2$ and
product of logs:
\begin{equation}
  f_{w_2} = \sum_{i=1}^{N_a} c_i \Li_2(a_i) +
  \sum_{i\le j=1}^{N_b} c_{i,j}\ln(b_i)\ln(b_j),
  \label{eq:ans2}
\end{equation}
with $c_{i}$ and $c_{i,j}$ yet-to-be-determined rational numbers. We
use the letters themselves as possible arguments for the logs, $b_i =
\alpha_i$.  For the $a_i$, we need to find arguments $a_i$ such that
both $a_i$ and $1-a_i$ factor on the original
alphabet~\cite{Duhr:2011zq}.  Using the
representation~\cref{eq:lettuv}, we find the candidates
\begin{equation}
  \{a_i\} = \left\{\sqrt{\frac{\alpha_{14}}{\alpha_5\alpha_{19}}},
  \sqrt{\frac{\alpha_{14}\alpha_{19}}{\alpha_5}},
  \sqrt{\frac{\alpha_5\alpha_{14}}{\alpha_{19}}},
  \sqrt{\alpha_5\alpha_{14}\alpha_{19}}\right\}.
\end{equation}
Introducing the short-hands
\begin{equation}
  r_1 = \sqrt{y_{r,2}},\quad
  r_2 = \sqrt{1+4z},\quad
  r_3 = \sqrt{y_{r,2}+4z},\quad
  r_4 = \sqrt{1+4 y_{r,2}z},
\end{equation}
it is easy to see that the candidates $a_i$ can be written as
\begin{equation}
  \{a_i\} =
  \left\{
  \frac{r_3(r_1 r_2+r_3)}{r_2(r_2+r_4)},\quad
  \frac{r_3(r_2-r_4)}{r_2(r_1 r_2+r_3)},\quad
  \frac{r_2(r_1 r_2+r_3)}{r_3(r_2+r_4)},\quad
  \frac{r_2(r_3-r_1 r_2)}{r_3(r_2+r_4)}
  \right\}.
\end{equation}
Note that $a_i\in [0,1]$, so $\Li_2$s of these arguments are
manifestly real.

To move forward, we compute the symbol of our ansatz~\cref{eq:lettuv}
and compare it against the one obtained from the iterated-integrals
representation of our solution\footnote{With our definitions, the
symbol of a polylogarithmic iterated integral is given by $\mathcal
S(I_{i_1,i_2,...,i_n}) = i_n\otimes ...\otimes i_2\otimes i_1$.},
using $u$ and $v$ as variables. We find that the $\{a_i\}$ and
$\{b_i\}$ arguments above are enough to fully match the symbol, and
fully determine the $c_i$, $c_{i,j}$ coefficients. Matching the
relative symbols is not enough to conclude that two functions are the
same. Indeed, they could differ by constants, possibly multiplying
lower-weight functions:
\begin{equation}
  2I_{W_1,W_{19}}+2I_{W_5,W_{19}}
  +2 I_{W_{14},W_{29}} = 
  f_{w2} + \sum d_i \ln(b_i) + e,
\end{equation}
with $d_i$ and $e$ weight-1 and weight-2 numbers, respectively. By
taking the derivative w.r.t. $y_{r}$ on both sides, we find that in
our cases all $d_{i}$ vanish. Finally, since $f_{w2}\to 0$ as we
approach our boundary point, we conclude that also $e$ vanishes, and
hence that our function is given entirely by $f_{w2}$.

We conclude this section by noting that in principle we could have
adopted a similar strategy (\ie write the final results in terms of
simple polylogarithms with arguments involving roots) also for the
RR$_1$, RV$_1$ cases discussed above. However, in those cases it was
straightforward to find a fast and robust GPL representation, so we
decided not to pursue this strategy there.
  
\subsection{Numerical evaluation: the elliptic case}
The numerical evaluation for the elliptic sector is entirely
different. In this case the integrals appearing in the cross section
do not admit a representation in terms of known functions for which
fast numerical evaluation routines exist. Our approach here is to
compute (accelerated) Frobenius series expansions around enough points
to achieve the desired precision in the physical region of
interest. In this paper we focus on charm production in the region
$Q\gtrsim 5~{\rm GeV}$, which implies $0<z\lesssim 1/10$. We then
found that the following two expansions are sufficient to obtain good
numerical control over our phase space:
\begin{itemize}
  \item threshold expansion followed by a small-mass expansion:
    $1-x_{r,3} \ll z \ll 1$;
  \item small-mass expansion away from threshold: $z \ll 1$,
    $1-x_{r,3} \gg z$.
\end{itemize}
We discuss them separately below. We will not give a full account of
how we computed the expansions above, since the strategy we followed
is for the most part identical to the one outlined in
\cref{sec:thresh_and_boundary}. The main difference compared to the
previous description is that we use the expansions of the master
integrals to compute a Frobenius expansion of the \emph{whole}
coefficient functions. We find that this drastically improves the
accuracy of our results when compared to computing the master
integrals numerically, and plugging the results into the unexpanded
expressions for the coefficient functions.  Expansions at the
coefficient function level also greatly decrease the number of
boundary constants needed, since only specific combinations of master
integrals appear.

For each kinematic point at which we evaluate our expanded cross
sections, whenever slow convergence is observed we employ a Shanks
transformation~\cite{Shanks}, which for a series $\{s_{n}\}_{n \in
  \mathbb{N}}$ (in our case $s_n$ would correspond to the expanded
coefficient functions truncated to order $n$) is defined as
\begin{equation}
  S(s_{n}) = \frac{s_{n+1}s_{n-1} - s_{n}^2}{s_{n+1} - 2 s_{n} + s_{n-1}} . 
\end{equation}
In general, we adopt the criterion of only using a Shanks
transformation if the relative variation given by the last order in
the cross-section series expansion, \ie $(s_n - s_{n-1})/s_{n-1}$, is
larger than $10^{-5}$.  We observe that this shields us from potential
numerical instabilities arising when the denominator in the equations
above becomes vanishingly small, which would require working at very
high numerical precision.

\subsubsection*{Threshold small-mass expansion}
This is the same expansion we computed in
\cref{sec:thresh_and_boundary} and it allows us to obtain accurate
results near and on threshold ($x_{r,3} = 1$).  The leading threshold
behaviour is $(1-x_{r,3})^2$, and it contains no logarithms of
$y_{r,3} = 1-x_{r,3}$. The coefficient functions in this region are
therefore very well behaved.  We obtained the results presented in
this publication by expanding in the variable $y_{r,3}$ to 13 orders
and truncating the subsequent small-mass expansion in the variable $z$
to 20 orders. Numerically evaluating such an expansion is effectively
instantaneous.

\subsubsection*{Small-mass expansion}
The small-mass expansion is a single-variable expansion in small
values of $z$. Contrary to the previous expansion, we do not need
to further expand around some fixed value of $x_{r,3}$. Indeed, once
$z$ is set to zero, the differential equation for the elliptic master
integrals in $x_{r,3}$ can be solved exactly in terms of harmonic
polylogarithms. We were further able to compute all boundary constants
up to transcendental weight 3 analytically in terms of $\pi$ and
$\zeta_3$, which further improves the accuracy and stability of our
results. We carried out this expansion to 15 orders in $z$.

In addition to the Shanks transform described above, for the small-mass
expansion we also employ a second acceleration technique
involving Bernoulli-like variables as discussed \eg in
refs~\cite{THOOFT1979365,Gehrmann:2001jv,Coro:2025vgn}.
Studying the differential equations for RR$_3$ and the master integral
coefficients in the cross section (before expanding at small $z$), we
find that all potential singularities at fixed $ x_{r,3} \in (0,1)$
arise for negative values of $z$, or for $z=0$.  In particular, the
singularity closest to $z=0$ corresponds to $z = z_s(x_{r,3}) = -(1 -
x_{r,3})/9 $, which for physical values of $x_{r,3}$ takes values in
$[-1/9,0]$.  Therefore, in addition to expanding to high order in $z$,
we employed a change of variables from $z$ to the Bernoulli-like
variable
\begin{equation}
  z^\prime = \log\left[1-\frac{z}{z_s(x_{r,3})}\right] = \log\left[1 +
    \frac{9 z}{1 - x_{r,3}} \right],
\end{equation}
which has the effect of mapping $z_s(x_{r,3})$, the pole of the cross
section nearest to $z=0$, to negative infinity while simultaneously
maintaining the large-mass limit at positive infinity and the small-mass
limit at $z=z^\prime=0$. In practice, after expanding the cross
section in $z$, we re-express the result in terms of $z^\prime$ via
the inverse transformation
\begin{equation}
  z = \frac{1 - x_{r,3}}{9}(e^{z^\prime}-1),
\end{equation}
and then re-expand to the same order in $z^\prime$. This leads to a
much faster convergence with respect to a standard expansion in $z$.
Also in this case, evaluating numerical results for the coefficient
functions is essentially instantaneous.


\section{UV and collinear renormalisation}
\label{sec:ren}
In this section, we describe the UV and collinear renormalisation of
our result. We renormalise the charm-quark mass on shell (OS), but adopt the
$\MSbar$ for the strong coupling $\alpha_s$. Also, since we are
interested in energies $Q\gtrsim m_c$, we work in a mixed scheme where
there are only 3 active flavours in the proton, but $\as$ is
renormalised with 4 flavours. To keep the discussion general, from now
on we will denote by $\nf$ the number of active flavours in the
proton, and by $\nh$ the number of additional (heavy) flavours that
contribute to the running of $\as$. To achieve this mixed scheme, we
first consider a theory with strictly three flavours, neglecting any
contributions coming from the emission of extra (real or virtual)
charm quarks. We then add the relevant extra charm diagrams and
perform the relevant scheme change.  These steps are discussed
separately in \cref{sec:uvren-nf,sec:cvren-nf,sec:ren-nh},
respectively. We have explicitly verified that this procedure agrees
with performing UV renormalisation directly in the mixed scheme, along
the lines of \eg
refs~\cite{Weinberg:1980wa,Ovrut:1980dg,Wetzel:1981qg,Bernreuther:1981sg,Bernreuther:1983zp,Bernreuther:1983ds}.
This provides an internal cross-check of our procedures.

\subsection{UV renormalisation in a three-flavour theory}
\label{sec:uvren-nf}
We start by writing the bare coefficient functions $c_{i,a}$ (we
suppress the number of final-state massive quarks at first)
\cref{eq:cdefs} in terms of bare quantities
\begin{equation}
  c_{i;a} = c_{i;a}^{(0),b}
  +\left(\frac{\asb\Sep}{2\pi}\right) c_{i;a}^{(1),b}
  +\left(\frac{\asb\Sep}{2\pi}\right)^2 c_{i;a}^{(2),b} +
  \mathcal O(\asb^3),
  \label{eq:cdefs_bare}
\end{equation}
where we have explicitly extracted factors of $\Sep = (4\pi)^\ep
e^{-\ep \gamma_E}$ (where $\gamma_E=0.577...$ is the Euler constant)
from the bare coefficient functions for convenience. We remind the
reader that $i\in(2,L,3)$ and $a$ is a flavour index. We also stress
that since we are neglecting additional emission of real or virtual
massive quarks at this stage, all contributions in
\cref{eq:cdefs_bare} stem from amplitudes involving exactly one
final-state charm quark. Finally, we note that beyond LO the bare
coefficient functions explicitly depend on the bare charm mass
$m_{q,b}$ through the heavy-quark propagator. We write this
schematically as
\begin{equation}
  c_{i;a}^{(k),b}=  
  c_{i;a}^{(k),b}(\{x\}; m_{q,b}).
\end{equation}
To renormalise our result, we multiply the coefficient functions by
the heavy-quark wave-function renormalisation factor
\begin{equation}
  Z_q = 1 +
  \left(\frac{\asb\Sep}{2\pi}C\right) z_q^{(1)}+
  \left(\frac{\asb\Sep}{2\pi}C\right)^2 z_q^{(2)}+
  \mathcal O(\asb^3),
  \label{eq:Zq-def}
\end{equation}
with $C = e^{\ep\gamma_E}\Gamma(1+\ep) m_q^{-2\ep}$ and\footnote{We
remind the reader that in this subsection we are neglecting
contributions coming from virtual charm loops, which will be discussed
in \cref{sec:ren-nh}.}
\begin{equation}
  \begin{split}
    z_q^{(1)} &= -\Cf\frac{(3-2\ep)}{2\ep(1-2\ep)},
    \\
    z_q^{(2)} &=
    \Cf^2\left(
    \frac{9}{8\ep^2}+\frac{51}{16\ep}+\frac{433}{32}-\frac{13\pi^2}{4}+
    4\pi^2\ln 2 - 6\zeta_3
    \right)+
    \\
    & + \Ca\Cf\left(-\frac{11}{8\ep^2}-\frac{101}{16\ep}-\frac{803}{32}+
    \frac{5\pi^2}{4}-2\pi^2\ln 2 + 3\zeta_3\right)+
    \\
    & + \Cf\nf\tr
    \left(\frac{1}{2\ep^2}+\frac{9}{4\ep}+\frac{59}{8}+\frac{\pi^2}{3}\right)
    +\mathcal O(\ep).
  \end{split}
\end{equation}
We also express the bare quark mass in terms of its on-shell
renormalised counterpart $m_q$ using
\begin{equation}
  m_{q,b} = Z_m^{\rm OS}\, m_q,\quad\quad Z_m^{\rm OS} =
  1 +
  \left(\frac{\asb\Sep}{2\pi}C\right) z_m^{(1)} + \mathcal O(\asb^2),
\end{equation}
with $z_m^{(1)} = z_q^{(1)}$. Schematically, this leads to
\begin{equation}
  \begin{split}
    c_{i;a}^{(k),b}(\{x\}; Z_m^{\rm OS} m_q) \equiv
    c_{i;a}^{(k),b}(\{x\}; m_q) +
    \left(\frac{\asb\Sep}{2\pi}C\right) z_m^{(1)} \times D_m\left[
      c_{i;a}^{(k),b}\right]+\mathcal O(\asb^2),
  \end{split}
\end{equation}
with
\begin{equation}
  D_m\left[c_{i;a}^{(k),b}\right] \equiv m_q\partial_{y}c_{i;a}^{(k),b}(\{x\}; y)\big|_{y\to m_q}.
\end{equation}
As it is well known, in practice computing the action of the operator
$D_m$ on the coefficient functions $c_{i;a}^{(k),b}$ effectively
amounts to calculating the relevant diagrams with squared heavy-quark
propagators.

After performing these manipulations, we express the bare coupling
$\asb$ in terms of its $\MSbar$ counterpart in a theory with $\nf$
flavours, $\asnf$, using
\begin{equation}
  \left(\mu_0^{2}\right)^{\ep}\,\asb\,\Sep = \left(\mu^{2}\right)^{\ep}\,
  \asnf \left[1-\left(\frac{\asnf}{2\pi}\right)\frac{\beta_{0;n_f}}{\ep}+
    \mathcal O(\asnf^2)\right],
\end{equation}
with $\mu$ the renormalisation scale, $\asnf \equiv \asnf(\mu^2)$ and
$\beta_{0;nf} = 11/6\, \Ca - 2/3\, \tr \nf$.\footnote{ Here and in
what follows, we use the subscript $\nf$ to stress that these
quantities are computed in a theory with $\nf=3$ light flavours,
without any effect coming from additional real or virtual emission of
massive quarks.}  Finally, we recall that the Larin-scheme replacement
\begin{equation}
  \gamma^\mu\gamma_5 \to \frac{i}{6} \ep^{\mu\nu\rho\sigma}
  \gamma_\nu\gamma_\rho\gamma_\sigma
\end{equation}
discussed in section~\ref{sec:bareci} implicitly breaks the
anticommutativity of $\gamma_5$ in $d$ dimensions, which leads to a
violation of the axial Ward identity.  To remedy this, one has to
perform an additional renormalisation of the axial current. For the
non-singlet current relevant for our calculation, this amounts to the
extra replacement~\cite{larin_scheme}
\begin{equation}
  \frac{i}{6} \ep^{\mu\nu\rho\sigma}
  \gamma_\nu\gamma_\rho\gamma_\sigma\to Z_5 Z^{\rm ax}_{\rm ns}\,
  \frac{i}{6} \ep^{\mu\nu\rho\sigma}\gamma_\nu\gamma_\rho\gamma_\sigma.
\end{equation}
To fix the renormalisation constants, one imposes that their anomalous
dimension vanishes
\begin{equation}
  \mu^2 \frac{d}{d\mu^2}\ln \left(Z_5 Z^{\rm ax}_{\rm ns}\right) = 0,
\end{equation}
which is equivalent to reinstating the axial Ward identities. To the
accuracy relevant for this work, this leads to~\cite{larin_scheme}
\begin{equation}
  Z_5 Z^{\rm ax}_{\rm ns} =
  1 + \left(\frac{\asnf}{2\pi}\right)z_{ax;\nf}^{(1)} +
  \left(\frac{\asnf}{2\pi}\right)^2 z_{ax;\nf}^{(2)} +
  \mathcal O(\asnf^3),
  \label{eq:larin}
\end{equation}
with
\begin{equation}
  z_{ax,\nf}^{(1)} = -2\Cf,\quad\quad
  z_{ax,\nf}^{(2)} =\frac{11}{2}\Cf^2 - \frac{107}{36}\Ca\Cf +
  \frac{1}{9}\Cf \tr \nf + \frac{\beta_{0;\nf}} {\ep}\Cf.
  \label{eq:larin_exp}
\end{equation}

Multiplying \cref{eq:cdefs_bare} by the relevant $Z$ factors as
described above, and expressing everything in terms of the OS charm
mass and $\MSbar$ strong coupling constant lead to the UV-renormalised
result
\begin{equation}
  c_{i;a} =
  \bar c_{i;a;n_f}^{(0)} + \left(\frac{\asnf}{2\pi}\right)
  \bar c_{i;a;n_f}^{(1)} +
  \left(\frac{\asnf}{2\pi}\right)^2
  \bar c_{i;a;n_f}^{(2)} + \mathcal O(\asnf^3),
\end{equation}
where the UV-renormalised coefficient functions
$\bar c_{i;a;\nf}^{(k)}$ read
\begin{equation}
  \begin{split}
    &\bar c_{i;a;n_f}^{(0)} = c_{i;a}^{(0),b},\\
    &\bar c_{i;a;n_f}^{(1)} = c_{i;a}^{(1),b} + 
    \left[C z_q^{(1)} + z_{ax;\nf}^{(1)}\delta_{i,3}\right]c_{i;a}^{(0),b},
    \\
    & \bar c_{i;a;n_f}^{(2)} = 
    c_{i;a;n_f}^{(2),b} +
    \left[
      C \left(z_m^{(1)}D_m + z_q^{(1)}\right) - \frac{\beta_{0;\nf}}{\ep} +
      z_{ax;\nf}^{(1)}\delta_{i,3}\right]  c_{i;a}^{(1),b} +
    \\
    &\quad\quad\quad + 
    \left[\left(-\frac{\beta_{0;\nf}}{\ep}
      +z_{ax;\nf}^{(1)}\delta_{i,3}\right)C z_{q}^{(1)} + C^2 z_{q}^{(2)} +
      z_{ax;\nf}^{(2)}\delta_{i,3}\right]c_{i;a;n_f}^{(0),b}.
  \end{split}
  \label{eq:cbar_res}
\end{equation}
We conclude this subsection with a comment on the $D_m
\left[c_{i;a}^{(1),b}\right]$ terms, \ie the ones coming from mass
renormalisation. As we briefly mentioned in \cref{sec:bcext}, they
contain spurious $y_r^{-2+k\ep}$ threshold double poles, which exactly
cancel analogous double poles present in the bare coefficient function
$c_{i,a}^{(2),b}$. In fact, the latter are bound to appear when the
amplitude is expressed in terms of the bare mass. To expose the
cancellation pattern, we have expressed the product of
$z_m^{(1)}\times D_m\left[ c_{i;a}^{(1),b}\right]$ in terms of a basis
of two-loop integrals used for $c_{i;a}^{(2),b}$, verifying that the
double poles vanish even before substituting any explicit expression
for the integrals.

\subsection{Collinear renormalisation in a three-flavour theory}
\label{sec:cvren-nf}
The $\bar c_{i;a;\nf}^{(k)}$ coefficient functions \cref{eq:cbar_res}
still contain collinear singularities. To absorb them, we consider the
convolution product $c_{i;a}\otimes f_{a}^{b}$ and express the bare
parton distribution function $f_a^b$ in terms of its
$\MSbar$-renormalised counterpart $f_{a;\nf}$, in a theory with
$\nf=3$ active flavours:
\begin{equation}
  (\mu_0^2)^{\ep} \Sep\, f^{b}_{a} =
  (\mu^2)^{\ep} Z^{\rm PDF}_{ab;\nf}\, \otimes f_{b;\nf},
\end{equation}
with
\begin{equation}
  Z_{ab,\nf}^{\rm PDF} = \delta_{ab}+
  \left(\frac{\asnf}{2\pi}\right)z^{{\rm PDF},(1)}_{ab;\nf}+
  \left(\frac{\asnf}{2\pi}\right)^2 z^{{\rm PDF},(2)}_{ab;\nf}+
  \mathcal O(\asnf^3),
\end{equation}
and
\begin{equation}
  z^{{\rm PDF},(1)}_{ab;\nf} = \frac{P_{ab}^{(0)}}{\ep},
  \quad\quad
  z^{{\rm PDF},(2)}_{ab;\nf} = \frac{1}{2\ep^2}
  \left(P_{ac}^{(0)}\otimes P_{cb}^{(0)} - \beta_{0;\nf} P_{ab}^{(0)}\right)+
  \frac{P^{(1)}_{ab}}{2\ep}.
\end{equation}
In these equations, $P_{ab}^{(0)/(1)}$ are the LO/NLO Altarelli-Parisi
splitting functions. Using these results, we can re-express our final
results in terms of fully-renormalised coefficient functions
\begin{equation}
  \left[c_{i;a;\nf}^{(0)}
    +\left(\frac{\asnf}{2\pi}\right)c_{i;a;\nf}^{(1)}
    +\left(\frac{\asnf}{2\pi}\right)^2 c_{i;a;\nf}^{(2)}
    +\mathcal O(\asnf^3)\right]\otimes f_{a;\nf}.
\end{equation}
The relation between the UV-renormalised coefficient functions $\bar
c_{i;a;\nf}$ \cref{eq:cbar_res} and the fully renormalised ones
$c_{i;a;\nf}$ is:
\begin{equation}
  \begin{split}
    &c_{i;a;\nf}^{(0)} = \bar c_{i;a;\nf}^{(0)},\\
    &c_{i;a;\nf}^{(1)} = \bar c_{i;a;\nf}^{(1)} +
    \bar c_{i;b;\nf}^{(0)}\otimes z^{{\rm PDF},(1)}_{ba;\nf},\\
    &c_{i;a;\nf}^{(2)} = \bar c_{i;a;\nf}^{(2)} +
    \bar c_{i;b;\nf}^{(1)}\otimes z^{{\rm PDF},(1)}_{ba;\nf}
    +\bar c_{i;b;\nf}^{(0)}\otimes z^{{\rm PDF},(2)}_{ba;\nf}.
  \end{split}
  \label{eq:cdefs_nf}
\end{equation}

To compute these convolutions, we require all LO splitting functions
$P^{(0)}_{ab}$ with $(ab) \in \{qq,qg,gq,gg\}$, as well as the NLO
splitting functions $P^{(1)}_{ab}$ with $(ab)\in \{q_i q_j,q_i\bar
q_j,qg\}$ (together with their charge conjugate $q\leftrightarrow \bar
q$). As it is customary in DIS, we organise the quark-(anti)quark NLO
splitting functions in terms of pure-singlet and non-singlet
contributions.  Specifically, we write
\begin{equation}
  P^{(1)}_{q_i q_j} = \delta_{ij} P^{{\rm NS},(1)}_{qq} +
  P^{{\rm PS},(1)}_{qq},
  \quad\quad
  P^{(1)}_{q_i \bar q_j} = \delta_{ij} P^{{\rm V},(1)}_{q\bar q} +
  P^{{\rm PS},(1)}_{qq},
\end{equation}
where we have used the fact that at this order the $qq$ and $q\bar q$
pure-singlet splitting functions are the same.  The expression for
these splitting functions, as well as the LO ones, is well-known, see
\eg ref.~\cite{Ellis:1996mzs}, so we will not report them here. For
the reader's convenience, we note that the relation between our
notation and the one in ref.~\cite{Ellis:1996mzs} is
\begin{equation}
  \begin{gathered}
     P^{{\rm NS},(1)}_{qq} = P^{V(1)}_{qq}\big|_{\rm ESW},\quad
    P^{{\rm V},(1)}_{q\bar q} = P_{q\bar q}^{V(1)}\big|_{\rm ESW},
    \\
    P^{{\rm PS},(1)}_{qq} = \left.\frac{P_{qq}^{(1)}-P_{qq}^{V(1)}-P_{q\bar q}^{V(1)}}{2\nf}\right|_{\rm ESW}.
\end{gathered}
\end{equation}

\subsection{Extra contributions proportional to $n_h$ and final
  results in the mixed scheme}
\label{sec:ren-nh}
In the previous subsection, we obtained renormalised results in a
theory with $\nf=3$ light flavours, and no additional contributions
coming from virtual and real charm quark emission. Here we describe
how to modify the $\nf=3$ coefficient functions in \cref{eq:cdefs_nf}
to account for these extra effects, to obtain our final results in a
scheme where we have only $\nf=3$ active flavours in the proton, but
we evolve $\as$ including an additional $\nh=1$ heavy quark. For this
discussion, we closely follow appendix A in
ref.~\cite{Behring:2020uzq}, and refer the reader to that work for
additional detail and derivations.

On top of the $\nf=3$ calculation, we have to consider the following
additional contributions
\begin{itemize}

\item virtual charm-loop corrections on top of massless ($q\to q'$)
  and massive ($s\to c$) transitions, \cfit VV$_0$ and VV$_1$ in
  section~\ref{sec:nnlo_structure};
  
\item real emission of a $c\bar c$ pair on top of massless $q\to q'$
  transitions, \cfit RR$_2$;

\item real-emission contributions with three charm (anti)quarks, \cfit RR$_3$;

\item heavy-quark contributions to the wave-function renormalisation
  constant of light ($Z_{lq}$) and heavy ($Z_{q,\nh}$) quarks, and of
  the gluon ($Z_{g}$). To the order required for this calculation,
  they read
  \begin{equation}
    \begin{split}
      Z_{lq} &= 1 + \left(\frac{\asb\Sep}{2\pi}C\right)^2 z^{(2)}_{lq} + \mathcal O(\asb^3),\\
      Z_{q,\nh} &= 1 + \left(\frac{\asb\Sep}{2\pi}C\right)^2 z^{(2)}_{q,\nh} + \mathcal O(\asb^3),
      \\
      Z_g &= 1 + \left(\frac{\asb\Sep}{2\pi}C\right) z^{(1)}_{g} + \mathcal O(\asb^2),
    \end{split}
  \end{equation}
  with\footnote{The renormalisation coefficients $z^{(2)}_{lq}$ and
  $z^{(1)}_{g}$ defined here correspond to $\tilde\Sigma_2(0)/C^2$ and
  $-\Pi_1(0)/C$ in ref.~\cite{Behring:2020uzq}, respectively.}
  \begin{equation}
    \begin{split}
      z_{lq}^{(2)} & = \Cf\tr\nh\left(\frac{3+\ep-2\ep^2}{2\ep(6+7\ep-13\ep^2-4\ep^3+4\ep^4)}\right),\\
      z_{q,\nh}^{(2)} &= \Cf\tr\nh \left(\frac{1}{\ep^2}+\frac{19}{12\ep}+\frac{1139}{72}-\frac{4\pi^2}{3}+\mathcal O(\ep)\right),
      \\
      z_{g}^{(1)} &= -\frac{2}{3\ep}\tr\nh.
    \end{split}
  \end{equation}
  In these equations, we explicitly added the $\nh$ subscript to $Z_q$
  to stress that we are only considering the heavy-quark-induced
  terms, and not the other contributions already present in the
  $\nf=3$ case, \cfit \cref{eq:Zq-def};

\item the relation between the $\MSbar$ strong coupling constant in a
  theory with only $\nf=3$ (massless) flavours (\ie $\asnf$ introduced
  in the previous subsection) and the one in the full theory with
  $\nf+\nh$ flavours:
  \begin{equation}
    \asnf = \as\left(1+\frac{\as}{2\pi} K_1 + \mathcal O(\as^2)\right),
  \end{equation}
  where $\as = \as(\mu^2)$ is the coupling in the full ($\nf+\nh$)
  theory, with $K_1$ defined through
  \begin{equation}
    K_1 = \frac{1}{\ep}\left(\beta_{0;\nf}-\beta_0\right) + C z^{(1)}_g =
    K'\ln\left(\frac{m_q^2}{\mu^2}\right)
    + \mathcal O(\ep),
    \label{eq:defK1}
  \end{equation}
  and $K'=\frac{2}{3}\tr\nh$, $\beta_0 = \frac{11}{6}\Ca -
  \frac{2}{3}\tr(\nf+\nh)$. From now on, quantities without an
  explicit $\nf$ subscript have to be interpreted in the full
  ($\nf+\nh$) theory;
  
\item the value of the axial renormalisation constants, \cfit
  \cref{eq:larin}, in the full theory. Fortunately, these can be
  easily obtained just by replacing $\nf\to\nf+\nh$ in the relevant
  expressions \cref{eq:larin_exp}:
  \begin{equation}
    \begin{split}
      & z_{ax}^{(1)} = z_{ax;\nf}^{(1)},\\ & z_{ax}^{(2)} =
      z_{ax;\nf}^{(2)} + \Delta z_{ax}^{(2)}, \quad \Delta
      z_{ax}^{(2)} = \frac{1}{9}\Cf\tr\nh-\frac{2}{3\ep}\Cf\tr\nh.
    \end{split}
  \end{equation}
\end{itemize}

In the remainder of this subsection, we briefly review how one can use
the ingredients above to obtain results in the full theory, following
ref.~\cite{Behring:2020uzq}. Up to NLO, there is no difference between
the coefficient functions for the theory with only $\nf=3$ flavours,
$c_{i;a;\nf}^{(i)}$, and the ones for the full theory,
$c_{i;a}^{(i)}$:
\begin{equation}
  c_{i;a}^{(i)} =   c_{i;a;\nf}^{(i)}\quad {\rm for~}i\in[0,1].
\end{equation}
To discuss the modifications required at NNLO, we consider different
partonic channels and processes involving a different number of charm
quarks in the final state separately.

\paragraph{No charm quarks in the final state,
  quark channel, $c_{i;q}^{(2),[0]}$:}
This situation is analogous to the one described in
ref.~\cite{Behring:2020uzq}, with the only difference being the
presence of the Larin-renormalised axial current in our
calculation. As explained in that reference, the result in the full
theory reads
\begin{equation}
  \label{eq:ciqvv0split}
  c_{i;q}^{(2),[0]} = c_{i;q;\nf}^{(2),[0]} + c_{i;q;\nh}^{(2),[0]}
  \equiv c_{i;q;\nf}^{(2),[0]} + K_1 c_{i;q}^{(1),[0]} +
  \Delta c_{i,q,{\rm VV}}^{(2),[0]},
\end{equation}
with $K_1$ defined in \cref{eq:defK1}. The extra $\Delta c$ piece is
just the properly renormalised, two-loop contribution stemming from
virtual charm loops, \cfit VV$_0$ in section~\ref{sec:nnlo_structure}.
It is defined as~\cite{Behring:2020uzq}
\begin{equation}
  \Delta c_{i;q,{\rm VV}}^{(2),[0]} =
  c_{i;q}^{{\rm VV}_0}
  +2 C^2 z_{lq}^{(2)} c_{i;q}^{(0),[0]}
  -C z_g^{(1)} c_{i;q}^{{\rm V}_0}
  +\delta_{i,3}c_{i;q}^{(0),[0]}
  \left((C z_g^{(1)}-K_1)z_{ax}^{(1)}+\Delta z_{ax}^{(2)}\right).
  \label{eq:def-DeltaVV0}
\end{equation}
In this equation, $c_{i;q}^{{\rm V}_0}$ is the \emph{fully
renormalised} virtual contribution to the NLO coefficient function. We
note that $\Delta c_{i;q,{\rm VV}}^{(2),[0]}$ \cref{eq:def-DeltaVV0}
is finite~\cite{Behring:2020uzq}. Furthermore, once normalised by the
corresponding (massless) Born coefficient function, the result must be
identical for the $i=2$ and $i=3$ cases (while the $i=L$ contribution
vanishes):
\begin{equation}
  \frac{\Delta c_{2;q,{\rm VV}}^{(2),[0]}}{c_{2;q}^{(0)}} =
  \frac{\Delta c_{3;q,{\rm VV}}^{(2),[0]}}{c_{3;q}^{(0)}}.
\end{equation}
We have explicitly verified that this is the case, and that the result
we obtain is identical (up to a suitable analytic continuation) to
eq.~(B.13) in ref.~\cite{Behring:2020uzq}.  This provides a strong
cross-check of our renormalisation procedure, and of our treatment of
$\gamma_5$ in the Larin scheme.

\paragraph{No charm quarks in the final state, gluon channel,
  $c_{i;g}^{(2),[0]}$:}
In this case, there are no additional Feynman diagrams to consider and
all the dependence on $\nh$ comes from the scheme change:
\begin{equation}
  \label{eq:cigvv0split}
  c_{i;g}^{(2),[0]} =   c_{i;g;\nf}^{(2),[0]} +  c_{i;g;\nh}^{(2),[0]}
  \equiv  c_{i;g;\nf}^{(2),[0]} + K_1\,c_{i;g}^{(1),[0]}.
\end{equation}

\paragraph{One charm quark in the final state, quark channel,
  $c_{i;q}^{(2),[1]}$:}
The situation is fully analogous to the 0-charm case discussed above.
Only in this case, we have to consider (double) virtual corrections to
massive $s\to c$ transitions. We obtain
\begin{equation}
  c_{i;q}^{(2),[1]} = c_{i;q;\nf}^{(2),[1]} + c_{i;q;\nh}^{(2),[1]}
  \equiv c_{i;q;\nf}^{(2),[1]} + K_1 c_{i;q}^{(1),[1]} +
  \Delta c_{i,q,{\rm VV}}^{(2),[1]},
\end{equation}
with
\begin{equation}
  \begin{split}
    \Delta c_{i;q,{\rm VV}}^{(2),[1]} &=
    c_{i;q}^{{\rm VV}_1}
    +C^2\left(z_{lq}^{(2)}+z_{q,\nh}^{(2)}\right) c_{i;q}^{(0),[1]}
    -C z_g^{(1)} c_{i;q}^{{\rm V}_1}
    \\
    &+\delta_{i,3}c_{i;q}^{(0),[1]}\left((C z_g^{(1)}-K_1)z_{ax}^{(1)}+
    \Delta z_{ax}^{(2)}\right).
  \end{split}
\end{equation}
Compared to \cref{eq:def-DeltaVV0}, here we have to consider the
two-loop virtual-charm corrections VV$_{1}$, as well as the
fully-renormalised one-loop contribution V$_1$. Also, we had to
replace $2z_{lq}^{(2)} \to z_{lq}^{(2)} + z_{q,\nh}^{(2)}$, as
appropriate for a massless-to-massive transition.

\paragraph{One charm quark in the final state, gluon channel,
  $c_{i;g}^{(2),[1]}$:}
The situation is fully analogous to the 0-charm final-state one:
\begin{equation}
  c_{i;g}^{(2),[1]} =   c_{i;g;\nf}^{(2),[1]} + K_1\,
  c_{i;g}^{(1),[1]}.
\end{equation}

\paragraph{Contributions with two or three charm quarks in the
  final state}
These contributions arise from additional $g\to c\bar c$ splitting on
top of massless and massive $q\to q'$ transitions. They are only
present in the quark channel, and are finite:
\begin{equation}
  c_{i;q}^{(2),[2]} = c_{i;q}^{{\rm RR}_{2}},\quad
  c_{i;q}^{(2),[3]} = c_{i;q}^{{\rm RR}_{3}}.
\end{equation}
The result for $c_{2,L;q}^{(2),[2]}$ was already available in the
literature~\cite{Blumlein:2016xcy} (see also
ref.~\cite{Klann:2026svr}). Our result is in perfect agreement with
it.

\subsection{Scale dependence of the final result}
In this section, we present relations that allow one to obtain results for
generic renormalisation and factorisation scales, given the result for the
coefficient function for $\mu_r=\mu_f=\mu_0$. As it is well known, they can
be obtained by solving the standard RGE equations
\begin{equation}
  \begin{split}
    & \frac{\mu_r^2\,d}{d\mu_r^2}
    \left[
      c_{i;a}\left(\as(\mu_r^2),\mu_r^2,\mu_f^2\right)\otimes f_{a;\nf}(\mu_f^2)
      \right]
    = 0, \\
    & \frac{\mu_f^2\,d}{d\mu_f^2}
    \left[
      c_{i;a}\left(\as(\mu_r^2),\mu_r^2,\mu_f^2\right)\otimes f_{a;\nf}(\mu_f^2)
      \right]=0,
    \end{split}
\end{equation}
together with
\begin{equation}
  \frac{1}{\as}
    \frac{\mu^2d\as}{d\mu^2} = -\beta_0 \left(
  \frac{\as}{2\pi}
  \right)+\mathcal O(\as^3),
\end{equation}
with $\as=\as(\mu^2)$ and
\begin{equation}
  \frac{\mu^2\,d f_{a;\nf}}{d\mu^2} =
  \left(
  \left(\frac{\asnf}{2\pi}\right)
  P^{(0)}_{ab}
  + 
  \left(\frac{\asnf}{2\pi}\right)^2
  P^{(1)}_{ab}+\mathcal O(\asnf^3)
  \right)\otimes f_{b;\nf},
\end{equation}
with $f_{i;\nf} = f_{i;\nf}(\mu^2)$ and $\asnf = \asnf(\mu^2)$. 

Although the solution of these equations is quite standard, we report it
here for the mixed scheme we are adopting. It reads
\begin{equation}
  \begin{split}
    &c_{i;a}^{(0)}(\mu_r^2,\mu_f^2) = c_{i;a}^{(0)},
    \\
    &c_{i;a}^{(1)}(\mu_r^2,\mu_f^2) =
    c_{i;a}^{(1)} - L_f \,c_{i;b}^{(0)}\otimes P^{(0)}_{ba},
    \\
    &c_{i;a}^{(2)}(\mu_r^2,\mu_f^2) =
    c_{i;a}^{(2)}- L_f \,c_{i;b}^{(0)}\otimes P^{(1)}_{ba}
    +\beta_0
    \left(
    L_r c_{i;a}^{(1)} + \frac{L_{f/r}^2-L_r^2}{2}
    c_{i;b}^{(0)}\otimes P^{(0)}_{ba}
    \right)
    +
    \\
    &\quad\quad\quad\quad
    +\frac{1}{2}
    \left(
    K'\left[(\ln(z)-L_f)^2-\ln^2 z\right] c_{i;b}^{(0)}
    +L_f^2 c_{i;d}^{(0)} \otimes P_{db}^{(0)}
    -2 L_f c_{i;b}^{(1)}
    \right)\otimes P^{(0)}_{ba},
  \end{split}
  \label{eq:scale}
\end{equation}
with $L_f = \ln(\mu_f^2/Q^2)$, $L_r = \ln(\mu_r^2/Q^2)$,
$L_{f/r} = \ln(\mu_f^2/\mu_r^2)$, $\ln(z) = \ln(m_q^2/Q^2)$,
and $c_{i;a}^{(k)} \equiv c_{i;a}^{(k)}(\mu_r^2=Q^2,\mu_f^2=Q^2)$.
The decoupling constant $K'$ is defined in \cref{eq:defK1}.

In our calculation, we have kept the scale generic but have not
differentiated between renormalisation and factorisation scale.  We
have explicitly checked that the $\ln(\mu^2/Q^2)$ dependence that we
obtain in our result agrees with the prediction \cref{eq:scale} if we
set $\mu_r^2=\mu_f^2=\mu^2$.


\section{Results}
\label{sec:results}
The procedure highlighted in the previous sections allowed us to
obtain exact analytic results for CC DIS coefficient functions
involving massive charm quarks in the final state up to NNLO, retaining the
exact mass dependence. Here we summarise the structure of our final
results, discuss the checks that we have performed to ensure they are
correct, and illustrate their flexibility by showing predictions for
hadronic structure functions to NNLO. We leave an in-depth
phenomenological study to the future.

The main new results of our work are the NNLO coefficient functions
$c_{i;a}^{(2),[n]}$. In particular, results with
$n=1$~\cite{Berger:2016inr,Gao:2017kkx} and $n=3$ charm quarks in the
final state were previously only known numerically.  We expressed all
our coefficient functions with $n=0,1,2$ in terms of manifestly-real
Goncharov polylogarithms, which can be evaluated very
efficiently. Results with $n=3$ involve elliptic structures. We have
obtained both a general formula in terms of Chen iterated integrals,
as well as a representation in terms of (accelerated) series expansions
which are suitable for phenomenological applications in the
perturbative region $Q\gtrsim5~{\rm GeV}$. All these results are
provided in computer-readable format in the ancillary files that
accompany this publication.

To test the correctness of our framework, we have explicitly checked
that our NLO results agree analytically with the ones in the
literature~\cite{Gluck:1996ve}. At NNLO, we have checked that the
0-mass and 2-mass coefficient functions, together with the
scheme-change between the pure $\nf=3$ calculation and the one in the
mixed scheme described in \cref{sec:ren}, agree with the literature as
well~\cite{Blumlein:2016xcy,Behring:2020uzq}. We have also repeated
our calculation setting $m_q=0$ from the outset, and compared our
results with the well-known NNLO massless coefficient functions,
finding full agreement. To further check our massive results, we have
compared them against the leading-power expansion in the region
$Q^2\gg m_q^2$ described in ref.~\cite{Blumlein:2014fqa}.\footnote{We
note that the asymptotic result~\cite{Blumlein:2014fqa} is organised
into coefficients $L_i$ and $H_i$ corresponding to our $c_i^{[0]} +
c_i^{[2]}$ and $c_i^{[1]} + c_i^{[3]}$, respectively. We also note
that there appears to be a typo in the gluonic $H_3^g$ coefficient of
that reference (eq. C.11), as it seems to be inconsistent with the
respective Mellin space result (eq. 3.47). We have found agreement
between our results and the Mellin result eq.~3.47 of
ref.~\cite{Blumlein:2014fqa}.} Specifically, we numerically evaluated
the relevant combinations described in
refs~\cite{Blumlein:2011zu,Blumlein:2014fqa} using our analytic
results, and compared with a numerical evaluation of the formulas in
those references, for generic values of $x$. In the asymptotic region
$m_q^2\ll Q^2$, \ie for $z\ll 1$, we found perfect numerical agreement
between the two results.

To test the flexibility of our result, we have implemented all our
coefficient functions in a FORTRAN code, and interfaced it with
LHAPDF~\cite{Buckley:2014ana} to compute hadronic predictions for the
CC structure functions $F^{W^\pm}_i$ up to NNLO. To validate our
implementation, we have performed extensive comparisons against the
NLO results presented in ref.~\cite{Buonocore:2024pdv}. We have found
perfect agreement for both fiducial cross-section results and
differential distributions.\footnote{To obtain sub-per-mille agreement
with ref.~\cite{Buonocore:2024pdv}, we had to set PDFs to zero
whenever they became negative and use a dedicated implementation of
the running of $\as$ instead of the standard LHAPDF one.} We then used
our implementation to compare our NNLO result against the numerical
one of refs~\cite{Berger:2016inr,Gao:2017kkx}. Although we obtained
qualitatively similar results at NNLO, we were not able to perform a
high-accuracy one-to-one comparison, since the source code for the
result of refs~\cite{Berger:2016inr,Gao:2017kkx} is not available. We
look forward to more dedicated benchmarking in the future, perhaps
within the framework of the PDF4LHC working group. As an alternative,
we once again compared our hadronic results against the asymptotic
expansions~\cite{Blumlein:2011zu,Blumlein:2014fqa}. For these
comparisons and for all the hadronic results shown in this section, we
used the \texttt{NNPDF40\_nnlo\_pch\_as\_01180\_nf\_3} PDFs
set~\cite{NNPDF:2021njg}, and set $m_q=1.5~{\rm GeV}$. Unless
specified otherwise, we set both the factorisation and renormalisation
scales to $\mu_r=\mu_f=Q$ for convenience.

Before presenting our results, we briefly comment on the quark
luminosities that enter our predictions. At this stage, we also
reinstate the full CKM dependence.\footnote{For our numerical
results, we use $|V_{ud}| = |V_{cs}| = \cos\theta_c$, $|V_{cd}| =
|V_{us}| = \sin\theta_c$, with $\cos\theta_c=0.97462$. In the text, we
leave the dependence on all the individual CKM matrix elements
explicit.} We introduce the notation
\begin{equation}
  \mathcal{F}_{i}^{W^\pm,(m),[n]} \equiv \sum_{a} \Bigl[
    \mathcal{L}_a^{W^\pm} \otimes c_{i;a}^{(m),[n]} \Bigr](x_{r,n}),
\end{equation}
where we stress that the convolution is in the rescaled $x_{r,n}$
variable, see \cref{sec:defs}. Here, $\mathcal L_a$ is a suitable
combination of parton distribution functions, that we will define
below.  We remind the reader that the index $n$ counts the number of
final-state charm and anti-charm quarks.  In terms of the $\mathcal
F_i$, the hadronic coefficient functions read
\begin{equation}
  F_{2,L}^{W^\pm,(m)} = \sum_n \xbj_{r,n} \mathcal F_{2,L}^{W^\pm,(m),[n]},
  \quad\quad
  F_{3}^{W^\pm,(m)} = \sum_n \mathcal F_{3}^{W^\pm,(m),[n]},
\end{equation}
while the full N$^k$LO result is
\begin{equation}
  F_{i}^{W^\pm, {\rm N}^k{\rm LO}} = \sum_{m=0}^{k}
  \left(\frac{\as}{2\pi}\right)^m F_i^{W^\pm,(m)}.
  \label{eq:pertExp}
\end{equation}

At LO, only the (anti) quark channel contributes:
\begin{equation}
  \mathcal F_{2,L}^{W^\pm,(0),[1]} = \LqORqbNSs\otimes c_{2,L;q}^{(0),[1]},
  \quad\quad
  \mathcal F_{3}^{W^\pm,(0),[1]} = \pm\LqORqbNSs\otimes c_{3;q}^{(0),[1]},
\end{equation}
where the $\pm$ in the $i=3$ contribution comes from relations like
\cref{eq:cqb_lo}. The non-singlet strange-quark luminosity is defined
as
\begin{equation}
  \LqNSs = |V_{cs}|^2 s + |V_{cd}|^2 d,
  \label{eq:LqNSs}
\end{equation}
and analogously for the anti-quark one. 
At NLO, the gluon channel starts contributing:
\begin{equation}
  \begin{split}
    &\mathcal F_{2,L}^{W^\pm,(1),[1]} = \LqORqbNSs\otimes c_{2,L;q}^{(1),[1]}
    +\LgHQ\otimes c_{2,L;g}^{(1),[1]},
    \\
    & \mathcal F_{3}^{W^\pm,(1),[1]} = \pm
    \left(\LqORqbNSs\otimes c_{3;q}^{(1),[1]}
    +\LgHQ\otimes c_{3;g}^{(1),[1]}\right),
  \end{split}
  \label{eq:calF_nlo}
\end{equation}
where the ``charmed'' gluon luminosity is defined as
\begin{equation}
  \LgHQ = \left(|V_{cs}|^2+|V_{cd}|^2\right) g.
  \label{eq:LgHQ}
\end{equation}
The $\pm$ sign in front of the gluon contribution in
\cref{eq:calF_nlo} is because in the $W^-$ channel there is an
anti-charm in the final state, see discussion at the end of
\cref{sec:nlo}.

At NNLO, luminosities are different in the $n=0,2$--masses and in the
$n=1,3$--masses cases.  For the former ($n=0,2$) case, we can write
the hadronic result as
\begin{equation}
  \begin{split}
    &\mathcal F_{2,L}^{W^\pm,(2),[0]+[2]} = \left(
    \mathcal L_{q(\bar q),{\rm NS}}^{(d)} +
    \mathcal L_{\bar q(q),{\rm NS}}^{(u)} \right)
    \otimes \left(c_{2,L;q;\nh}^{(2),[0]} + c_{2,L;q}^{(2),[2]}
    \right)
    +\mathcal L^{(u)}_g\otimes c_{2,L;g}^{(2),[0]}
    \\
    &\mathcal F_{3}^{W^\pm,(2),[0]+[2]} = \pm \left(
    \mathcal L_{q(\bar q),{\rm NS}}^{(d)} -
    \mathcal L_{\bar q(q),{\rm NS}}^{(u)} \right)
    \otimes \left(c_{3;q;\nh}^{(2),[0]} + c_{3;q}^{(2),[2]}
    \right),
  \end{split}
  \label{eq:calF02}
\end{equation}
where all the coefficient functions have been discussed in
\cref{sec:ren-nh}. The luminosities read
\begin{equation}
  \begin{gathered}
  \mathcal L_{q,{\rm NS}}^{(d)} = |V_{ud}|^2 d + |V_{us}|^2
  s,\quad\quad \mathcal L_{q,{\rm NS}}^{(u)} = \left(|V_{ud}|^2 +
  |V_{us}|^2 \right) u, \\
  \mathcal L_{g}^{(u)} = \left(|V_{ud}|^2 +
  |V_{us}|^2 \right) g,
  \end{gathered}
  \label{eq:LqNSd}
\end{equation}
and analogously for anti-quarks. For $n=1,3$ final-state charm quarks, much
of the complexity comes from the $c_{i;q}^{(2),[1]}$ term. Indeed, at
NNLO we have to consider non-singlet quark ($s\to c$, ``$q_{\rm
  NS}$'') and antiquark ($\bar s \to c$, ``$\bar q_{{\rm NS}_{\bar q
    q}}$''), as well as pure-singlet (``$q_{\rm PS}$'')
contributions.\footnote{For representative diagrams of each
contribution, see \cref{sec:nnlo_structure}.} We clearly tag them here,
as well as in our ancillary files. We obtain
\begin{equation}
  \begin{split}
    &\mathcal F_{2,L}^{W^\pm,(2),[1]+[3]} =
    \LqORqbNSs \otimes \left(c_{2,L;q_{\rm NS}}^{(2),[1]}+
    c_{2,L;q}^{(2),[3]}\right) +
    \LqbORqNSs \otimes c_{2,L;\bar q_{{\rm NS}_{\bar q q}}}^{(2),[1]}
    +\\
    &\quad\quad\quad\quad\quad\quad\quad\quad
    \mathcal L^{(c)}_{q_{\rm PS}}
    \otimes c_{2,L;q_{\rm PS}}^{(2),[1]}
    +
    \mathcal L^{(c)}_{g}
    \otimes c_{2,L;g}^{(2),[1]},
    \\
    &\mathcal F_{3}^{W^\pm,(2),[1]+[3]} =\pm\bigg[
      \LqORqbNSs \otimes \left(c_{3;q_{\rm NS}}^{(2),[1]}+
    c_{3;q}^{(2),[3]}\right) +
    \LqbORqNSs \otimes c_{3;\bar q_{{\rm NS}_{\bar q q}}}^{(2),[1]}
    +\\
    &\quad\quad\quad\quad\quad\quad\quad\quad\quad
    \mathcal L^{(c)}_{q_{\rm PS}}
    \otimes c_{3;q_{\rm PS}}^{(2),[1]}
    +
    \mathcal L^{(c)}_{g}
    \otimes c_{3;g}^{(2),[1]}\bigg],
  \end{split}
  \label{eq:calF13_split}
\end{equation}
where the pure-singlet quark distribution is defined as
\begin{equation}
  \mathcal L^{(c)}_{q_{\rm PS}} =
  \left(|V_{cd}|^2+|V_{cs}|^2\right)
  \left[
    (u+\bar u) + (d+\bar d) + (s + \bar s)
    \right].
\end{equation}

\begin{figure}
  \centering
  \includegraphics[width=0.32 \textwidth]{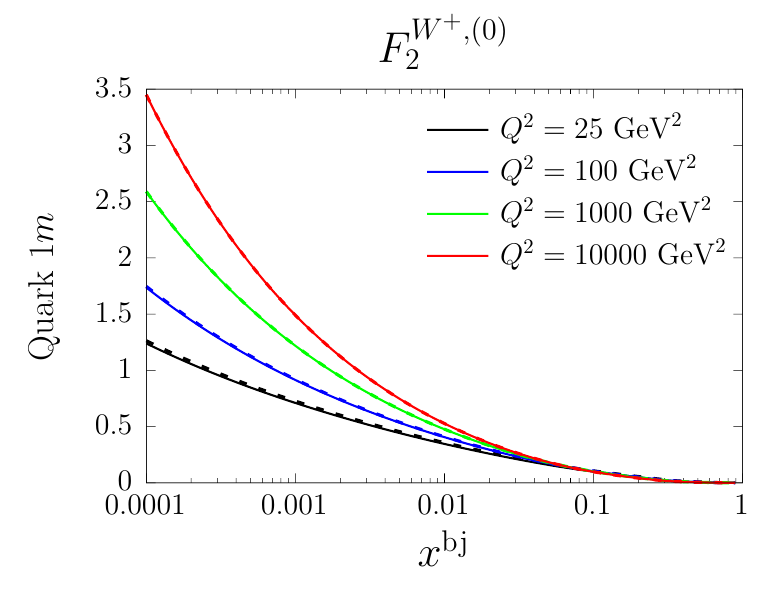}
  \includegraphics[width=0.32 \textwidth]{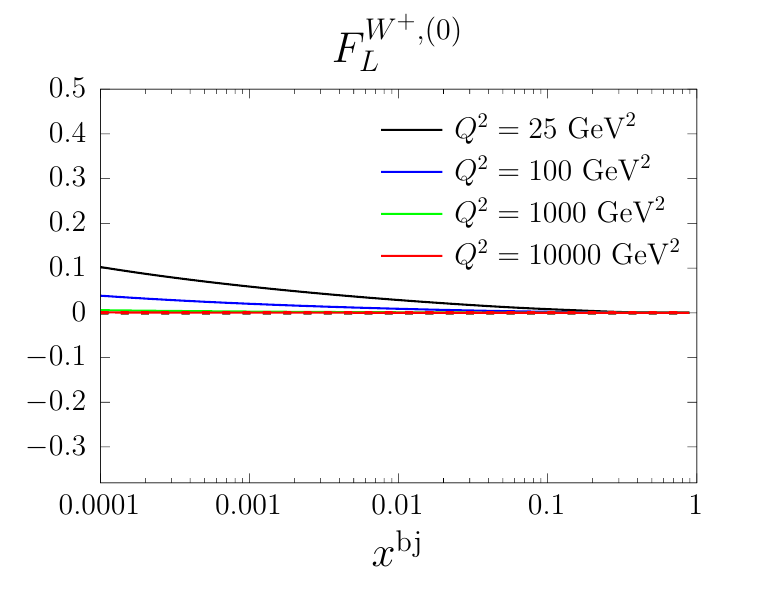}
  \includegraphics[width=0.32 \textwidth]{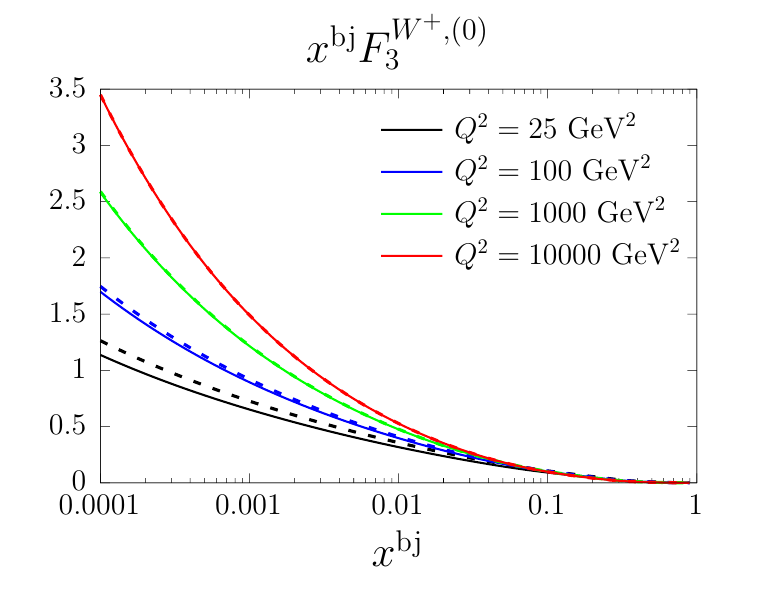}
  \caption{Heavy-flavour LO structure functions $F_{i}^{W^+,(0)}$ for
    different values of $Q^2$ and $\mu_f = \mu_r \equiv \mu = Q$. The
    solid lines are the result with full mass dependence, while the
    dashed lines correspond to the leading-power expansion in
    $\frac{m_q^2}{Q^2}$. The $1m$ label indicates that we consider
    contributions with exactly one massive charm in the final
    state. These contributions are proportional to the $\LqNSs$
    luminosity \cref{eq:LqNSs}. See text for details.}
  \label{fig:lo_comparison}
\end{figure}

\begin{figure}
  \centering
  \includegraphics[width=0.32 \textwidth]{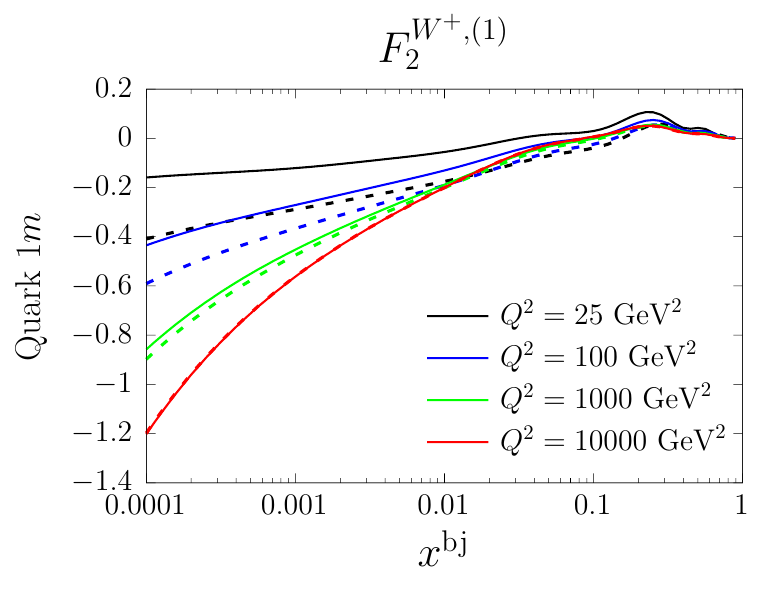}
  \includegraphics[width=0.32 \textwidth]{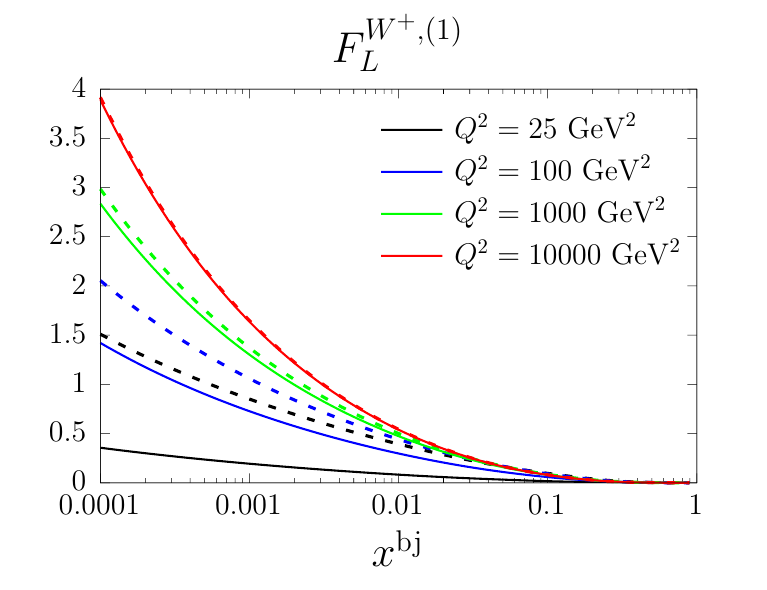}
  \includegraphics[width=0.32 \textwidth]{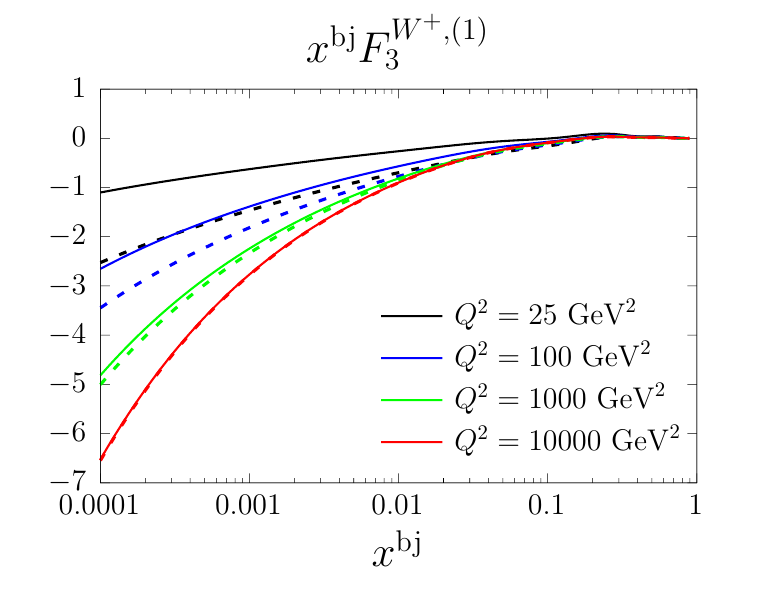}
  \caption{Same as \cref{fig:lo_comparison}, but at NLO. Only the
    quark channel is included in these plots. We remind the reader
    that we denote with ``$(1)$'' the coefficient of $\as/(2\pi)$ in
    the perturbative expansion, see \cref{eq:pertExp}.}
  \label{fig:nloq_comparison}
\end{figure}

\begin{figure}
  \centering
  \includegraphics[width=0.32 \textwidth]{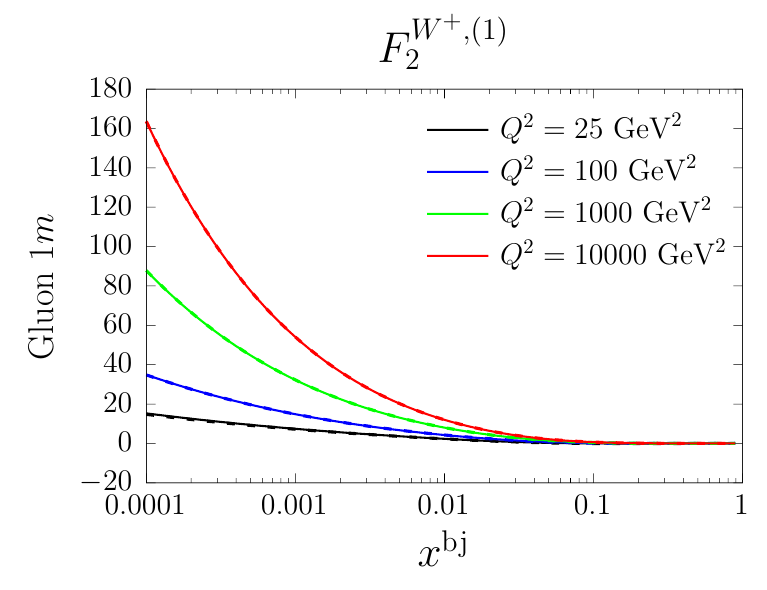}
  \includegraphics[width=0.32 \textwidth]{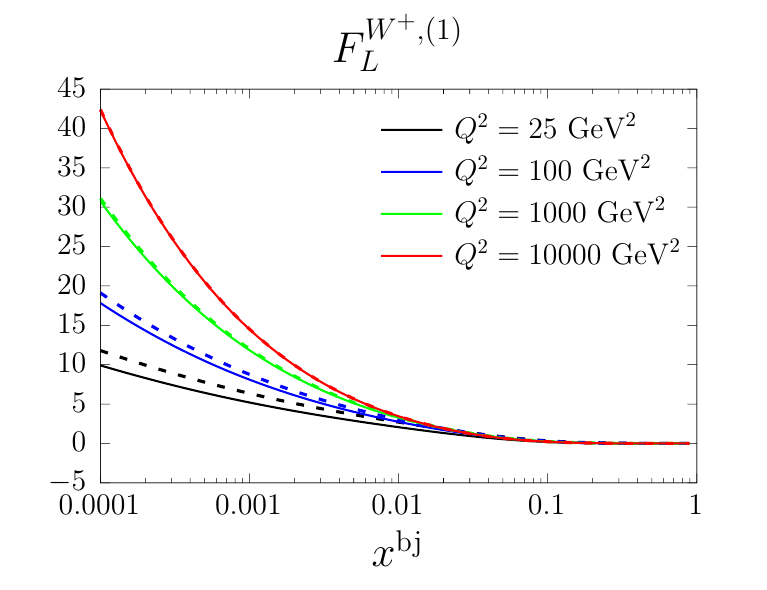}
  \includegraphics[width=0.32 \textwidth]{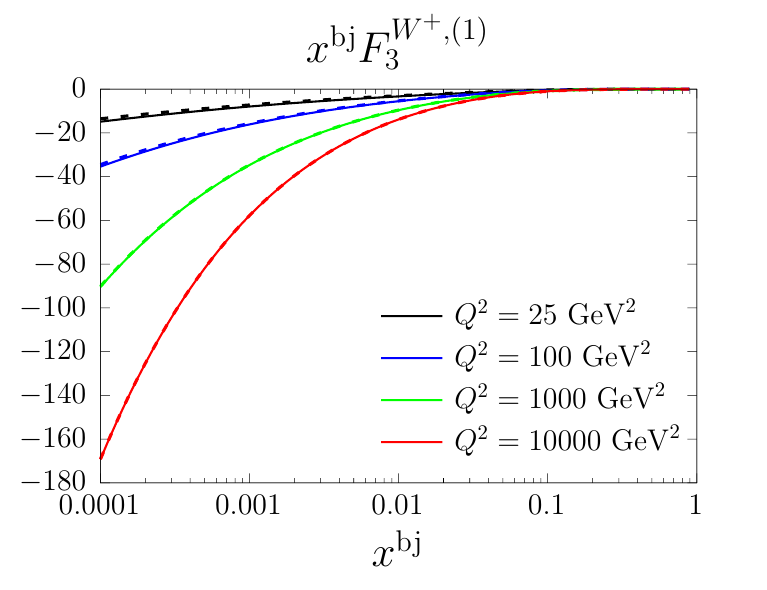}
  \caption{Same as \cref{fig:nloq_comparison}, but for the gluon channel
    instead. The contributions shown here are proportional to the
    $\LgHQ$ luminosity \cref{eq:LgHQ}, see text for details.}
  \label{fig:nlog_comparison}
\end{figure}

\begin{figure}
  \centering
  \includegraphics[width=0.32 \textwidth]{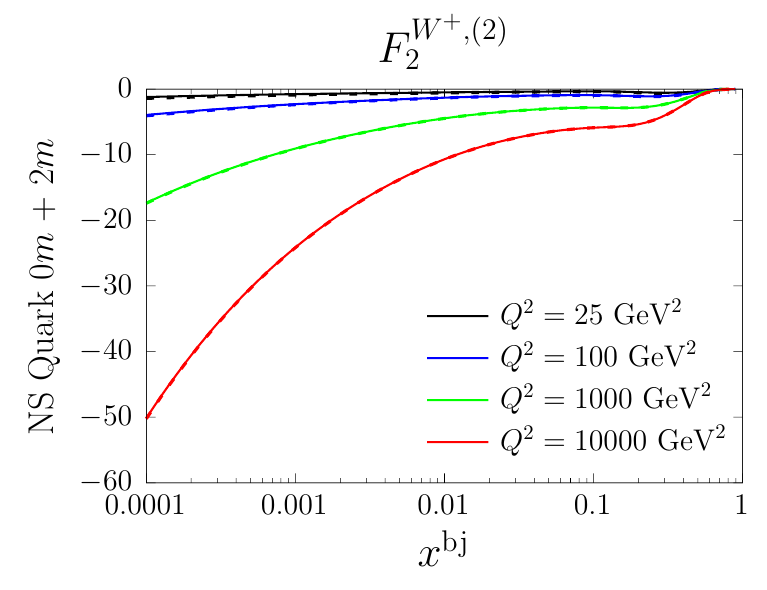}
  \includegraphics[width=0.32 \textwidth]{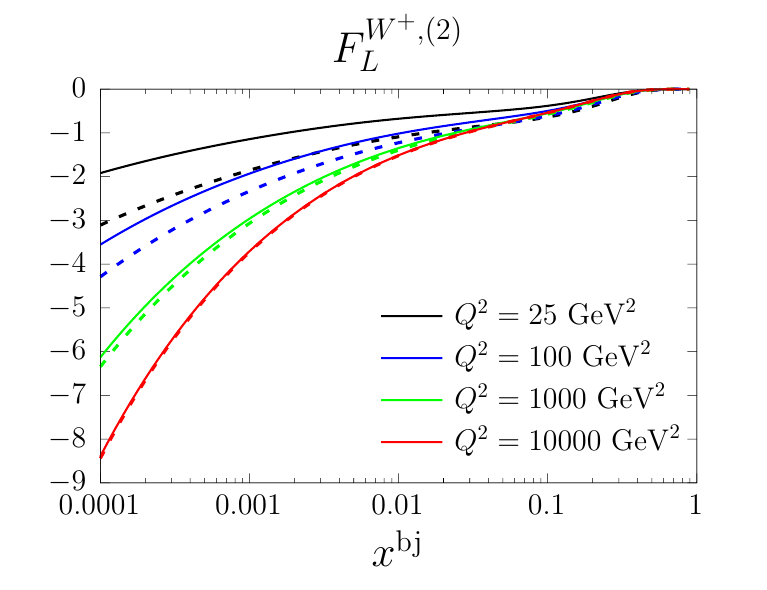}
  \includegraphics[width=0.32 \textwidth]{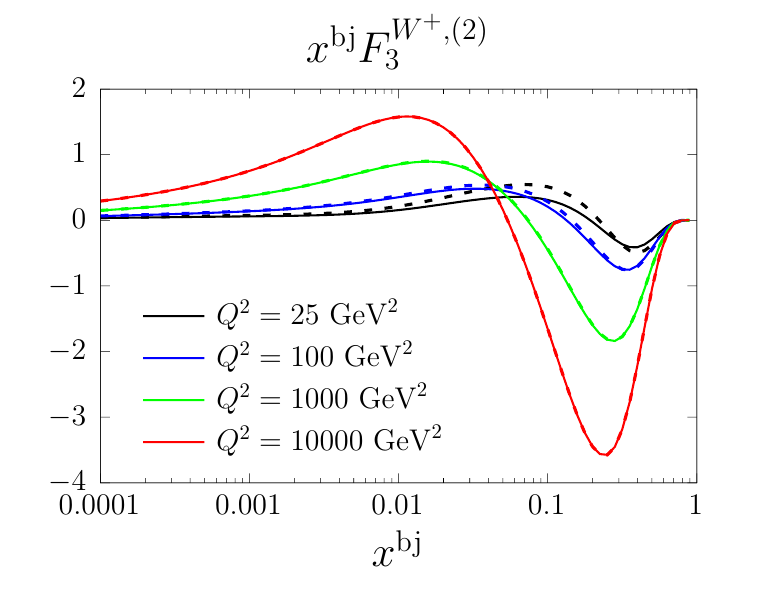}
  \\
  \includegraphics[width=0.32 \textwidth]{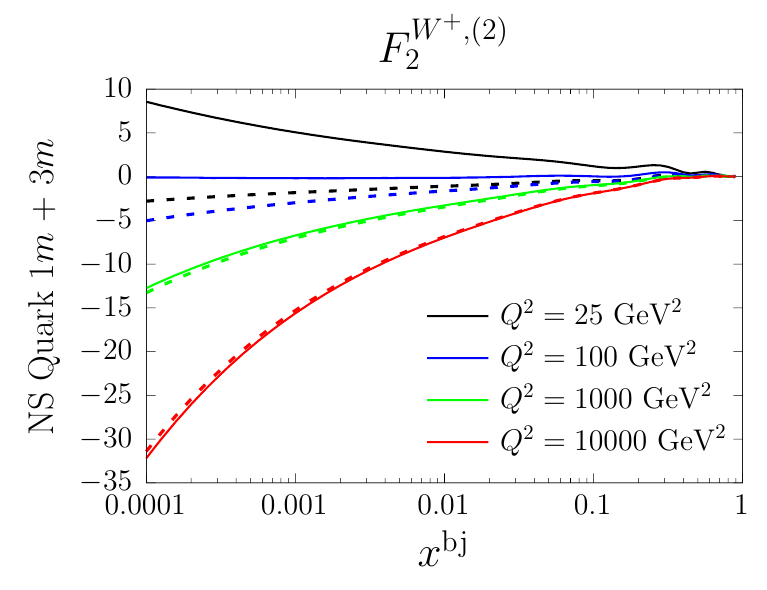}
  \includegraphics[width=0.32 \textwidth]{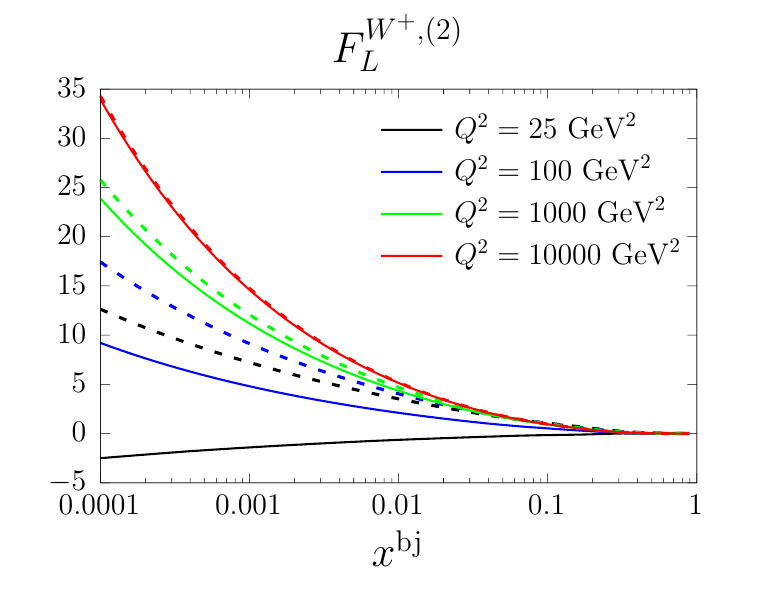}
  \includegraphics[width=0.32 \textwidth]{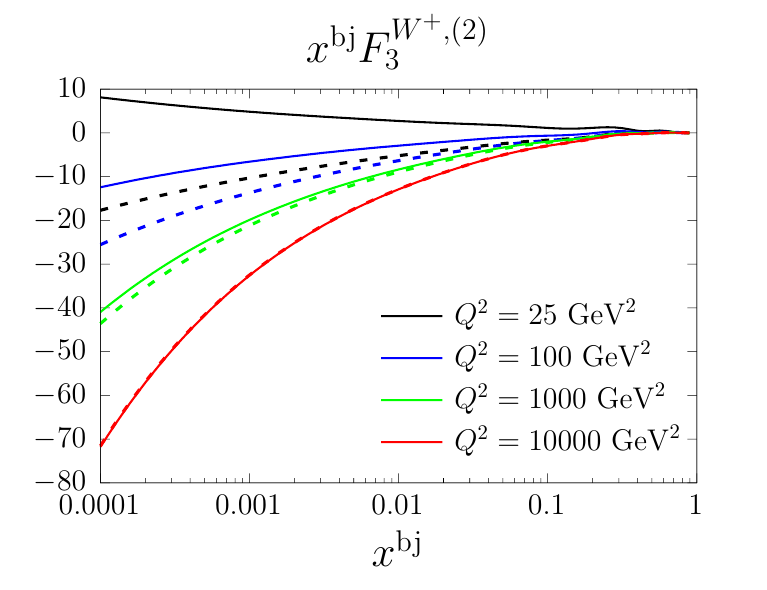}
  \\
  \includegraphics[width=0.32 \textwidth]{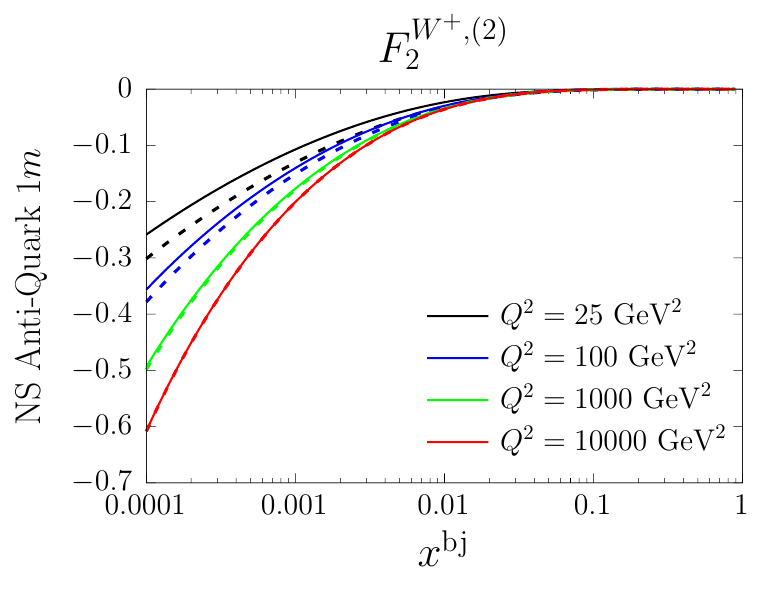}
  \includegraphics[width=0.32 \textwidth]{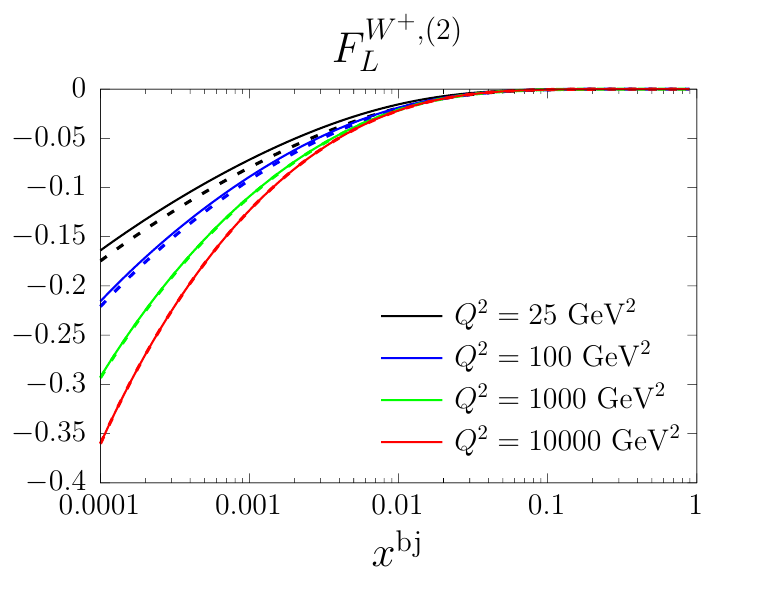}
  \includegraphics[width=0.32 \textwidth]{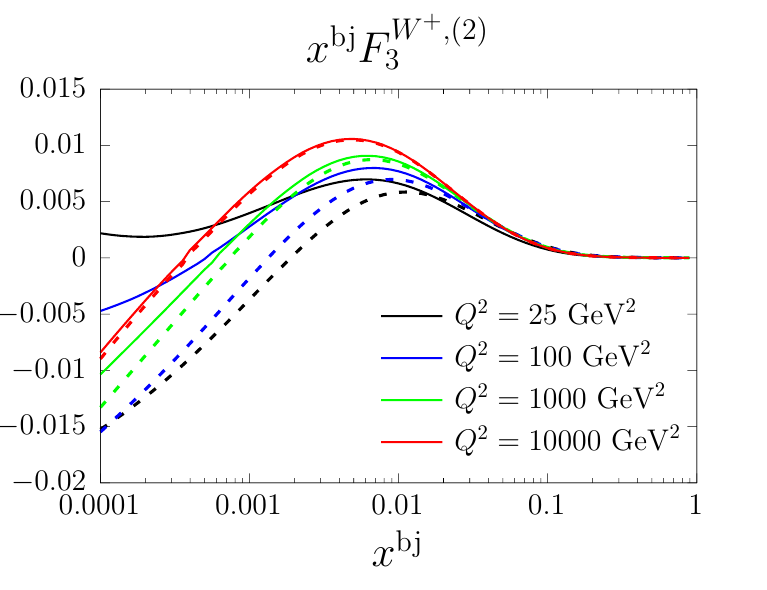}
  \\
  \includegraphics[width=0.32 \textwidth]{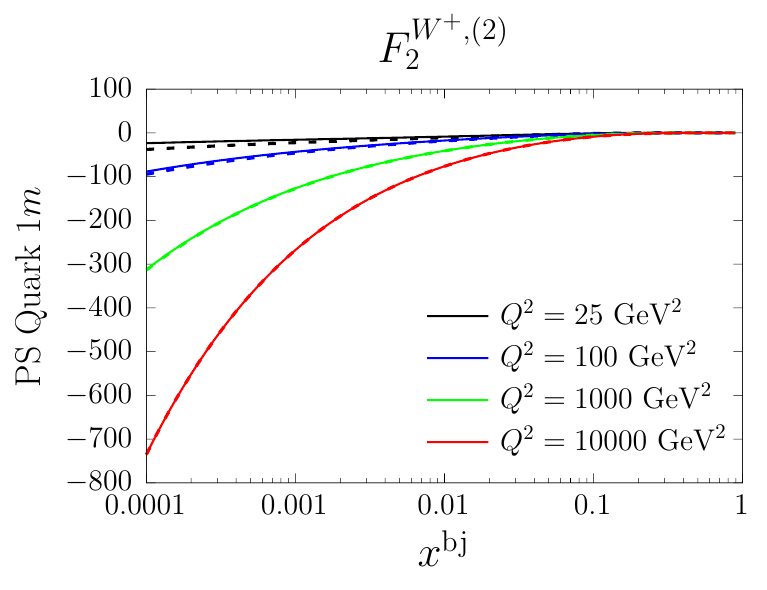}
  \includegraphics[width=0.32 \textwidth]{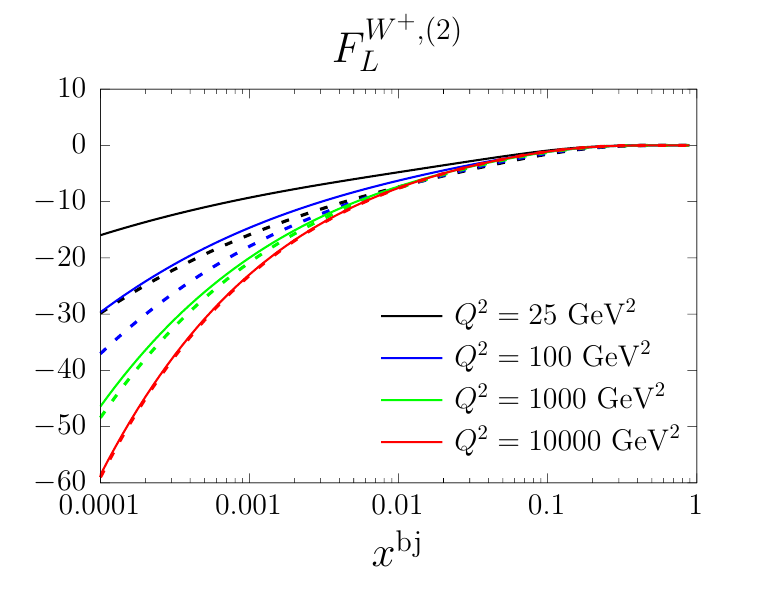}
  \includegraphics[width=0.32 \textwidth]{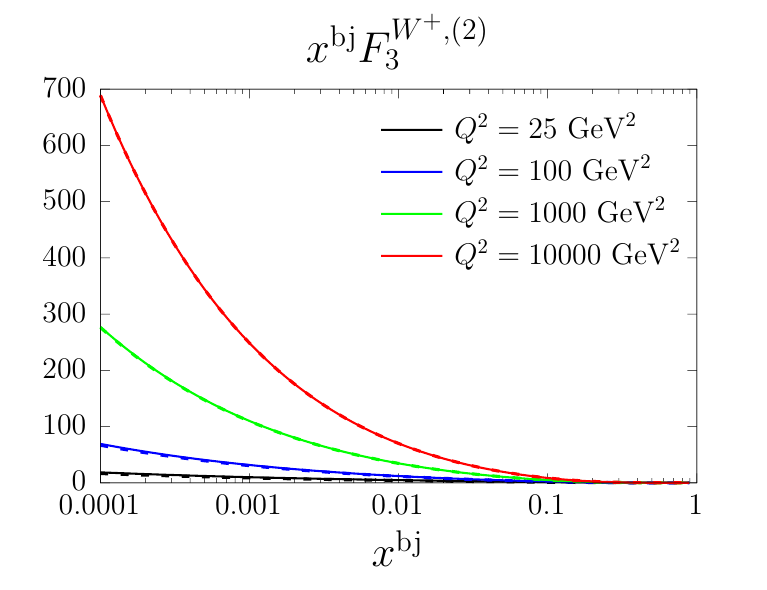}    
  \caption{Same as \cref{fig:lo_comparison}, but at NNLO, for the
    quark channel only. First row: contribution with either no
    charm quarks or a single $c\bar c$ pair in the final state, \ie no
    net charm flavour, see \cref{eq:calF02}. Note that these are
    massive corrections to ``$d\to u$'' CC DIS transitions, and as
    such are proportional to the $\mathcal L_{q,\rm NS}^{(u,d)}$
    luminosities, \cfit \cref{eq:LqNSd}. Second row: genuine NNLO
    corrections to the ``$s\to c$'' process. These entail
    contributions with exactly one charm quark in the final state
    ($1m$) and with an additional $c\bar c$ pair ($3m$).  These
    contributions are proportional to the same luminosity as LO, \cfit
    \cref{eq:LqNSs}. Third row: same as the second row, but for
    ``$\bar s \to c$'' transitions. Fourth row: same as the second
    row, but for the pure-singlet channel, \cfit
    \cref{eq:calF13_split}. We stress that all contributions apart
    from the second row ($1m)$ enter at NNLO for the first time. See
    text for details.}
  \label{fig:nnloq_comparison}
\end{figure}

\begin{figure}
  \centering
  \includegraphics[width=0.32 \textwidth]{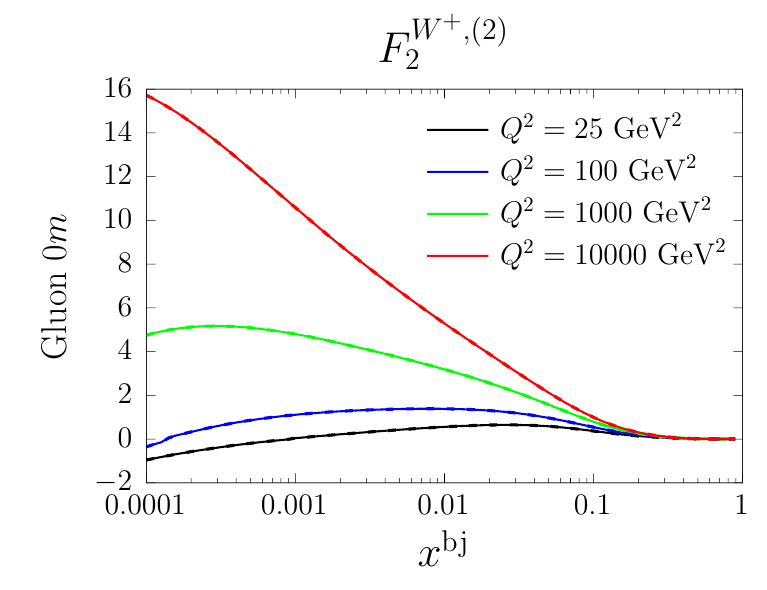}
  \includegraphics[width=0.32 \textwidth]{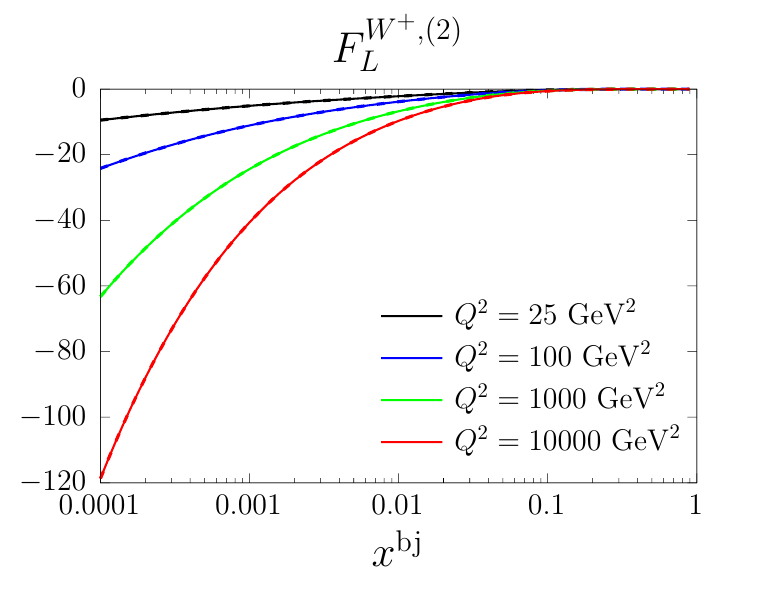}
  \\
  \includegraphics[width=0.32 \textwidth]{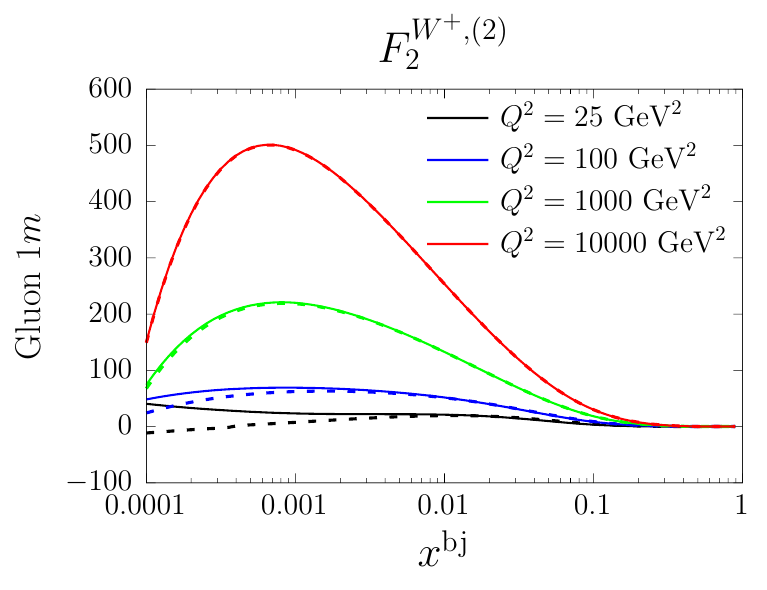}
  \includegraphics[width=0.32 \textwidth]{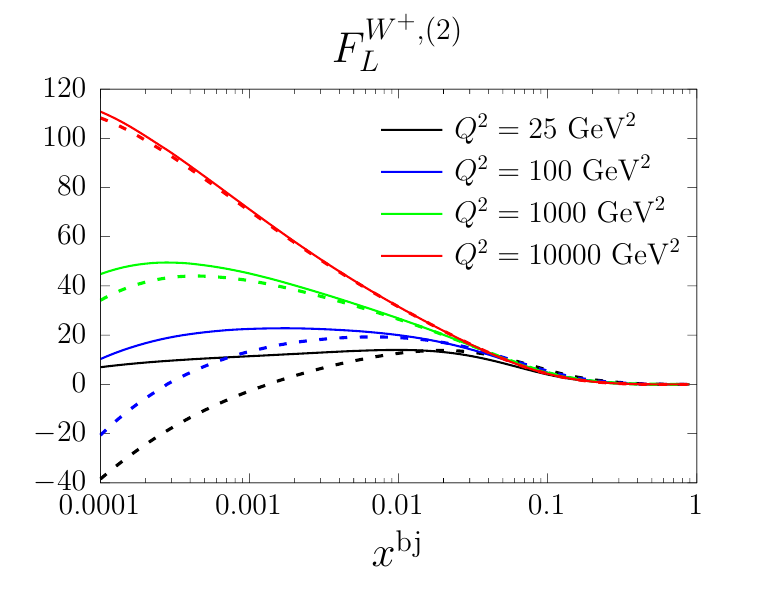}
  \includegraphics[width=0.32 \textwidth]{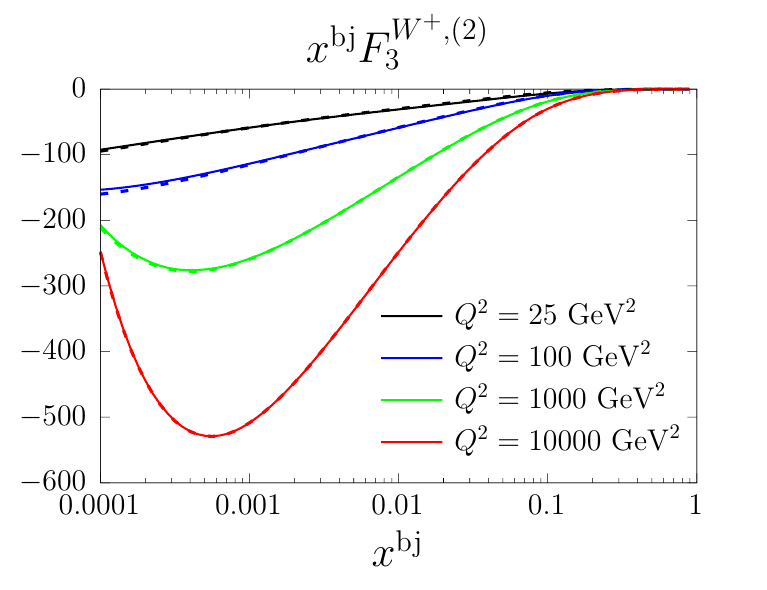}
  \caption{Same as \cref{fig:nnloq_comparison}, but for the gluon channel.
    Note that $F_3$ vanishes for massless transitions, see text for details.}
  \label{fig:nnlog_comparison}
\end{figure}

\begin{figure}
  \centering
  \includegraphics[width=0.32 \textwidth]{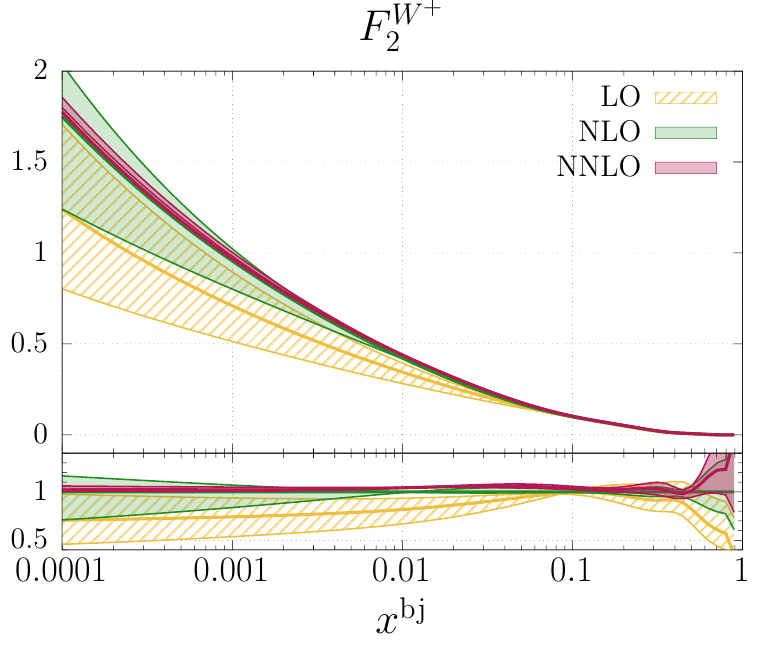}
  \includegraphics[width=0.32 \textwidth]{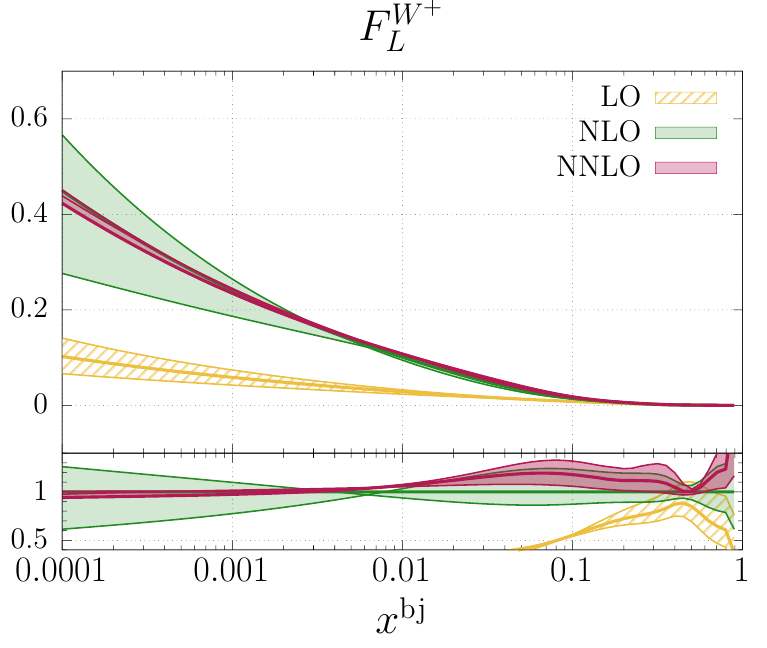}
  \includegraphics[width=0.32 \textwidth]{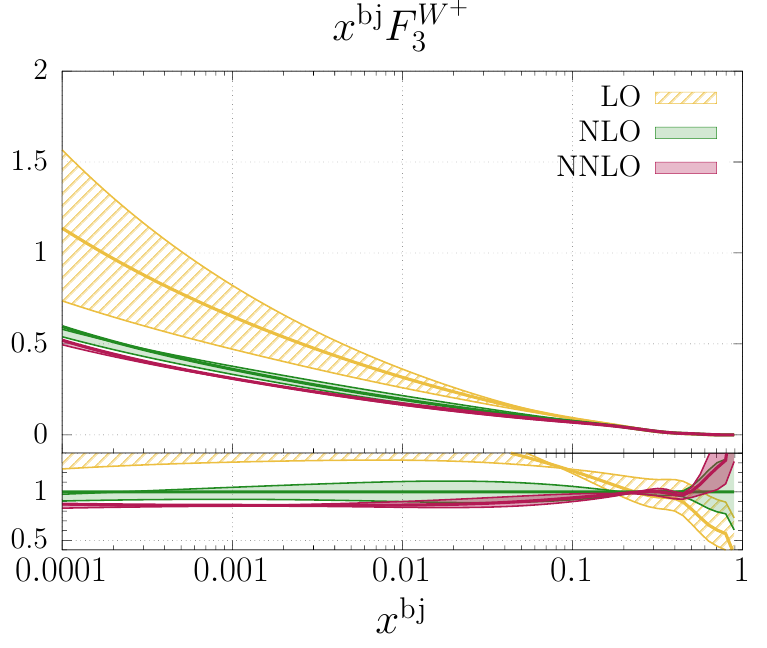}
  \caption{Upper pane: result for charm-induced effects in the
    structure functions for $Q^2=25~{\rm GeV}^2$. The error band is
    obtained by varying the renormalisation and factorisation scales
    by a factor 2 around $\mu = Q$, see text for details. Lower pane:
    ratio to NLO.  Note the different normalisation of the $y$-axis
    for $F_L$.  We consider three active flavours in the proton but
    evolve the strong coupling with four flavours, see text for
    details. We take the specific value of $\alpha_s(\mu)$ from the
    four-flavour set of ref.~\cite{NNPDF:2021njg}.}
  \label{fig:bands_25}
\end{figure}

We now present our numerical results for the comparison with the
asymptotic expansion of refs~\cite{Blumlein:2011zu,Blumlein:2014fqa}.
For definiteness, we focus on $W^+$.
We start from the LO contribution $F_i^{W^+,(0)}$, shown in
\cref{fig:lo_comparison}. We plot the structure functions for four
different values of $Q^2$, ranging from $Q^2=(5~{\rm GeV})^2$, where
the ratio $m_q^2/Q^2$ is arguably not very small, to the asymptotic
region $Q^2 = (100~{\rm GeV})^2$.\footnote{We stress that these
comparisons are intended only as asymptotic checks of the validity of
our calculation.  Indeed, for very large $Q^2$ using predictions with
four active quark flavours is clearly not phenomenologically
appropriate.}  For $F_{2,3}$, we see that for all $Q^2$ the asymptotic
expansion (dashed lines) almost perfectly reproduces the exact result
(solid lines). This is expected, as in these cases the only difference
between leading-power and exact results is the slow rescaling
$x\to x_{r}$, \cfit \cref{sec:defs,sec:lo}. The slightly worse
behaviour for $F_3$ is because we are multiplying it by $\xbj$ and not
(as it is implicitly done for $F_2$) by the rescaled $\xbj_r$, see
\cref{sec:defs}.  The situation is different for $F_L$, since in this
case the leading-power result is zero, \cfit \cref{eq:cq_lo}.

At NLO, the pattern is different, see
\cref{fig:nloq_comparison,fig:nlog_comparison}. Indeed, in this case
we see that in the quark channel at moderate $Q^2$ the exact and
asymptotic results differ significantly, especially in the low-$\xbj$
region. Nevertheless, we observe that for asymptotically large $Q^2$
our result nicely agrees with the leading-power one. In the gluon
channel (\cref{fig:nlog_comparison}), the situation is instead quite
similar to the LO case. For completeness, we remind the reader that at
leading power in the $z=m_q^2/Q^2\to 0$ expansion, there is no
$\ln(z)$ dependence in the quark channel at NLO, while there is for
the gluon channel.

We now move to NNLO. In \cref{fig:nnloq_comparison}, we focus on the
quark channel. In the first row, we show the result with $n=0,2$, \ie
results where there is no net charm flavour in the final state. The
pattern here is very similar to NLO. 
The situation is very different for the $n=1,3$ contributions, \ie the
contributions where there is a net charm flavour in the final
state. Here, in the non-singlet channel there are significant
differences between our exact result and the asymptotic expansion,
both in the quark ($q_{\rm NS}$, second row) and anti-quark ($\bar
q_{{\rm NS}_{\bar q q}}$, third row) cases. However, also here the
exact and asymptotic results converge to each other for $Q^2\gtrsim
1000~{\rm GeV}^2$, which gives us confidence in the correctness of our
result. For the non-singlet quark case, at low $Q^2$ the asymptotic
expansion is qualitatively different from the exact result. We also
note that the anti-quark channel is numerically smaller than the quark,
which can be explained by it being proportional to the sub-leading
colour combination $\Cf(\Ca-2\Cf)$. 
In the pure-singlet channel (fourth row), the differences are
less marked and the asymptotic and exact result converge to each other
for $Q^2\gtrsim 100~{\rm GeV}^2$. The behaviour here is qualitatively
similar to the NLO gluon one \cref{fig:nlog_comparison}, which is to
be expected.

We finally discuss the gluon channel at NNLO,
\cref{fig:nnlog_comparison}. In the first row, we plot contributions
with no charm quarks in the final state. Here, all the impact of the
charm comes through the scheme change discussed in \cref{sec:ren-nh},
which is by construction proportional to $\ln z$ (with our scale
choice). As a consequence, the asymptotic result is exact in this case.
This is evident from our plot, which gives us confidence about our
implementation of the scheme change. Note that we do not show $F_3$,
since it is zero in the massless case.
In the second row we show corrections with exactly one charm quark in
the final state. Interestingly, at least in our setup, the asymptotic
expansion seems to work very well for $F_3$, slightly worse for $F_2$
and significantly worse for $F_L$. However, in all cases, for
large-enough $Q^2$ the exact and asymptotic expansions agree. 

Having validated our coefficient functions, we now present one final
result for the complete structure functions, summed over all the
relevant channels. They are shown in \cref{fig:bands_25}. The error
bands are obtained by varying $\mu_r=\mu_f = \mu \in
\{Q/2,Q,2Q\}$. Results are shown for $Q^2=25~{\rm GeV}^2$, where mass
effects are non-negligible. In these plots, the bulk of the result
comes from contributions with exactly one charm quark in the final
state.
We see that in all cases NLO corrections are significant, as it is
well known. The exceedingly large NLO corrections for $F_L$ are not
surprising, as $F_L$ is zero at LO in the $m_q\to 0$ limit. We see
that including the NNLO results computed here leads to moderate
corrections, but significantly reduces the uncertainty coming from
scale variation. We stress that these results are for illustration
only: we leave a thorough phenomenological investigation of their
impact for CC DIS to future investigations.


\section{Conclusions and outlook}
\label{sec:conclusions}
In this paper, we have documented our analytic computation of NNLO QCD
corrections to heavy-quark production in charged-current DIS.
Compared to results available in the literature, we retain the exact
dependence on the charm-quark mass, which allows for a reliable
description of the intermediate/low $Q^2$ region. We worked in a mixed
scheme where there are only three active quarks in the proton, but the
emission of additional virtual and real massive charm quarks is
allowed.  At NNLO, this led us to consider cases with at most three
massive quarks in the final state. Our results contain both
polylogarithmic and elliptic structures. We expressed the former in
terms of manifestly-real Goncharov polylogarithms, which allow for a
fast and reliable numerical evaluation. For the latter, we have
presented our result in a formal way in terms of Chen iterated
integrals. We have also shown how to express them in terms of
(accelerated) series expansions. In this respect, we have found that it
was enough to consider deep expansions around both the threshold and
the massless limit to cover the whole $Q\gtrsim 5~{\rm GeV}$
kinematic region. All these results are available in computer-readable
format in the ancillary files that accompany our publication. They
have passed non-trivial self-consistency checks. We have also
extensively validated them against approximations available in the
literature, namely against the leading-power results in the asymptotic
$Q^2\gg m_q^2$ limit. 
Together with the NNLO results for initial-state massive
charm~\cite{Kudashkin:2026uaf}, the calculation reported here provides
all missing ingredients for a full NNLO analysis of CC DIS, retaining
exact charm-quark mass dependence.

There are several possible applications of our results.  First, it
would be very interesting to study their phenomenological impact for 
CC DIS. From a more theoretical point of view, they could help
elucidate the structure of asymptotic expansions beyond leading power,
as well as the structure of mass corrections in various kinematic
configurations (\eg around the threshold or in the high-energy
limit). Our framework could be easily extended to also cover the cases
where lepton masses should not be neglected, along the lines of
ref.~\cite{Buonocore:2024pdv}.  Similarly, we envision that it should
be possible to extend our results to the case where both charm and
bottom quarks are present, taking into account their (different)
masses.  In this case, one would also have to extend the validity
region of the series expansions discussed in \cref{sec:MIs}. This will
most likely require introducing more expansion points, which, however,
we expect could be treated with techniques analogous to the ones
discussed here.  Another possible line of investigation would be to
extend our result to less inclusive scenarios, \eg by taking into
account charm fragmentation. We leave these interesting research
avenues to the future.

\acknowledgments We thank Lorenzo Tancredi for many insightful
discussions on several topics relevant for this calculation, and
Christoph Nega for discussions about the canonical basis. We are
grateful to the NNPDF collaboration, and in particular to Felix Hekhorn
and Juan Rojo for encouragement. We also thank Felix Hekhorn, together
with Jun Gao, for help with comparisons against results in the
literature. Finally, we are grateful to Juan Rojo, Felix Hekhorn and
Cesare Mella for comments on the manuscript. This research was
supported by the UKRI Frontier Research Grant programme, underwriting
the ERC Consolidator Grant \textsc{precSM} (UKRI946), by the Science
and Technology Facilities Council (STFC) under grant ST/X000761/1, and
by the Lectureship programme of Wadham College. The work of G.G. was
also supported by the ETH Zurich Postdoctoral Fellowship programme.


\appendix

\section{Master Integrals}
\label{app:appmi}

In this appendix, we provide a list of all the master integrals needed
for the result of this paper. In the figures below, vector bosons,
massless partons and massive quarks are depicted by a wavy line, a
solid line, and a solid double line, respectively.  We use the
topology definitions in \cref{eq:topodefs} and make use of the index
notation defined in \cref{sec:nlo}. At NNLO, we do not draw diagrams
that are simple products of one-loop Feynman integrals, but report
these in the relevant captions for completeness.

\newlength{\GMBlockW} \setlength{\GMBlockW}{3.3 cm}
\newlength{\GMBlockH} \setlength{\GMBlockH}{3.25 cm}
\newlength{\GMGap} \setlength{\GMGap}{0.8em}

\ExplSyntaxOn
\seq_new:N \l__gm_seq

\NewDocumentCommand{\GMBlock}{m m o o}{%
  \begin{minipage}[c][\GMBlockH][c]{\GMBlockW}
    \centering
    \input{#1}
    \vspace{\GMGap}
    \scalebox{.6}{
    \(
    \begin{aligned}  
    &  #2
      \IfValueT{#3}{\\ &  #3}
      \IfValueT{#4}{\\ &  #4}
    \end{aligned}
    \)
    }
  \end{minipage}%
}

\seq_new:N \l__gm_seq_trim
\cs_new:Npn \__gm_item_or_blank:n #1
  {
    \int_compare:nNnTF {#1} <= { \seq_count:N \l__gm_seq }
      { \seq_item:Nn \l__gm_seq {#1} }
      { } 
  }
\ExplSyntaxOff

\setlength{\arrayrulewidth}{0.9pt}

\begin{table}[h]
\centering
\begin{tabular}{|T|T|T|T|}
  \hline
  \GMBlock{./tikz/R/TBoxmc13_101}{\Op{fBox}{1_c,0,1^m_c}}&
  \GMBlock{./tikz/R/TBoxm2c13_111}{\Op{fBox}{1_c,1^m,1^m_c}}&
  \GMBlock{./tikz/V/TBoxcvm_011}{\Op{fBox}{0,1,1^m}}&
  \GMBlock{./tikz/V/TBoxcvm_001}{\Op{fBox}{0,0,1^m}}\\
  \hline
\end{tabular}
\caption{Pre-canonical master integrals relevant for $\mathrm{R}_1$
  and $\mathrm{V}_1$.}
\label{tab:lapr1v1}
\end{table}

\begin{table}[h]
\centering
\begin{tabular}{|T|}
  \hline
  \GMBlock{./tikz/VV/VBm3cv0_1010110}{\Op{VB}{1,0,2^m,0,1,1^m,0}}[\Op{VB}{1,0,2^m,0,2,1^m,0}]
  \\
  \hline
\end{tabular}
\caption{Pre-canonical master integrals relevant for $\mathrm{VV}_0$.
  We do not plot the integrals $\Op{VB}{0,0,1^m,0,0,1^m,0}$ and
  $\Op{VB}{1,0,1^m,0,1,0,0}$, which can be written in terms of
  products of one-loop integrals.}
\label{tab:lapvv0}
\end{table}

\begin{table}[h]
\centering
\begin{tabular}{|T|T|T|}
  \hline
  \GMBlock{./tikz/VV/VBm4cvm_0010110}{\Op{VB}{0,0,1,0,1^m,1,0}}&
  \GMBlock{./tikz/VV/VBm4cvm_0110100}{\Op{VB}{0,1,1,0,1^m,0,0}}[\Op{VB}{0,2,1,0,1^m,0,0}]&
  \GMBlock{./tikz/VV/VBm6cvm_0010110}{\Op{VB}{0,0,1^m,0,1^m,1^m,0}}\\
  \hline
  \GMBlock{./tikz/VV/VBm4cvm_1010110}{\Op{VB}{1,0,1,0,1^m,1,0}}&
  \GMBlock{./tikz/VV/VBm6cvm_0111001}{\Op{VB}{0,1^m,1^m,2,0,0,1}}[\Op{VB}{0,1^m,2^m,1,0,0,1}][\Op{VB}{0,2^m,2^m,1,0,0,1}]&
  \GMBlock{./tikz/VV/VBm6cvm_1010011}{\Op{VB}{1,0,1^m,0,0,2^m,1}}[\Op{VB}{1,0,2^m,0,0,1^m,1}]\\
  \hline
  \GMBlock{./tikz/VV/VBm6cvm_1010110}{\Op{VB}{1,0,1^m,0,2^m,1^m,0}}[\Op{VB}{1,0,2^m,0,1^m,1^m,0}]&
  \GMBlock{./tikz/VV/VBm5cvm_0111101}{\Op{VB}{0,1,1,1,1^m,0,2^m}}[\Op{VB}{0,1,1,1,2^m,0,1^m}]&
  \GMBlock{./tikz/VV/VNPm2cvm_1011111}{\Op{VNP}{1,0,1,1,1^m,1,1^m}}\\
  \hline
\end{tabular}
\caption{Pre-canonical master integrals relevant
  for $\mathrm{VV}_1$.  We do not plot the integrals
  $\Op{VB}{0,0,0,0,2^m,0,2^m}$,
  $\Op{VB}{1,0,0,0,2^m,0,2^m}$, and $\Op{VB}{2,2,0,0,1^m,0,1^m}$,
  which can be written in terms of products of one-loop integrals.}
\label{tab:lapvv1}
\end{table}

\begin{table}[h]
\centering
\begin{tabular}{|T|T|T|}
  \hline
  \GMBlock{./tikz/RV/VBm4c45_0111100}{\Op{VB}{0,1,1,1_c,1^m_c,0,0}}&
  \GMBlock{./tikz/RV/TVBm9c45_0111110}{\Op{VB}{0,1^m,1,1^m_c,1_c,1^m,0}}[\Op{VB}{0,1^m,1,2^m_c,1_c,1^m,0}]&
  \GMBlock{./tikz/RV/TVNPm3c67_0101111}{\Op{VNP}{0,1^m,0,1,1,1^m_c,1_c}}\\
  \hline
  \GMBlock{./tikz/RV/VBm2c45_0011111}{\Op{VB}{0,0,1^m,1_c,1^m_c,1^m,1}}&
  \GMBlock{./tikz/RV/VBm3c67_1010011}{\Op{VB}{1,0,2^m,0,0,1^m_c,1_c}}[\Op{VB}{2,0,1^m,0,0,1^m_c,1_c}][\Op{VB}{1,0,2^m,0,0,2^m_c,1_c}]&
  \GMBlock{./tikz/RV/VBm5c45_0111101}{\Op{VB}{0,1,1,1_c,1^m_c,0,1^m}}[\Op{VB}{0,1,1,1_c,1^m_c,0,2^m}][\Op{VB}{0,1,1,1_c,2^m_c,0,1^m}]\\
  \hline
  \GMBlock{./tikz/RV/VBm6c67_1010111}{\Op{VB}{1,0,1^m,0,1^m,1^m_c,1_c}}[\Op{VB}{1,0,1^m,0,2^m,1^m_c,1_c}][\Op{VB}{1,0,2^m,0,1^m,1^m_c,1_c}]&
  \GMBlock{./tikz/RV/VNPm2c67_0011111}{\Op{VNP}{0,0,1,1,1^m,1_c,1^m_c}}&
  \GMBlock{./tikz/RV/TVNPm3c67_0111111}{\Op{VNP}{0,1^m,1^m,1,1,1^m_c,1_c}}\\
  \hline
  \GMBlock{./tikz/RV/VNPm2c67_1011111}{\Op{VNP}{1,0,1,1,1^m,1_c,1^m_c}}&
  \GMBlock{./tikz/RV/TVBpqm2c15_1111111}{\Op{VBpq}{1_c,1,1,1,1^m_c,1^m,1^m}}&
  \GMBlock{./tikz/RV/TVBpqm2c26_1111111}{\Op{VBpq}{1,1_c,1,1,1^m,1^m_c,1^m}}\\
  \hline
  & \GMBlock{./tikz/RV/VBpqm1c15_1111111}{\Op{VBpq}{1_c,1^m,1^m,1,1^m_c,1,1}} & \\
  \hline
\end{tabular}
\caption{Pre-canonical master integrals relevant for
  $\mathrm{RV}_1$. We do not plot the integrals
  $\Op{VB}{0,1^m,0,1_c,1^m_c,0,0}$, $\Op{VB}{0,0,0,1_c,1^m_c,1,1^m}$,
  $\Op{VB}{0,1,0,1_c,1^m_c,0,1^m}$,
  $\Op{VB}{0,1^m,1^m,0,0,1^m_c,1_c}$,
  $\Op{VB}{0,1^m,0,1_c,1^m_c,1^m,1}$,
  $\Op{VB}{0,1^m,0,1,1^m,1^m_c,1_c}$,
  $\Op{VB}{1,1^m,0,0,1^m,1^m_c,1_c}$ and
  $\Op{VB}{1^m,1^m,0,1^m_c,1_c,1^m,1}$, which can be written in terms
  of products of one-loop integrals.  }
\label{tab:laprv1}
\end{table}

\begin{table}[h]
\centering
\begin{tabular}{|T|T|T|T|}
  \hline
  \GMBlock{./tikz/RR/VBm1c356_0010110}{\Op{VB}{0,0,1_c,0,1^m_c,1_c,0}}[\Op{VB}{0,0,2_c,0,1^m_c,1_c,0}]&
  \GMBlock{./tikz/RR/VBm4c356_1010110}{\Op{VB}{1,0,1_c,0,1^m_c,1_c,0}}&
  \GMBlock{./tikz/RR/HBm1c347_0111101}{\Op{HB}{0,1^m,1^m_c,1_c,1^m,0,1_c}}[\Op{HB}{0,1^m,1^m_c,1_c,1^m,0,2_c}][\Op{HB}{0,1^m,2^m_c,1_c,1^m,0,1_c}]&
  \GMBlock{./tikz/RR/THNPm3c245_0101111}{\Op{HNP}{0,1^m_c,0,1_c,1_c,1^m,1}}\\
  \hline
  \GMBlock{./tikz/RR/TVBm9c347_1011011}{\Op{VB}{1^m,0,1_c,1^m_c,0,1^m,1_c}}[\Op{VB}{1^m,0,1_c,1^m_c,0,2^m,1_c}]&
  \GMBlock{./tikz/RR/TVNPm3c156_1110110}{\Op{VNP}{1_c,1^m,1^m,0,1_c,1^m_c,0}}[\Op{VNP}{1_c,1^m,2^m,0,1_c,1^m_c,0}]&
  \GMBlock{./tikz/RR/VBm2c347_0011111}{\Op{VB}{0,0,1^m_c,1_c,1^m,1^m,1_c}}&
  \GMBlock{./tikz/RR/VBm5c356_1010111}{\Op{VB}{1,0,1_c,0,1^m_c,1_c,1^m}}[\Op{VB}{1,0,1_c,0,1^m_c,1_c,2^m}][\Op{VB}{1,0,1_c,0,2^m_c,1_c,1^m}]\\
  \hline
  \GMBlock{./tikz/RR/VBm7c356_1010110}{\Op{VB}{1^m,0,1^m_c,0,1_c,2_c,0}}[\Op{VB}{1^m,0,2^m_c,0,1_c,1_c,0}][\Op{VB}{2^m,0,2^m_c,0,1_c,1_c,0}]&
  \GMBlock{./tikz/RR/VNPm2c347_0011111}{\Op{VNP}{0,0,1_c,1_c,1^m,1,1^m_c}}&
  \GMBlock{./tikz/RR/HNPm1c245_0111101}{\Op{HNP}{0,1^m_c,1^m,1_c,1_c,0,2}}[\Op{HNP}{0,1^m_c,2^m,1_c,1_c,0,1}]&
  \GMBlock{./tikz/RR/THNPm3c245_0111111}{\Op{HNP}{0,1^m_c,1^m,1_c,1_c,1^m,1}}\\
  \hline
  \GMBlock{./tikz/RR/TVNPm3c347_0111111}{\Op{VNP}{0,1^m,1^m_c,1_c,1,1^m,1_c}}&
  \GMBlock{./tikz/RR/VNPm2c156_1011111}{\Op{VNP}{1_c,0,1,1,1^m_c,1_c,1^m}}&
  \GMBlock{./tikz/RR/VNPm2c347_1011111}{\Op{VNP}{1,0,1_c,1_c,1^m,1,1^m_c}}&
  \GMBlock{./tikz/RR/HBm1c347_1111111}{\Op{HB}{1,1^m,1^m_c,1_c,1^m,1,1_c}}\\
  \hline
  \GMBlock{./tikz/RR/THBm2c347_1111111}{\Op{HB}{1,1,1_c,1_c,1^m,1^m,1^m_c}}&
  \GMBlock{./tikz/RR/TVBpqm2c136_1111111}{\Op{VBpq}{1_c,1,1_c,1,1^m,1^m_c,1^m}}&
  \GMBlock{./tikz/RR/TVBpqm2c235_1111111}{\Op{VBpq}{1,1_c,1_c,1,1^m_c,1^m,1^m}}&
  \GMBlock{./tikz/RR/VBpqm1c136_1111111}{\Op{VBpq}{1_c,1^m,1^m_c,1,1^m,1_c,1}}\\
  \hline
\end{tabular}
\caption{Pre-canonical master integrals relevant for $\mathrm{RR}_1$.}
\label{tab:laprr1}
\end{table}

\begin{table}[h]
\centering
\begin{tabular}{|T|T|}
  \hline
  \GMBlock{./tikz/RR/VBm3c356_0010110}{\Op{VB}{0,0,1^m_c,0,1_c,1^m_c,0}}[\Op{VB}{0,0,2^m_c,0,1_c,1^m_c,0}]&
  \GMBlock{./tikz/RR/VBm3c356_1010110}{\Op{VB}{1,0,2^m_c,0,1_c,1^m_c,0}}[\Op{VB}{1,0,2^m_c,0,2_c,1^m_c,0}]\\
  \hline
\end{tabular}
\caption{Pre-canonical master integrals relevant for $\mathrm{RR}_2$.}
\label{tab:laprr2}
\end{table}

\begin{table}[h]
\centering
\begin{tabular}{|T|T|T|}
  \hline
  \GMBlock{./tikz/RR/VBm2c356_0010110}{\Op{VB}{0,0,1^m_c,0,1^m_c,1^m_c,0}}[\Op{VB}{0,0,2^m_c,0,1^m_c,1^m_c,0}]&
  \GMBlock{./tikz/RR/VBm6c356_1010110}{\Op{VB}{1,0,1^m_c,0,1^m_c,1^m_c,0}}[\Op{VB}{1,0,2^m_c,0,1^m_c,1^m_c,0}]&
  \GMBlock{./tikz/RR/VBm2c356_0011111}{\Op{VB}{0,0,1^m_c,1,1^m_c,1^m_c,1}}\\
  \hline
  \GMBlock{./tikz/RR/VBm6c356_1010111}{\Op{VB}{1,0,1^m_c,0,1^m_c,1^m_c,1}}[\Op{VB}{1,0,1^m_c,0,2^m_c,1^m_c,1}][\Op{VB}{1,0,2^m_c,0,1^m_c,1^m_c,1}]&
  \GMBlock{./tikz/RR/VBpqm1c235_0111101}{\Op{VBpq}{0,1^m_c,1^m_c,1,1^m_c,0,1}}[\Op{VBpq}{0,1^m_c,2^m_c,1,1^m_c,0,1}][\Op{VBpq}{0,2^m_c,1^m_c,1,1^m_c,0,1}]&
  \GMBlock{./tikz/RR/VBpqm1c235_1111111}{\Op{VBpq}{1,1^m_c,1^m_c,1,1^m_c,1,1}}\\
  \hline
\end{tabular}
\caption{Pre-canonical master integrals relevant for $\mathrm{RR}_3$.}
\label{tab:laprr3}
\end{table}

\clearpage

\bibliographystyle{JHEP}
\bibliography{references.bib}

\printindex

\end{document}